\documentclass[onecolumn,amsmath,amssymb,12pt,superscriptaddress,nofootinbib]{revtex4}
\pdfoutput=1

\usepackage[latin1]{inputenc}
\usepackage[english]{babel}
\usepackage{amssymb}
\usepackage{amsmath}
\usepackage{amsthm}
\usepackage[]{graphicx}
\usepackage[]{subfigure}
\usepackage{tensor}
\usepackage{color}
\usepackage{cancel}
\usepackage{setspace}
\usepackage{fancyhdr}
\usepackage[bookmarks,linktocpage, colorlinks=true, plainpages = false, citecolor = blue,  linkcolor=blue, urlcolor = blue, filecolor = blue]{hyperref} 

\newcommand{\smallWidthLeft}{222.5pt}
\newcommand{\smallWidthRight}{212.5pt}
\newcommand{\thirdWidthLeft}{160.33pt}
\newcommand{\thirdWidthRight}{152.33pt}

\newcommand{\fullWidth}{270pt}

\begin{document}

\allowdisplaybreaks
\begin{titlepage}

\title{On the No-Boundary Proposal \\ for \\ Ekpyrotic and Cyclic Cosmologies \vspace{.3in}}

\author{Lorenzo Battarra}
\email[]{lorenzo.battarra@aei.mpg.de}
\author{Jean-Luc Lehners}
\email[]{jlehners@aei.mpg.de}

\affiliation{Max--Planck--Institute for Gravitational Physics (Albert--Einstein--Institute), 14476 Potsdam, Germany}

\begin{abstract}

\vspace{.3in}
\noindent The no-boundary proposal provides a compelling theory for the initial conditions of our universe. We study the implications of such initial conditions for ekpyrotic and cyclic cosmologies. These cosmologies allow for the existence of a new type of ``ekpyrotic instanton'', which describes the creation of a universe in the ekpyrotic contraction phase. Remarkably, we find that the ekpyrotic attractor can explain how the universe became classical. In a cyclic context, in addition to the ekpyrotic instantons there exist de Sitter-like instantons describing the emergence of the universe in the dark energy phase. Our results show that typically the ekpyrotic instantons yield a higher probability. In fact, in a potential energy landscape allowing both inflationary and cyclic cosmologies, the no-boundary proposal implies that the probability for ekpyrotic and cyclic initial conditions is vastly higher than that for inflationary ones.
\end{abstract}
\maketitle

\end{titlepage}

\tableofcontents

\section{Introduction}

In cosmology we always ask: ``and what happened before?''. This inevitably leads to the question of the origin of our universe. In the hot big bang model, the singularity theorems of Hawking and Penrose demonstrate the existence of an initial singularity, given certain assumptions about the energy conditions of matter in the universe \cite{Hawking:1969sw}. Inflationary cosmology does not remove this initial singularity, rather it moves it somewhat further into the past, as a theorem of Borde, Guth and Vilenkin shows \cite{Borde:2001nh}. Even in cyclic cosmologies, initial conditions are typically required: this is because each cycle must grow in size in order to avoid a build-up of entropy (as originally discussed by Tolman \cite{Tolman}) - then, starting from any finite patch, at a finite time in the past its size was smaller than Planckian and initial conditions must be provided\footnote{Assuming an infinite universe from the beginning appears unattractive to us, as this requires an infinite amount of tuning in the infinite past - this can hardly be considered a physical explanation of the state of the universe.}. In all cases, an ignorance of initial conditions is vexing, as everything follows from these initial conditions. For this reason, we will explore the consequences of what is perhaps the most appealing proposal for the initial conditions of the universe, namely the no-boundary proposal of Hartle and Hawking \cite{Hawking:1981gb,Hartle:1983ai}.

The no-boundary proposal is formulated in the path integral approach to quantising gravity, and in fact only makes sense in the semi-classical approximation. Thus, a crucial assumption that we will implicitly have to make is that this semi-classical approximation to quantum gravity is enough to understand the origin of the universe. The no-boundary proposal then posits that in the path integral one should only sum over geometries that are regular in the past, and do not contain any boundary there. The path integral can in fact be evaluated by the saddle point approximation, and the geometries that arise in this manner are so-called ``fuzzy instantons'', i.e. they are {\it complex} four-dimensional geometries interpolating between a final hypersurface and a rounded-off region satisfying appropriate regularity conditions in the ``past''. If the instanton field configuration is real not only on the final hypersurface, but also over a sufficiently large interval to its past, we can say that a real, Lorentzian, classical universe has emerged. Otherwise, if the fields remain complex-valued, the universe remains in a fully quantum state. One of the great merits of the no-boundary proposal is in fact that it shows how and when a classical universe can emerge from a quantum state. In this sense the no-boundary proposal can answer an important, but seldom-asked question, namely why the universe appears classical at all, given that the fundamental laws of nature are quantum laws\footnote{Many papers have discussed another interesting question, namely how cosmological quantum fluctuations can turn into classical temperature fluctuations in the cosmic background radiation \cite{Albrecht:1992kf,Polarski:1995jg,Kiefer:2007zza,Martin:2012ua,Battarra:2013cha}. In all of those studies, the background is however assumed to be classical from the outset.}.

Until recently, the no-boundary proposal has been studied exclusively in the inflationary context \cite{Hartle:2007gi,Hartle:2008ng,Hartle:2010vi,Hertog:2013mra,Hwang:2014vba}. These studies led to the following claims: 

1. Inflation is needed in order to obtain a classical universe. 

2. Inflation must last for longer than a certain minimal number of e-folds (typically a few) in order to obtain a classical universe. 

3. In a potential landscape, plateau models of inflation are preferred over power-law potentials. 

4. A small number of e-folds is generically preferred - however, if in addition one rewards long phases of inflation by weighting their probability by the physical volume that they generate, the prediction is that inflation should begin right at the onset of the (slow-roll) eternal inflationary regime. 

In the present paper, we will review most of these results and then go on to study the no-boundary proposal in the context of ekpyrotic and cyclic cosmologies \cite{Khoury:2001wf,Steinhardt:2001st,Lehners:2008vx}. These cosmologies provide interesting alternatives to inflation, as they can solve the standard cosmological puzzles (such as the flatness and horizon puzzles) while generating nearly scale-invariant and nearly-Gaussian density fluctuations during a phase of slow ekpyrotic contraction \cite{Lehners:2013cka,Li:2013hga,Qiu:2013eoa,Fertig:2013kwa,Ijjas:2014fja}. In these models, the ekpyrotic phase is followed by a bounce into the expanding radiation dominated phase of the hot big bang model. Cyclic models not only provide a radically different view of the past, but also of the future evolution of our universe. Our findings in the ekpyrotic/cyclic context lead us to update all of the statements enumerated above - in particular, we will show the following:

1. Ekpyrotic instantons exist (we recently presented these solutions in \cite{Battarra:2014xoa}). These describe the creation of a universe which, as it becomes classical, emerges in an ekpyrotic contraction phase. Thus, the ekpyrotic phase can also render the universe classical.

2. Just as in the inflationary context, the general preference is for small magnitudes of the potential $|V(\phi)|$. However, while for inflation this corresponds to a preference for a small number of e-folds, in the ekpyrotic context this corresponds to a preference for a large number of e-folds of contraction.

3. In a cyclic universe including a dark energy phase, ekpyrotic and de Sitter-like instantons co-exist. The latter ones describe the emergence of a classical universe in the dark energy phase, which is then followed by the ekpyrotic contraction phase. The probability for ekpyrotic instantons, where the universe emerges in the ekpyrotic phase, is however vastly higher in general.

4. In a potential energy landscape containing both inflationary and cyclic regions, ekpyrotic and cyclic initial conditions are vastly likelier than inflationary ones. In other words, the no-boundary wavefunction is dominated by ekpyrotic instantons.

\begin{figure}[]
\centering
\begin{minipage}{\fullWidth}
\includegraphics[width=\fullWidth]{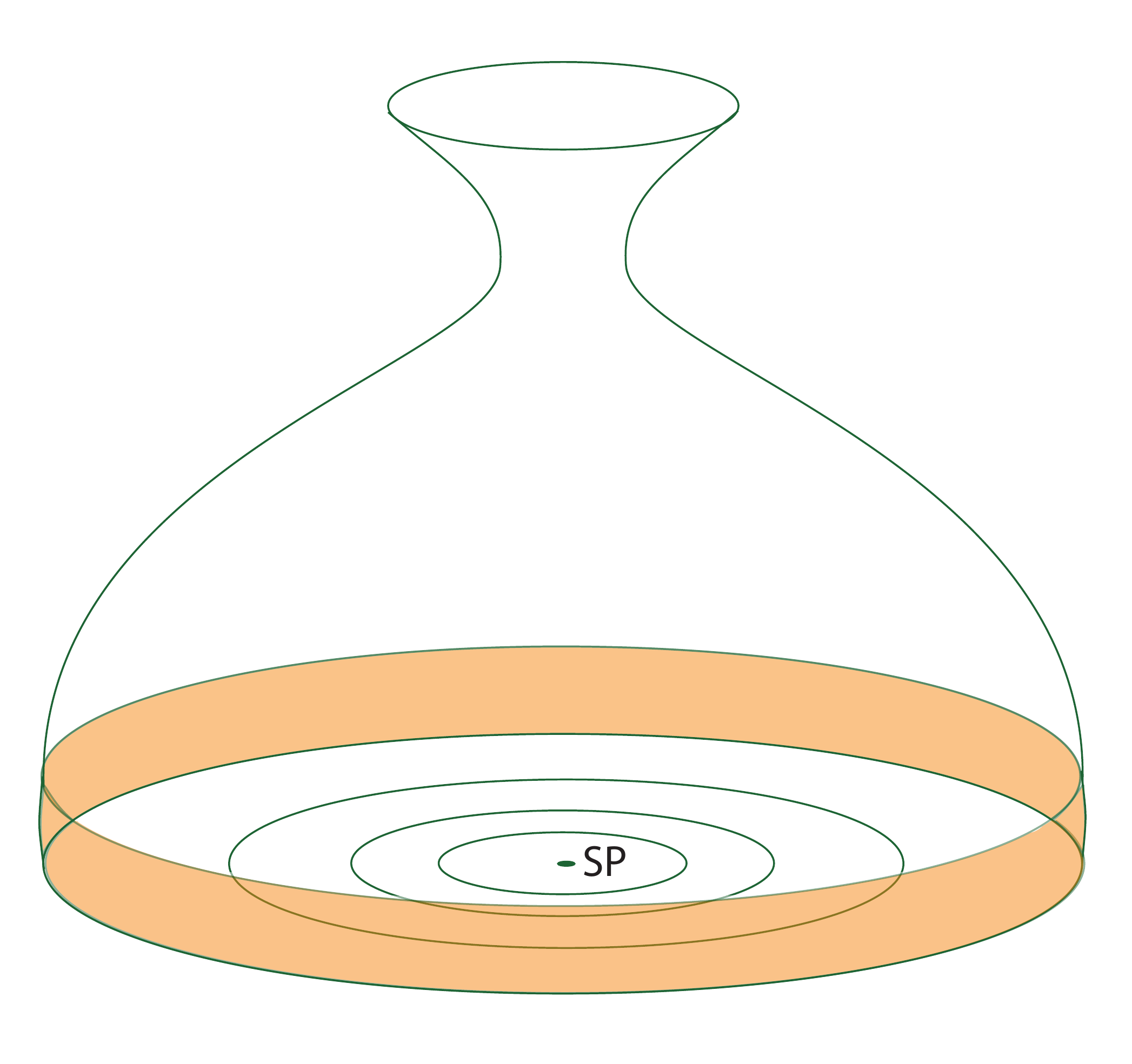}
\end{minipage}%
\caption{\label{fig:carafe} A schematic picture of ekpyrotic instantons: they have a shape reminiscent of a wine decanter - a flat bottom (in fact a portion of Euclidean flat space), followed by a region where both the scale factor and the scalar field are fully complex (the orange shaded region), and finally the ekpyrotic contracting phase during which a real Lorentzian history is reached. For illustrative purposes we have added a bounce at the end of the ekpyrotic phase (at the top of the carafe).}
\end{figure}

Thus, we have found that the no-boundary proposal is not only compatible with ekpyrotic and cyclic cosmologies, but in fact strongly favours the corresponding initial conditions. Note that this result concerns the origin and initial conditions of the universe. The full history relating these initial conditions to the present state of the universe depends on the subsequent dynamics (and may involve various phases of cosmological evolution) -- in this regard, it will be important to extend the present study by adding a bounce to the end of the ekpyrotic phase\footnote{To date, no fully understood model of a cosmic bounce exists yet. This is an active research area -- see for instance \cite{Turok:2004gb,Buchbinder:2007ad,Creminelli:2007aq,Easson:2011zy,Koehn:2013upa,Battarra:2014tga,Battefeld:2014uga,Bars:2011aa,Bars:2013qna,Bars:2014iba}.}. Our work highlights the importance of having an attractor mechanism in order for the universe to become classical - in this respect, the ekpyrotic attractor is equally well suited than the inflationary one. These results significantly widen the scope of the no-boundary proposal, and provide a new view on its implications for the origin of our universe. 

The plan of the paper is as follows: in section \ref{section:review} we will review the no-boundary proposal. We will then review its implications for inflation in section \ref{section:inflation} -- people familiar with the no-boundary proposal may skip this section. In section \ref{section:ekpyrotic} we will present a detailed study of ekpyrotic fuzzy instantons, which interpolate between the no-boundary regularity conditions at the ``bottom'' of the instanton and a real, Lorentzian, ekpyrotic contraction phase. We will put a special emphasis on the way in which the ekpyrotic attractor mechanism causes these instantons to become classical and Lorentzian. As we will show in section \ref{section:cyclic}, in a cyclic universe these ekpyrotic instantons co-exist with de Sitter-like instantons in which the universe emerges in the dark energy phase. We will contrast the properties of the two types of instantons and discuss their relative probabilities. We will close with a discussion in section \ref{section:Discussion}.

\section{Review of the No-Boundary Proposal} \label{section:review}

The no-boundary proposal is formulated in the Euclidean approach to quantum gravity. We will consider gravity minimally to coupled to a single scalar field, so that the Euclidean action reads
\begin{eqnarray}
S_E = -i \, S &=& -i \int d ^4 x  \sqrt{-g} \left( \frac{R}{2 \kappa^2} - \frac{1}{2} g ^{\mu \nu} \partial _{\mu} \phi\, \partial _{\nu} \phi - V( \phi) \right) \\ &=& - \int d ^4 x  \sqrt{g} \left( \frac{R}{2 \kappa^2} - \frac{1}{2} g ^{\mu \nu} \partial _{\mu} \phi\, \partial _{\nu} \phi - V( \phi) \right)\;,
\end{eqnarray}
where the Wick rotation is defined by $ \sqrt{- |g|} = -i\, \sqrt{|g|}$ \footnote{Note that the sign of the Wick rotation is a matter of choice. We have chosen the same Wick rotation as the one applicable in ordinary quantum field theory. The opposite sign choice would lead to a reversed hierarchy of probabilities later on in Eq. \eqref{eq:prob} and thus to different conclusions.}. We will simplify our setting further by restricting ourselves to the metric Ansatz
\begin{eqnarray} \label{eq:metric}
ds ^2 & = & N ^2( \lambda) d \lambda ^2 + a ^2( \lambda) d \Omega _3 ^2 \;,
\end{eqnarray}
where $N(\lambda)$ represents the lapse function, $a$ the scale factor of the universe and $d\Omega_3^2$ is the metric on a unit 3-sphere. With the scalar field $\phi$ depending only on the ``time'' coordinate $\lambda,$ the mini--superspace Euclidean action then becomes
\begin{equation}
S_E = \frac{6 \pi ^2}{ \kappa^2} \int d \lambda N \left( - a \frac{ \dot{a} ^2 }{N ^2} - a + \frac{ \kappa^2 a ^3}{3} \left( \frac{1}{2} \frac{ \dot{ \phi} ^2}{N ^2} + V \right) \right) \;,
\end{equation}
where $\; \dot{} \equiv d/d \lambda$. Allowing complex functions of $ \lambda$, the integral can be interpreted as a contour integral in the complex plane, with $d\tau \equiv N d \lambda$,
\begin{equation} \label{eq:complexAction}
S_E = \frac{ 6 \pi ^2}{ \kappa^2} \int d \tau \left( - a a ^{\prime 2} - a + \frac{ \kappa^2 a ^3}{3} \left( \frac{1}{2} \phi ^{\prime 2} + V \right) \right) \;,
\end{equation}
where $\; ^{\prime} \equiv d/d \tau$. The invariance of the action functional with respect to changes of the complex contour generalises Euclidean time--reparameterisation invariance and imposes the constraint equation (Friedmann equation)
\begin{equation}
a ^{\prime 2}  =  1 + \frac{ \kappa^2 a ^2}{3} \left( \frac{1}{2} \phi ^{\prime 2} - V \right) \;.
\end{equation}
The field equations read
\begin{eqnarray}
\phi '' + 3 \frac{ a'}{a} \phi' - V_{, \phi} & = & 0 \;,\\
a'' + \frac{ \kappa^2 a}{3} \left( \phi ^{\prime 2} + V \right) & = & 0 \;.
\end{eqnarray}
Using the Friedmann equation, the on--shell action can be simplified to
\begin{equation} \label{eq:onshellAction}
S_E ^{inst} =  \frac{ 4 \pi ^2}{ \kappa^2} \int d \tau\, \left(- 3 a + \kappa^2 a ^3 V \right) \;.
\end{equation}

The no-boundary wavefunction is defined via
\begin{equation}
\Psi( b, \chi) = \int_{\cal{C}} \delta a \delta \phi \, e^{-S_E(a,\phi)}  \;,
\end{equation}
where $b$ and $\chi$ are the (late-time) values of the scale factor and scalar field of interest, and where one only sums over paths (4-geometries) $\cal{C}$ that are regular and rounded-off in the past (this requirement has already partly been taken into account in the mini-superspace approximation by specialising to line elements containing a 3-sphere, see Eq. \eqref{eq:metric}). In practice, the above integral can be evaluated by the saddle point approximation,
\begin{equation}
\Psi( b, \chi)  \sim \sum e^{- S_E(b, \chi)} \;,
\end{equation}
where $S_E(b, \chi)$ is the Euclidean action of a \textit{complex} instanton solution $(a(\tau), \phi( \tau))$ of the action \eqref{eq:complexAction}, satisfying the following:
\begin{itemize}
\item $a(0) = 0$ and the solution is regular there (\textit{no boundary}) - regularity then implies that we must also require $a'(0)=1$ and $\phi'(0)=0.$ Thus, instantons can be labelled by the (generally complex) value of the scalar field $\phi_{SP}$ at the ``South Pole'' $\tau=0.$ 
\item There exists a point $ \tau_f$ in the complex $\tau$ plane where $(a, \phi) = (b, \chi),$ with $b,\chi \in \mathbb{R}$ being the arguments of the wavefunction. The Euclidean action $S_E(b, \chi)$ is evaluated along any path joining $\tau= 0$ to $\tau= \tau_f,$ where the choice of path is irrelevant as long as the instanton presents no singularities/branch cuts in the complex plane.
\end{itemize}

As is evident from the metric Ansatz \eqref{eq:metric}, a classical, Lorentzian universe corresponds to $a$ and $\phi$ taking real values, with $d\tau = i dt$ evolving in the purely imaginary direction (and $t$ being the real, physical time coordinate). In that case, as is clear from Eq. \eqref{eq:complexAction} or \eqref{eq:onshellAction}, only the imaginary part of the action keeps changing, while the real part has reached a constant value. Given that 
\begin{equation}
\Psi^\star \Psi \sim e^{-2 Re(S_E)}, \label{eq:prob}
\end{equation}
this suggests that we can regard $e^{-2 Re(S_E)}$ as the relative (un-normalised) probability density for the particular classical history implied by this instanton. Note that this notion of the emergence of a classical history from a complex instanton fits well with the standard notion of WKB classicality, as the wavefunction $\Psi \sim e^{-S_E}$ approaches a constant/slowly-varying amplitude while its phase keeps varying rapidly. By contrast, when no classical Lorentzian history is reached and the real part of the action keeps changing, no meaningful notion of probability can be defined.

As always, it helps to consider a simple example. The best-known instanton corresponds to the situation where the potential is a positive constant $V=3H^2$, and the real, Lorentzian solution corresponds to de Sitter space with the scalar field being constant. The Euclidean version of de Sitter space is a 4-sphere, and the famous Hawking instanton corresponds to running the contour from $\tau=0$ to $\tau = \pi/(2H)$ first (in Planck units $\kappa=1$), with $a=\frac{1}{H} \sin(H \tau)$. This part of the instanton corresponds to half of the Euclidean de Sitter 4-sphere. One then continues the contour in the imaginary direction by defining $\tau \equiv \pi/(2H) + i t,$ so that along that part of the contour $a = \frac{1}{H} \sin(\pi/2 + i H t) = \frac{1}{H} \cosh(H t).$ Thus one has glued de Sitter space in its closed slicing at the waist of the hyperboloid onto half of the Euclidean 4-sphere. The real part of the Euclidean action only varies along the first, horizontal part of the contour, with 
\begin{equation} \label{eq:deSitter}
S_{E, \,half \, S^4} = - \frac{ 12 \pi ^2}{\kappa^4 V } \;,
\end{equation}
where we have reinstated $\kappa.$ This leads to the well-known formula
\begin{equation}  \label{eq:usualFormula}
\Psi \approx \textrm{exp} \left( \frac{12 \pi ^2}{\kappa^4 V} \right) \;.
\end{equation}
Even though this is a highly simplified context, one important feature of the no-boundary proposal is immediately highlighted: small values of the potential are preferred over large values. As we will see shortly, this remains true when the scalar field is dynamical too.

Before proceeding, let us briefly remark that in order to perform the numerical calculations, it is often useful to re-scale the action. Starting again from the Euclidean action
\begin{equation}
S_E = - \int d ^4 x  \sqrt{g} \left( \frac{R}{2 \kappa^2} - \frac{1}{2} g ^{\mu \nu} \partial _{\mu} \phi\, \partial _{\nu} \phi - V( \phi) \right) \;,
\end{equation}
we can re-scale (for an arbitrary constant $ V_0$)
\begin{equation} \label{eq:scaling}
\phi  \equiv  \kappa ^{-1} \bar{ \phi} \;, \quad
V  \equiv  V_0 \bar{V} \;,\quad
g_{\mu\nu} \equiv  \kappa^{-2} V_0 ^{-1} \bar{ g}_{\mu\nu} \;.
\end{equation}
This leads to
\begin{equation}
S_E = -\frac{1}{ \kappa^4 V_0} \int d ^4x \sqrt{ \bar{ g}} \left( \frac{ \bar{R}}{2} - \frac{1}{2} \bar{g} ^{\mu \nu} \partial _{\mu} \bar{ \phi} \partial _{\nu} \bar{ \phi} - \bar{V} \right) \;,
\end{equation}
where the scalar field is now measured in Planck units and the overall scale of the potential appears out front. Incidentally, this re-scaling also explains the functional dependence on the potential in Eqs. \eqref{eq:deSitter} and \eqref{eq:usualFormula}.

\section{Review of Implications for Inflation} \label{section:inflation}

The no-boundary proposal has been extensively studied in the context of inflation, see in particular \cite{Hartle:2007gi,Hartle:2008ng,Hertog:2013mra}. Here we will review the main results, with a special emphasis on the way in which the universe becomes classical. 

\subsection{Attracted to a Classical Universe}

The no-boundary proposal has the potential to explain how a classical universe can arise from an initial quantum state. An interesting question is whether it is always possible to predict a classical universe in this context. Previous studies have concluded that the answer to this question is negative: in general a classical universe is not predicted - rather, special conditions have to be met. More precisely, it has been claimed that an inflationary phase (exceeding a certain minimal number of e-folds) is required in order to obtain a classical universe. In the present section, we will reproduce this result in some detail, as we will be interested in the analogous question for ekpyrotic models later on. 

In analogy with the setting discussed by Hartle, Hawking and Hertog (HHH) \cite{Hartle:2007gi,Hartle:2008ng}, we consider a potential consisting of a cosmological constant added to a mass term for $\phi,$
\begin{equation} \label{eq:potinfl}
V = \frac{ 3 H ^2}{ \kappa^2} \left(1 + \frac{1}{2} m ^2 \phi ^2 \right) \;,
\end{equation}
where $H$ is a constant here. The first goal is to find the classical histories. In the context of inflation, HHH found that the value of the scalar field at the South Pole 
\begin{equation}
\phi_{SP} = |\phi_{SP}| e^{i\theta}
\end{equation} 
must be precisely tuned: for each value of the modulus $|\phi_{SP}|,$ HHH numerically found (at most) one specific value of the angle $\theta$ such that there is a vertical line in the complex $\tau$ plane of the corresponding instanton solution on which $a$ and $\phi$ are approximately real. For the potential \eqref{eq:potinfl} this relies on the fact that, in general, when $y \equiv \textrm{Im}(\tau)$ becomes large the scalar field becomes small and
\begin{eqnarray}
a & \simeq & a_0 e^{- i H \tau} \;,\\ \label{eq:fit1}
\phi & \simeq & \phi _{+} e^{ i \Delta _{+} \tau} + \phi _{-} e^{ i \Delta _{-} \tau} \;,\\
\Delta _{\pm} & = & \frac{3H}{2} \left(1 \pm \sqrt{1 - \frac{4 m ^2}{3\kappa^2}} \right) \;.
\end{eqnarray}
Adjusting $X \equiv \textrm{Re}(\tau)$ one can therefore find a line where $a$ is essentially real:
\begin{equation} \label{eq:realA}
\textrm{Im}(a_0 e^{- i H X}) = 0 \;.
\end{equation}
The asymptotic reality of the scalar field requires
\begin{eqnarray} \label{eq:cond1}
\frac{3H^2}{\kappa^2} m ^2 < 9H^2/4: & & \textrm{Im}( \phi _{-} e^{ i \Delta _{-} X}) = 0 \;, \\ \label{eq:cond2}
\frac{3H^2}{\kappa^2}m ^2 > 9H^2/4: & & \phi _{-} e^{ i \Delta ^{*} X} = \phi _{+} ^{*} e^{ - i \Delta ^{*} X} \;,
\end{eqnarray}
where conventionally
\begin{equation}
\Delta _{+} \equiv \Delta = \frac{3H}{2} \left(1 + i \sqrt{ \frac{4 m ^2}{3\kappa^2}-1} \right) \;.
\end{equation}
If conditions \eqref{eq:cond1} or \eqref{eq:cond2} are satisfied in the respective cases for some values of $|\phi_{SP}|$ and $ \theta$, then along the $ \textrm{Re}(\tau) = X$ line, the real part of the action approaches a constant, because the imaginary parts of $ a$ and $ \phi$ decay fast enough. So for all the values $( b _{y}, \chi_ y) = \textrm{Re}( a( X + i y), \phi( X + i y))$ the no-boundary wavefunction (approximately) has the same module. Given that the imaginary part is very small, the set $( b _{y}, \chi_ y)$ describes a classical trajectory, to which the no-boundary wavefunction attributes a probability.

Note that \eqref{eq:cond1} corresponds to a single real condition, while \eqref{eq:cond2} corresponds to one complex or two real conditions. Given that we have two optimisation parameters $X$ and $ \theta$, and keeping in mind that one of these is used already in Eq. \eqref{eq:realA}, it becomes clear that in addition one needs a dynamical mechanism that allows \eqref{eq:cond2} to be (at least approximately) satisfied. This is the inflationary attractor mechanism, which only works if $|\phi_{SP}|$ is large enough.

\begin{figure}[htbp]
\begin{minipage}{\smallWidthLeft}
\includegraphics[width=\smallWidthRight]{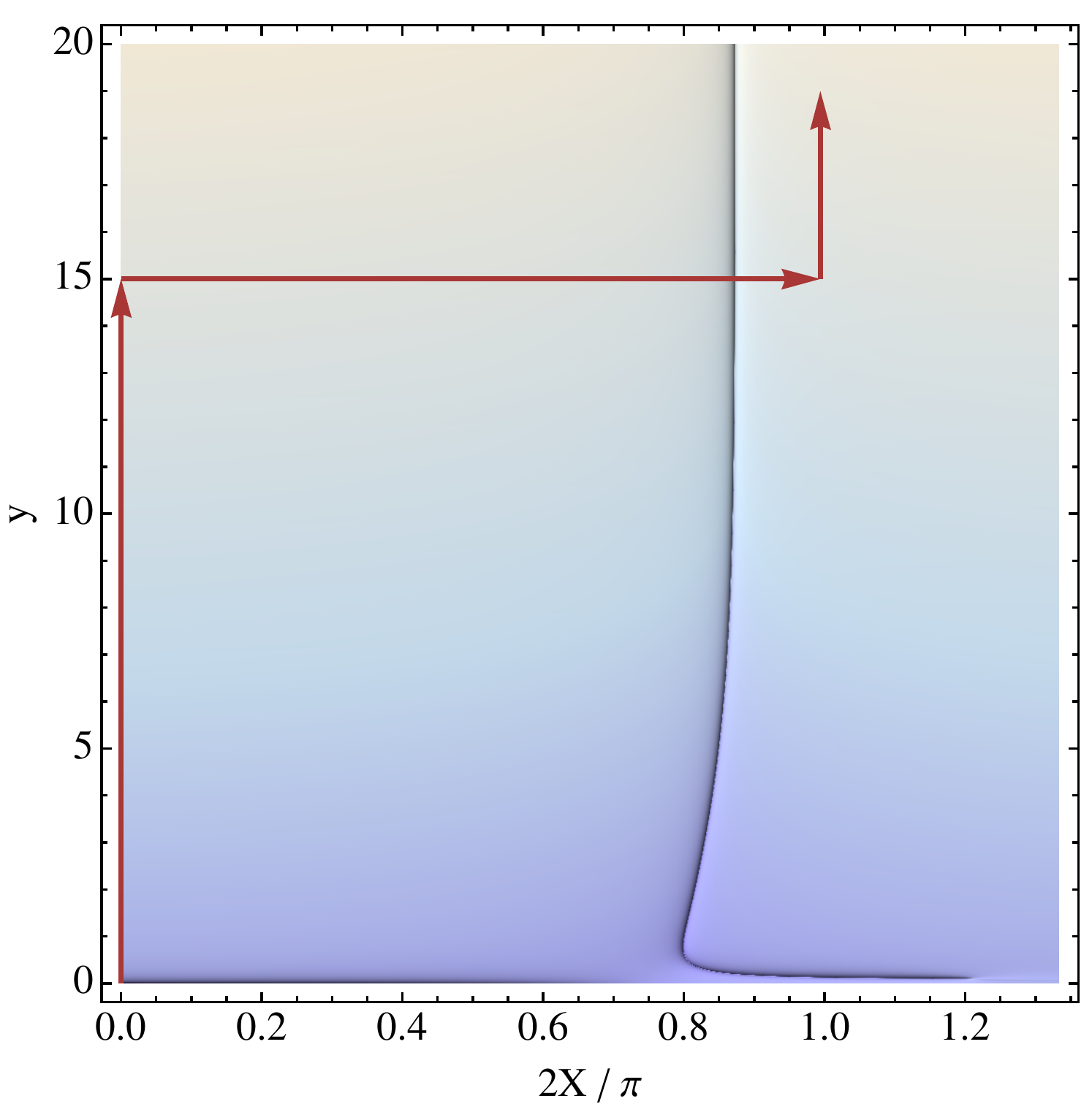}
\end{minipage}%
\begin{minipage}{\smallWidthRight}
\includegraphics[width=\smallWidthRight]{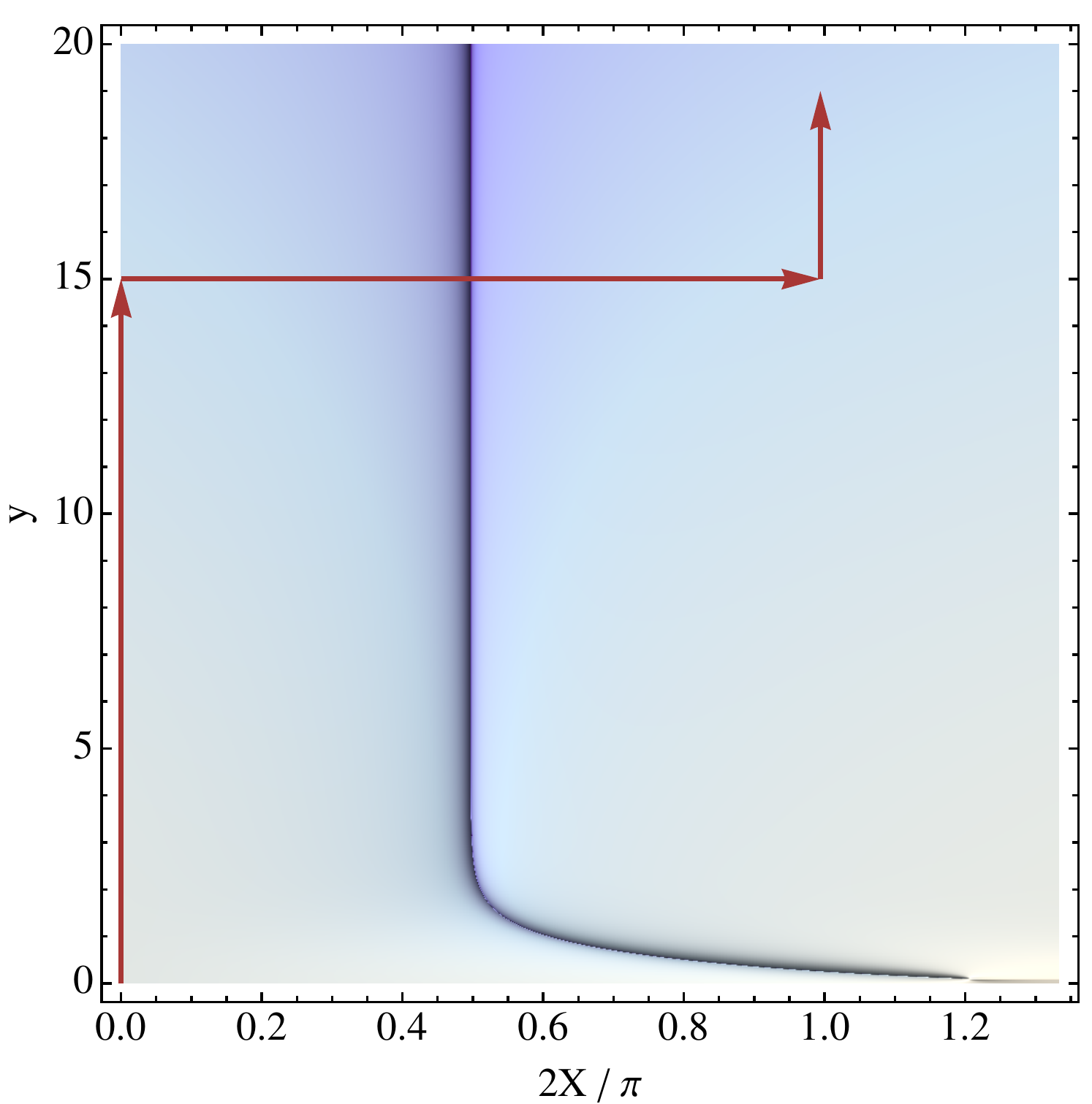}
\end{minipage}%
\caption{ \label{fig:one} An example with $H^2=1/3,$ $m^2 = 3\kappa^2/16$, $ |\phi_{SP}| = 1$, $ \theta = - 1/8.$ The left panel shows a density plot of $ \log{ | \textrm{Im}(a)|}$, so that the dark line corresponds to the locus where $a$ is real. The right panel shows the analogous plot for the scalar field. The real $a$ and $ \phi$ vertical lines are then made to coincide by optimising the value of $ \theta,$ see Figure \ref{fig:two}.} \mbox{}\vspace{1cm} \\
\begin{minipage}{\smallWidthLeft}
\includegraphics[width=\smallWidthRight]{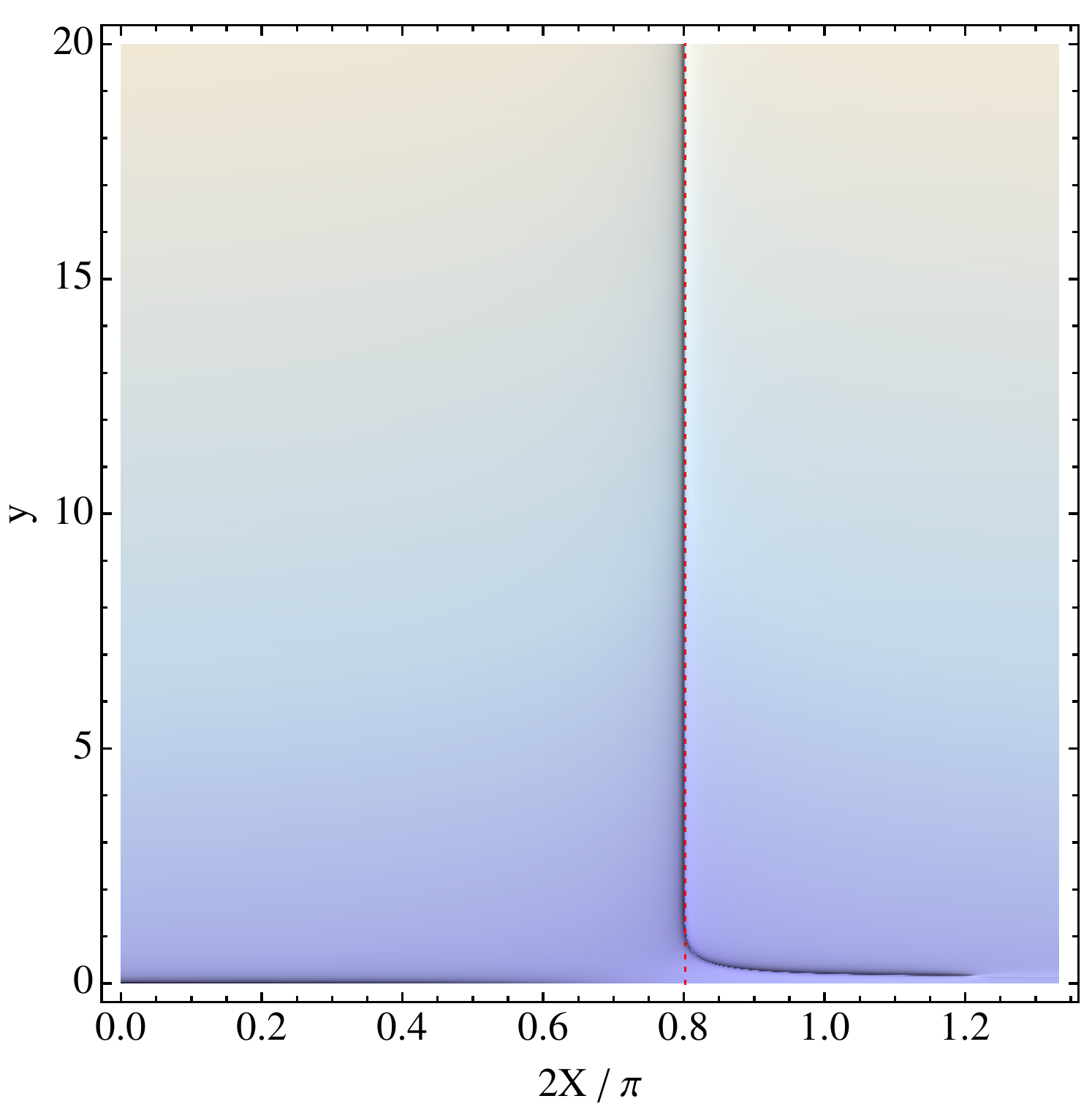}
\end{minipage}%
\begin{minipage}{\smallWidthRight}
\includegraphics[width=\smallWidthRight]{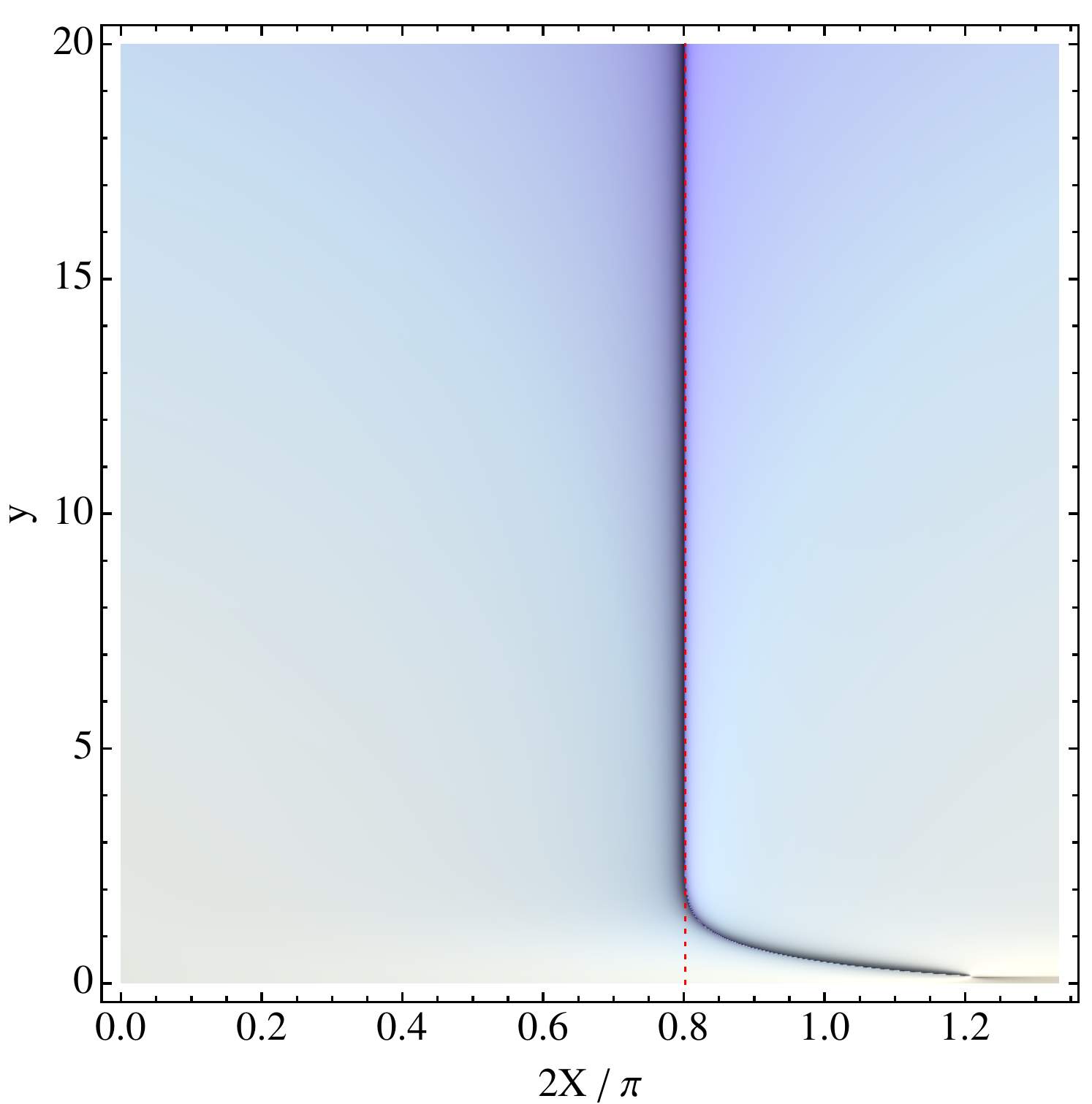}
\end{minipage}%
\caption{ \label{fig:two} Now $ \theta = \theta_{optimal} \approx -0.2896.$ The real $ a$ and real $ \phi$ lines now (asymptotically) coincide and a classical history is obtained.}
\end{figure}

For general values of $|\phi_{SP}|$ and $ \theta$, when $ y$ is large the scalar field becomes very small (though complex-valued) and $a$ is approximately real along a vertical line. Our strategy is to fix $|\phi_{SP}|$ and to solve the field equations along the contour in Figure \ref{fig:one}, with the first segment reaching the point $i\,y_{opt}$ where $y_{opt}$ is some constant. We then fit the behaviour of $a$ and $\phi$ along the last segment of the contour. Finally, we optimise the variables $X$ and $ \theta$ in order to solve the constraints \eqref{eq:realA} and \eqref{eq:cond1} (or either the real or the imaginary part of \eqref{eq:cond2} when $ m ^2 > 3\kappa^2/4$).

When $ m ^2 < 3\kappa^2/4$ the procedure is entirely self-consistent and one finds classical histories for every value of $|\phi_{SP}|$. The only subtle point is that $y_{opt}$ should be large enough for the scalar field to be very small up there. The optimisation then amounts to aligning the $ \textrm{Im}(a) = 0$ and $ \textrm{Im}( \phi) = 0$ lines, which are asymptotically vertical, see Figs. \ref{fig:one} and \ref{fig:two}.

When $m ^2 > 3\kappa^2/4$ one needs the attractor mechanism in order to obtain classical histories. Indeed, the reality of $a$ and $ \phi$ on the $ \textrm{Re}( \tau) = X$ line imposes three real conditions according to (\ref{eq:cond1},\ref{eq:cond2}):
\begin{eqnarray}
\textrm{Im}(a_0 e^{- iHX}) & = & 0 \;,\\
\textrm{Re} \left( \phi _{+} e^{ i \Delta X} - \phi  _{-} ^{*} e^{- i \Delta X} \right) & = & 0 \;,\\
\textrm{Im} \left( \phi _{+} e^{ i \Delta X} - \phi  _{-} ^{*} e^{- i \Delta X} \right) & = & 0 \;.
\end{eqnarray}
In general, we expect these three equations to have multiple solutions which depend on the values of $ \gamma=\tan(\theta)$ and $ |\phi_{SP}|$ \footnote{Here, we momentarily use the notation of HHH that the phase of $\phi_{SP}$ is denoted by $\gamma=\tan(\theta),$ so that one may compare more easily with \cite{Hartle:2008ng}.}:
\begin{eqnarray}
\textrm{Im}(a_0 e^{- iHX}) & = & 0 \quad \Longrightarrow \quad X = X_a(|\phi_{SP}|, \gamma)\;,\\
\textrm{Re} \left( \phi _{+} e^{ i \Delta X} - \phi  _{-} ^{*} e^{- i \Delta X} \right) & = & 0 \quad \Longrightarrow \quad X = X_R( |\phi_{SP}|, \gamma)\;,\\
\textrm{Im} \left( \phi _{+} e^{ i \Delta X} - \phi  _{-} ^{*} e^{- i \Delta X} \right) & = & 0 \quad \Longrightarrow \quad X = X_I( |\phi_{SP}|, \gamma) \;.
\end{eqnarray}
For fixed $ |\phi_{SP}|$, it will generally be possible to optimise the value of $ \gamma$ such that two of them coincide, e.g. $X_a$ and $X_R$:
\begin{equation}
X_a ( |\phi_{SP}|, \gamma_R( |\phi_{SP}|)) = X_R ( |\phi_{SP}|, \gamma_R( |\phi_{SP}|)) \;.
\end{equation}
The inflationary attractor mechanism then ensures that, \textit{for sufficiently large} $|\phi_{SP}|$, one has:
 \begin{equation} \label{eq:attractor}
 X_I ( |\phi_{SP}|, \gamma_R(|\phi_{SP}|)) \simeq X_a = X_R \;.
 \end{equation}
 and the scalar field is approximately real. This is explained by HHH in \cite{Hartle:2008ng} in terms of the slow-roll solution that is reached at large $y$. 

\begin{figure}[t]
\begin{minipage}{\thirdWidthLeft}
\includegraphics[width=\thirdWidthRight]{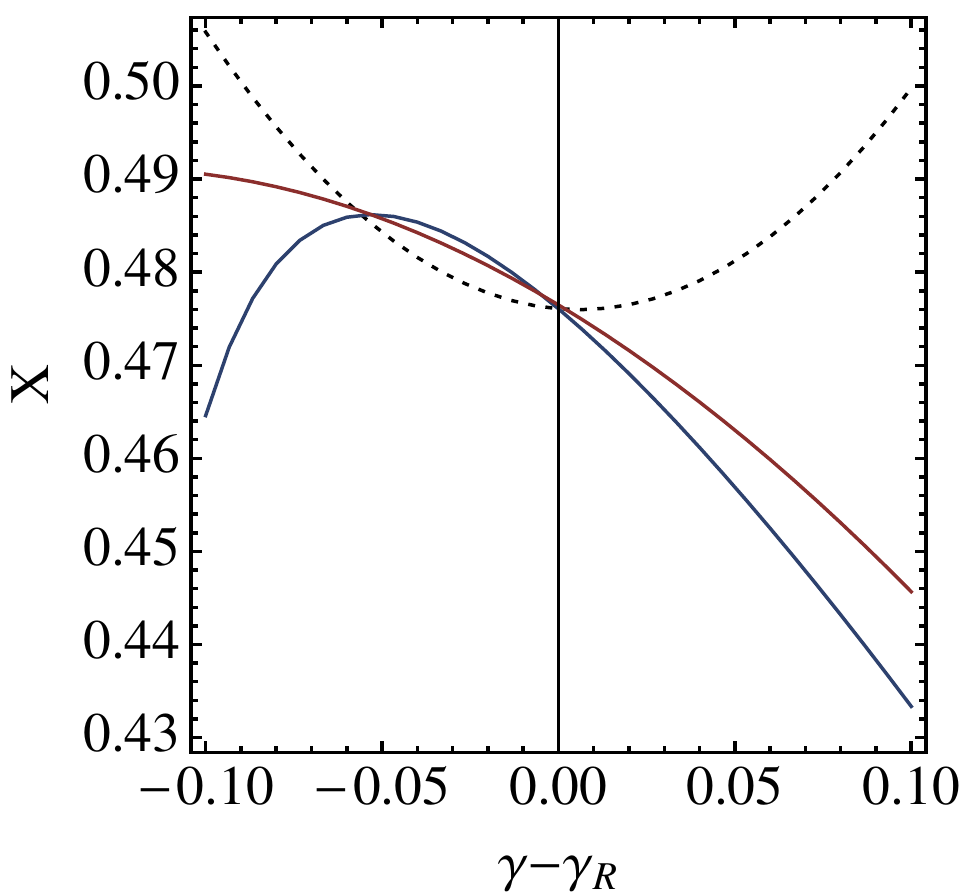}
\end{minipage}%
\begin{minipage}{\thirdWidthLeft}
\includegraphics[width=\thirdWidthRight]{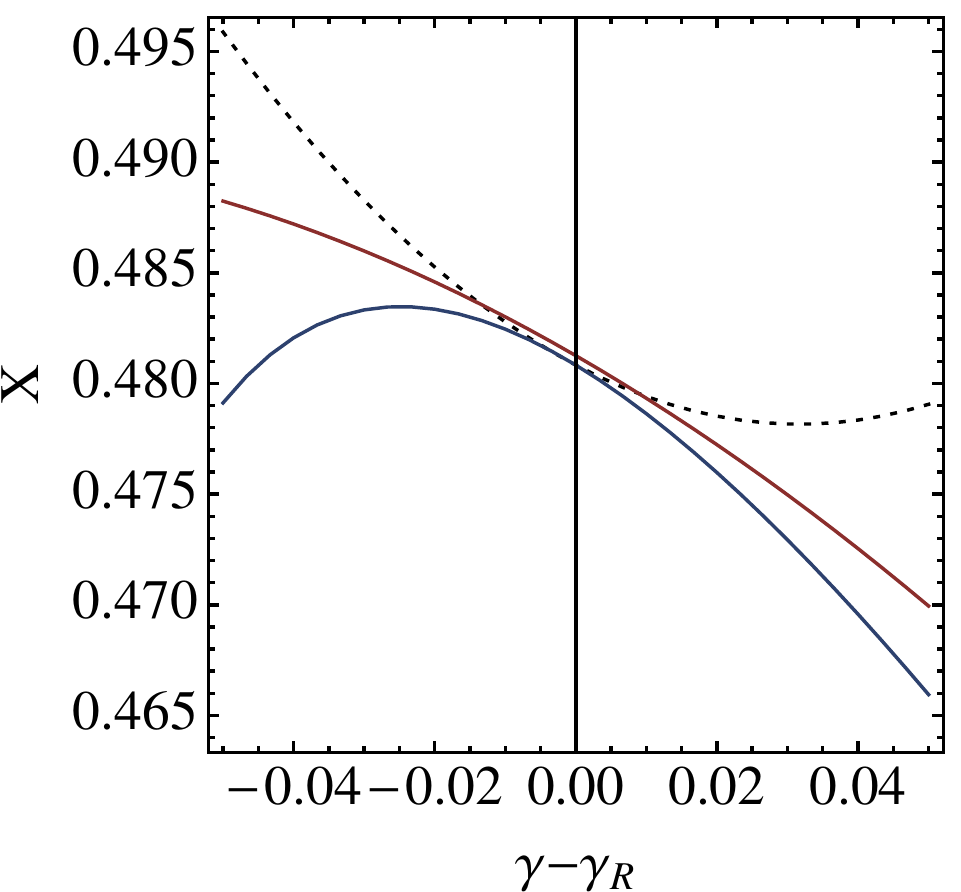}
\end{minipage}%
\begin{minipage}{\thirdWidthRight}
\includegraphics[width=\thirdWidthRight]{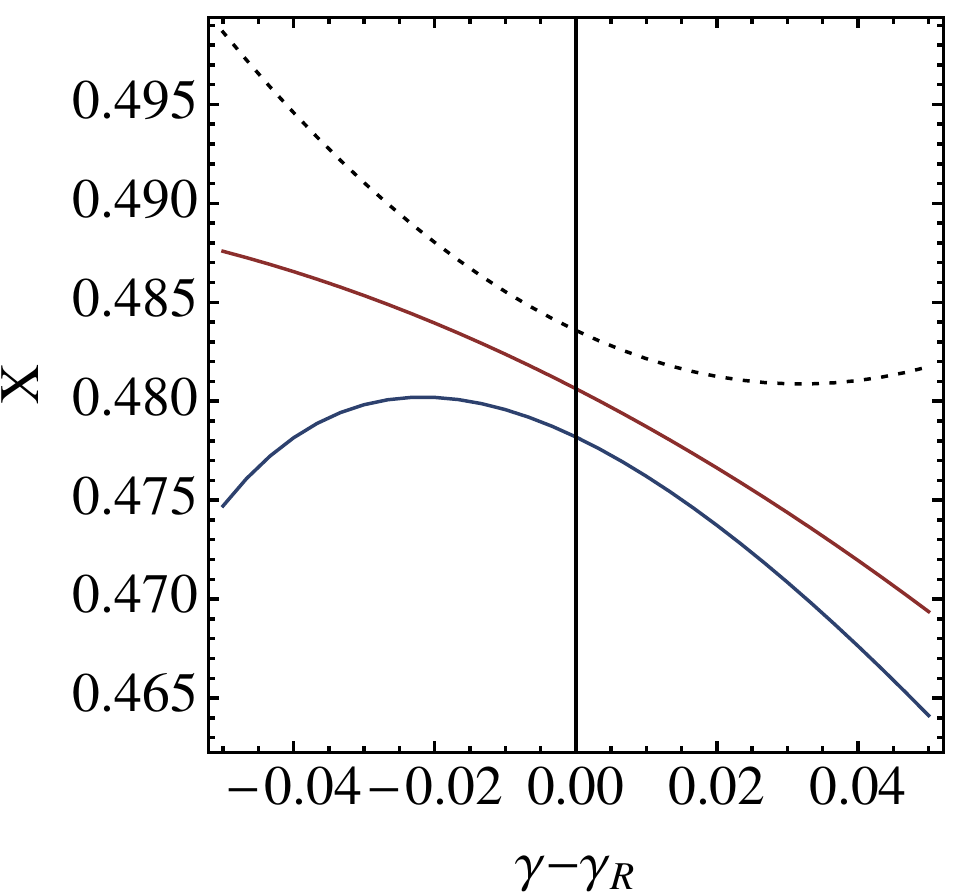}
\end{minipage}%
\caption{ \label{fig:criticality}Values of $X_a$, $X_R$ and $X_I$ for $m ^2 = 3\kappa^2$ and $|\phi_{SP}|$ resp. slightly above, equal and slightly below the critical value $ \phi_c$, conventionally defined as the value where the solution $ \gamma_R$ disappears. Note that at the critical value (middle panel) there is already a noticeable difference between $ \gamma_R$ and $ \gamma_I$.}
\end{figure}

When $ |\phi_{SP}|$ is decreased towards the minimum of the potential, the equality \eqref{eq:attractor} holds less and less precisely. In fact, there is a pretty sharp transition around a critical value $ \phi_{c}$ below which the reality conditions cannot be satisfied anymore, see Fig. \ref{fig:criticality}. Intuitively, this can be understood as saying that when $|\phi_{SP}|$ is too small, the inflationary phase becomes too short for the attractor mechanism to work efficiently. This result indicates that a certain minimal number of e-folds is required in order to render the universe classical. The precise number depends on the detailed shape of the potential, but is typically of order a few \cite{Hartle:2008ng}. We should point out a slight difference between the results of HHH and our treatment here: as is evident from the left panel of Fig. \ref{fig:criticality}, we find that the reality conditions are typically satisfied for two separate values of $\gamma.$ Hence, there exists a second instanton at the same value of $|\phi_{SP}|,$ and which also leads to a classical universe -- however, this second instanton has a higher Euclidean action, and hence is sub-dominant in the path integral.

\subsection{A Toy Inflationary Landscape}

\begin{figure}[]
\centering
\begin{minipage}{\fullWidth}
\includegraphics[width=\fullWidth]{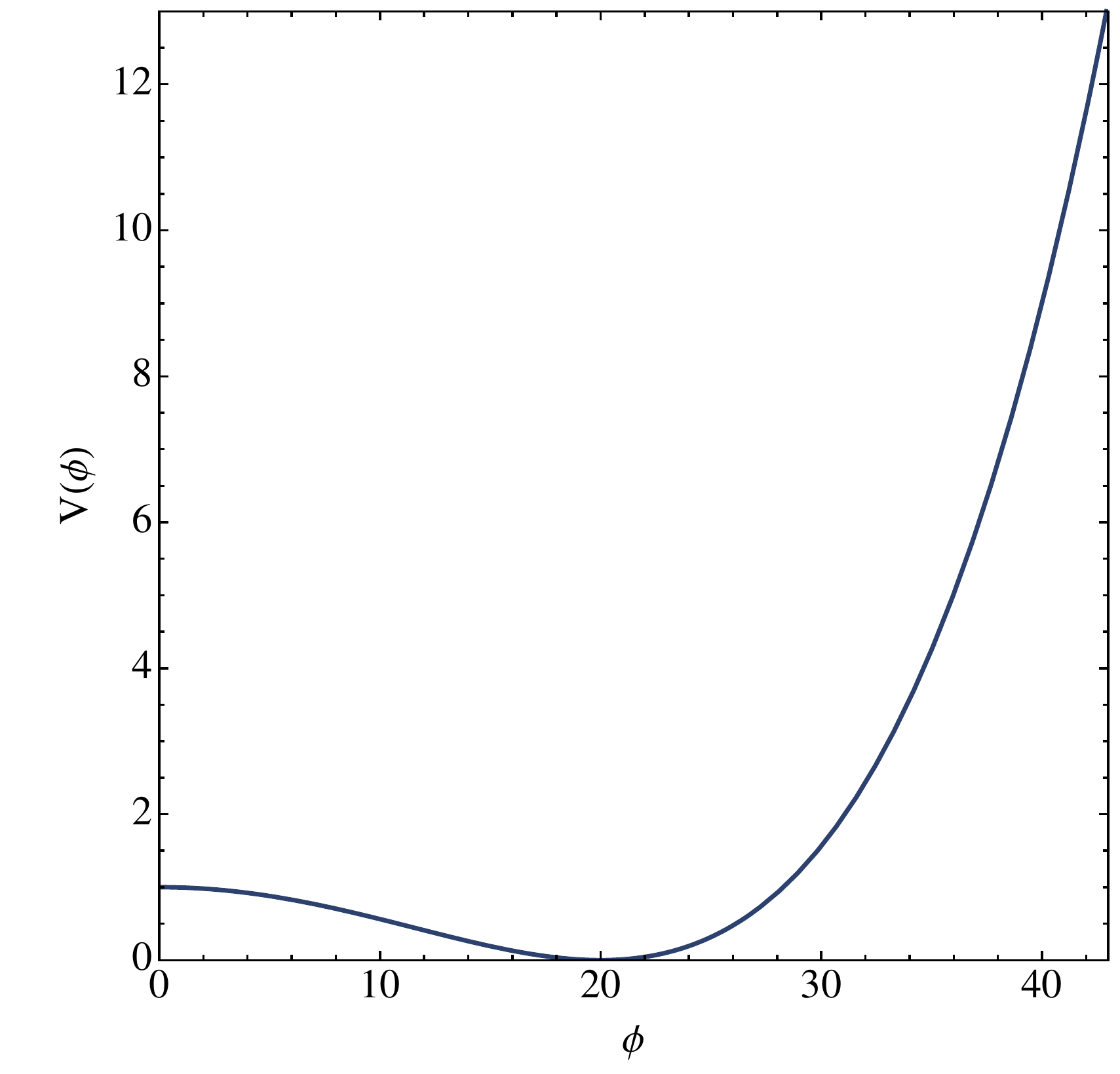}
\end{minipage}%
\caption{\label{fig:compPotential} A toy inflationary landscape potential consisting of a plateau region on the left and a plower-law region on the right, where we have arbitrarily normalised the potential so that $V(0)=1$. This simplified setting allows us to illustrate the implications of the no-boundary proposal for a generic purely inflationary landscape.}
\end{figure}

For illustration, we will consider a toy potential energy landscape, where our vacuum can be reached from two inequivalent inflationary regions (see Fig. \ref{fig:compPotential})
\begin{equation} \label{eq:HiggsPotential}
V(\phi) = \frac{1}{v^4} (\phi^2- v^2)^2\;.
\end{equation}
Near $\phi=0$ inflation proceeds along a plateau region of the potential (whose broadness is parameterised by a constant $v$), while for large $\phi$ the potential follows a power-law behaviour ($V \sim \frac{1}{v^4} \phi^4$). This setting allows us to ask whether it is more likely to be in a classical universe that originated from the plateau rather than from the power-law region (see also \cite{Ijjas:2013vea,Hertog:2013mra}).

\begin{figure}[]%
\begin{minipage}{\smallWidthLeft} \flushleft
\includegraphics[width=\smallWidthRight]{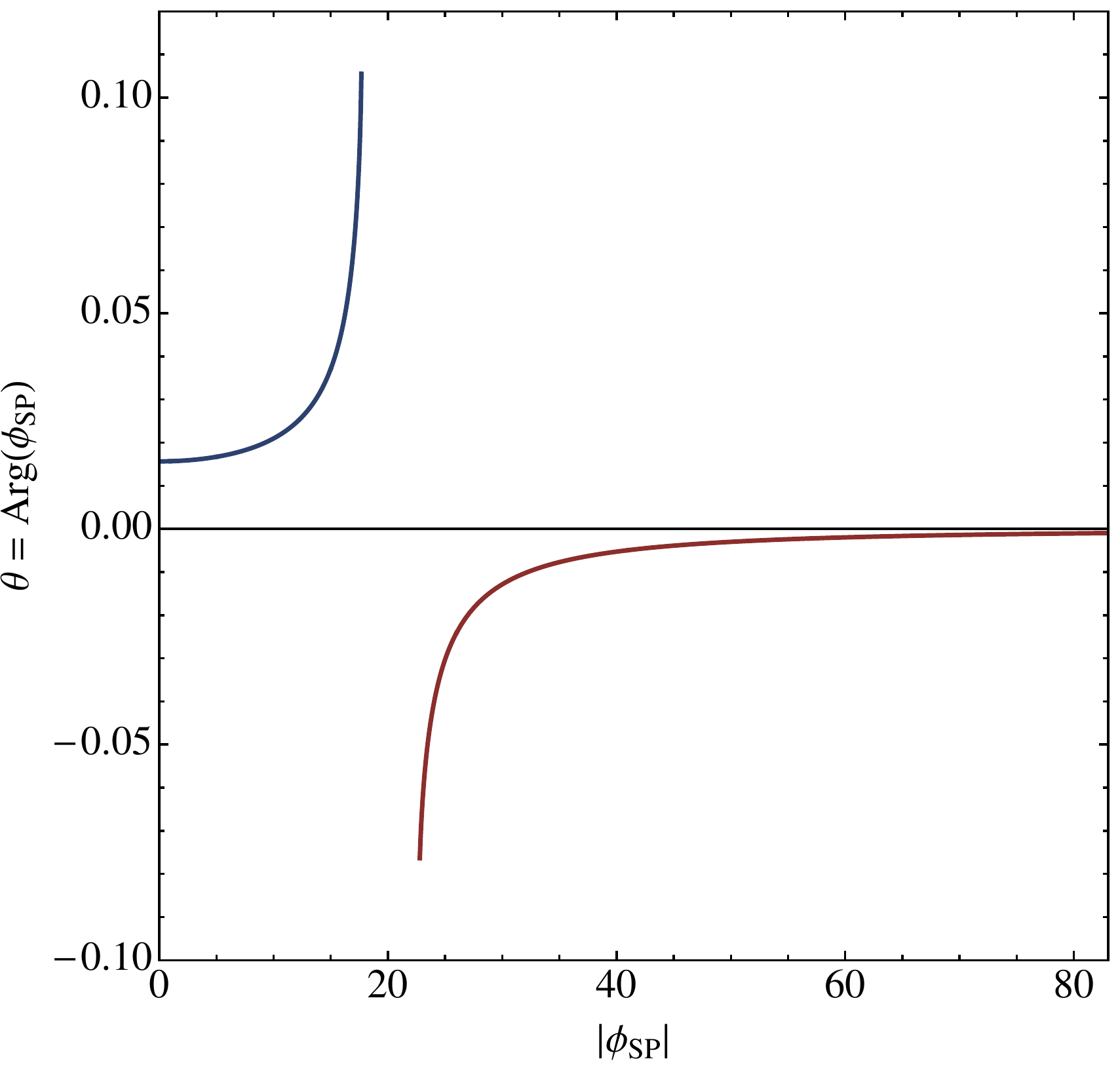}
\end{minipage}%
\begin{minipage}{\smallWidthRight} \flushleft
\includegraphics[width=\smallWidthRight]{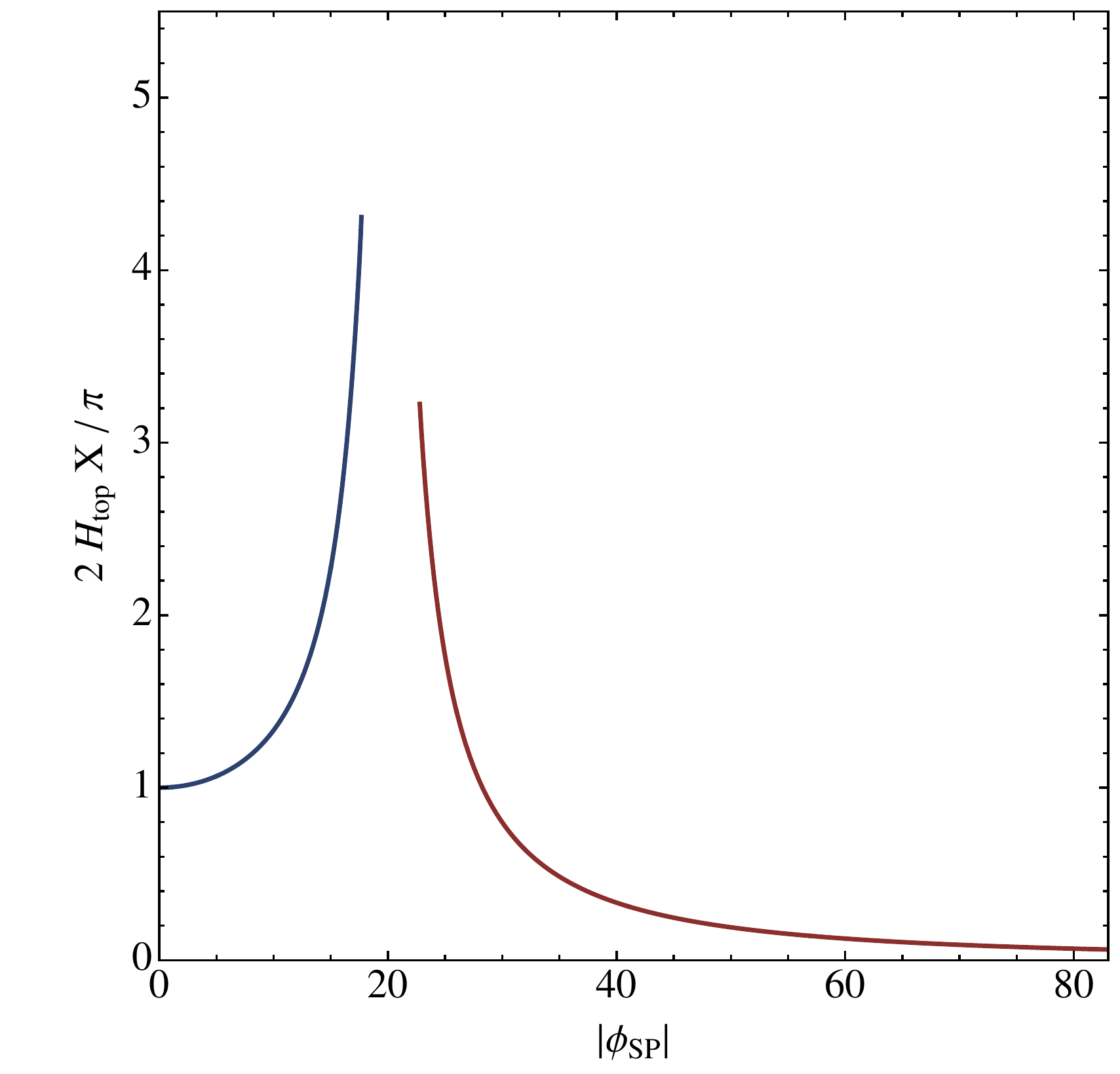}
\end{minipage}%
\caption{\label{Fig:Higgs1} Plots of the optimised phase $\theta$ at the South Pole (left panel) and the location $X$ of the vertical line along which classicality is reached (right panel). For our numerical example, we have used $v=20$ (in reduced Planck units $\kappa=1$) and $X$ is expressed in terms of $H_{top} \equiv \sqrt{V(\phi=0)/3} = 1/\sqrt{3}.$ All results can be re-scaled to any desired overall scale of the potential using Eqs. \eqref{eq:scaling}.}
\end{figure}

\begin{figure}[]%
\begin{minipage}{\smallWidthLeft} \flushleft
\includegraphics[width=\smallWidthRight]{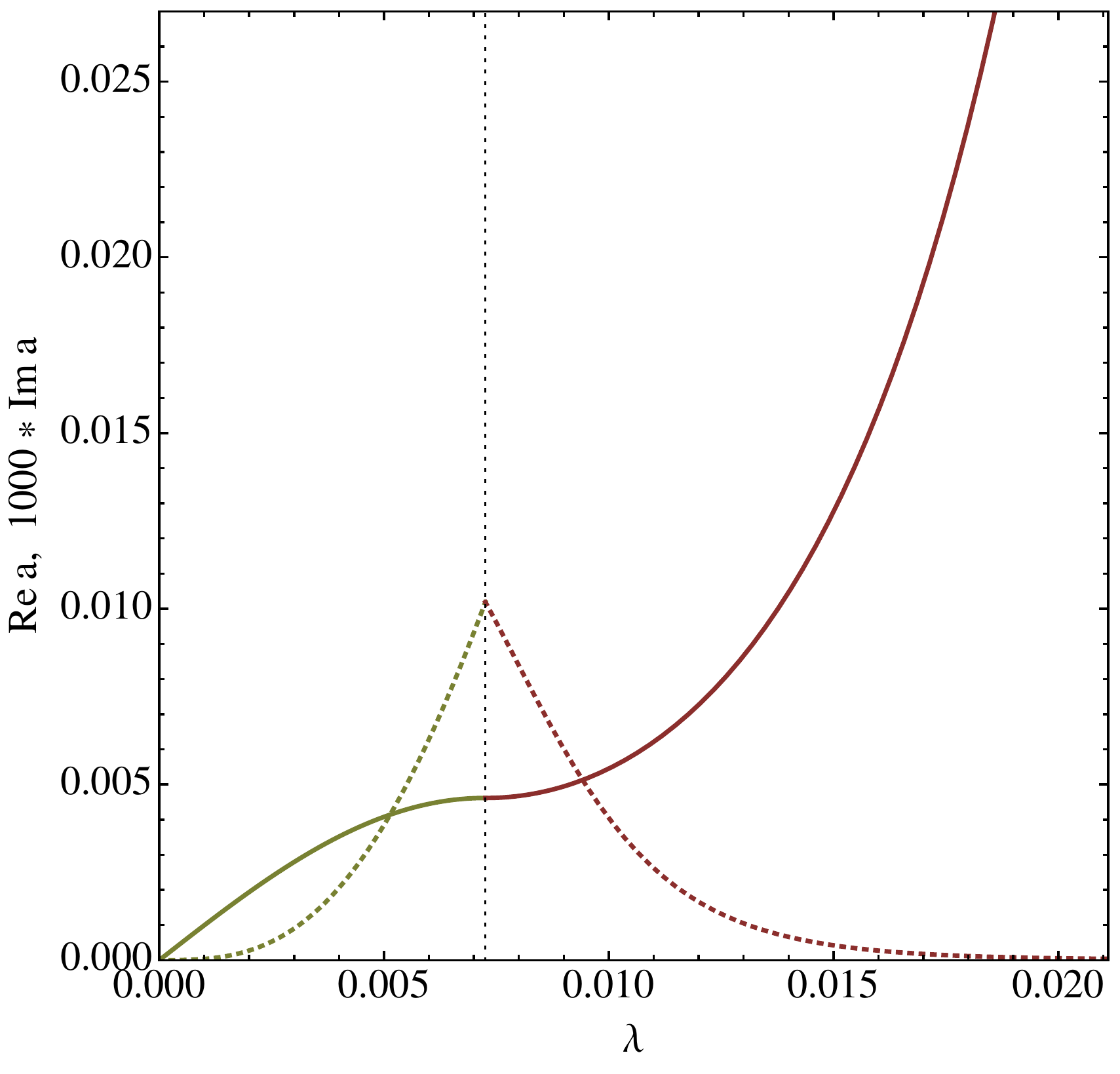}
\end{minipage}%
\begin{minipage}{\smallWidthRight} \flushleft
\includegraphics[width=\smallWidthRight]{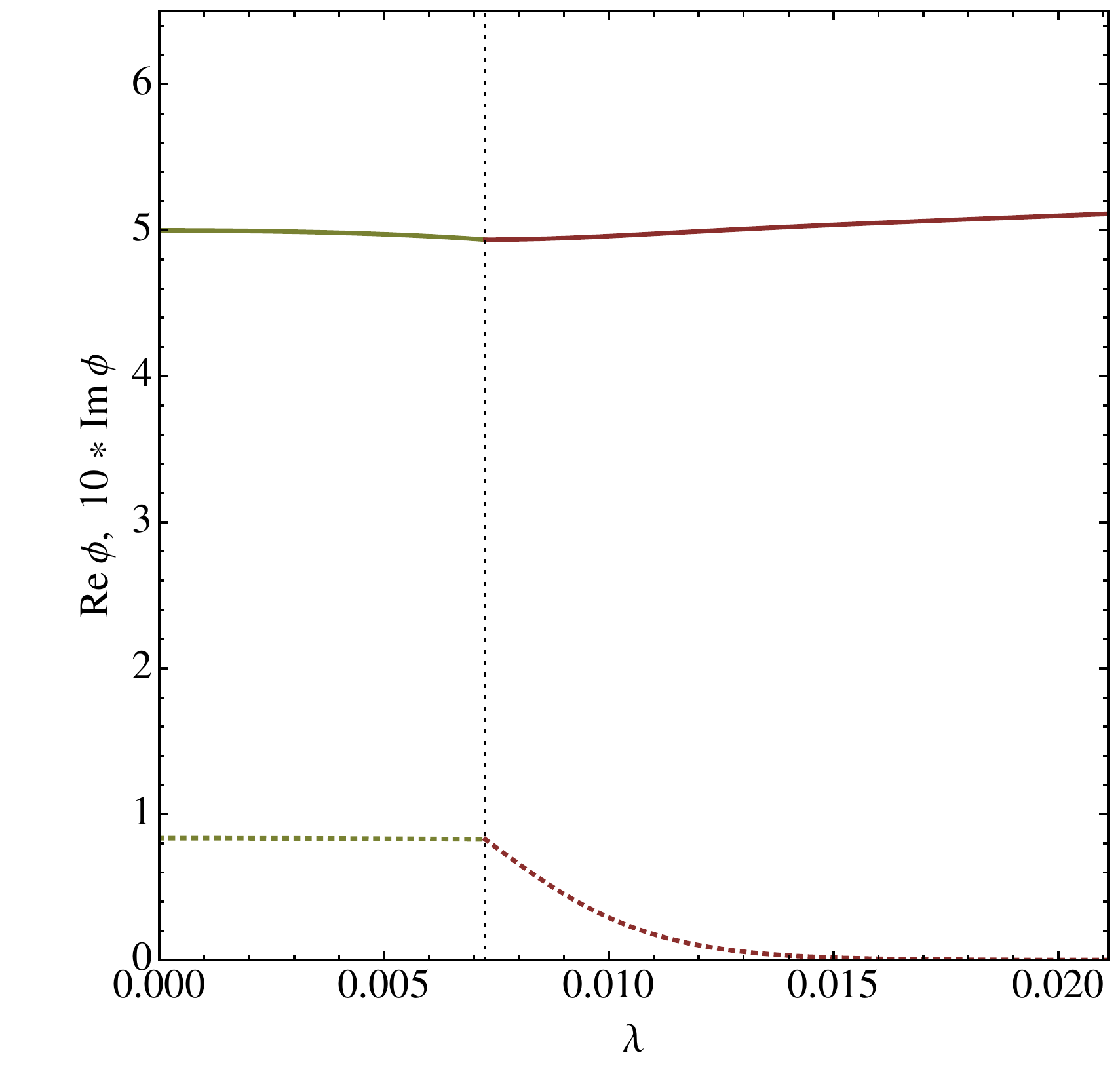}
\end{minipage}%
\caption{\label{Fig:Higgsaphi} Example of an inflationary instanton on the plateau region, with  $v=20$ and $|\phi_{SP}|=5, \theta \approx 0.01671$. For this example, the integration contour is chosen in the ``standard'' way, from the origin out to $\tau = X$ with $X/\sqrt{3} \approx 1.5707$ and then up along the imaginary $\tau$ direction, along the vertical line where classicality is reached. The left panel shows the real (solid line) and imaginary (dotted line) parts of the scale factor $a$, with the imaginary part being amplified $1000$ times. The right panel shows the analogous plot for the scalar field $\phi,$ with the imaginary part (dotted line) magnified $10$ times. As the universe becomes classical, $a$ starts growing exponentially while $\phi$ takes a value that is approximately equal to $|\phi_{SP}|$ and then slowly rolls down the potential.}
\end{figure}

The instantons on both sides of the potential minimum are all similar in character. In order to study them, we use standard integration contours running from the origin of the $\tau$ plane horizontally out to $Re(\tau_f) = X,$ and from there vertically up to the final position $\tau_f.$ In Fig. \ref{Fig:Higgs1} we plot the optimised values of the phase $\theta$ at the South Pole, as well as the corresponding values of $X,$ such that a classical history is reached along the $Re(\tau) = X$ line. Fig. \ref{Fig:Higgsaphi} provides an example of such an inflationary instanton. At the South Pole, the scalar field contains a small imaginary part, which dies off along the vertical part of the contour. In a similar manner, the scale factor becomes increasingly real along the vertical contour, and the universe starts growing exponentially. The scalar field $\phi$ approximately takes the value $|\phi_{SP}|$ when classicality is reached, which can be seen as a consequence of slow-roll. 

Note that, as Fig. \ref{Fig:Higgs1} shows, these instantons cease to exist when $|\phi_{SP}|$ approaches the potential minimum at $\phi=20.$ In this case, the inflationary phase is too short to cause both $a$ and $\phi$ to become real, and no classical, Lorentzian history is reached. Thus, in the inflationary context a minimum number of e-folds (of order a few) is needed in order for the no-boundary state to lead to a classical universe.

\begin{figure}[]
\centering
\begin{minipage}{\fullWidth}
\includegraphics[width=\fullWidth]{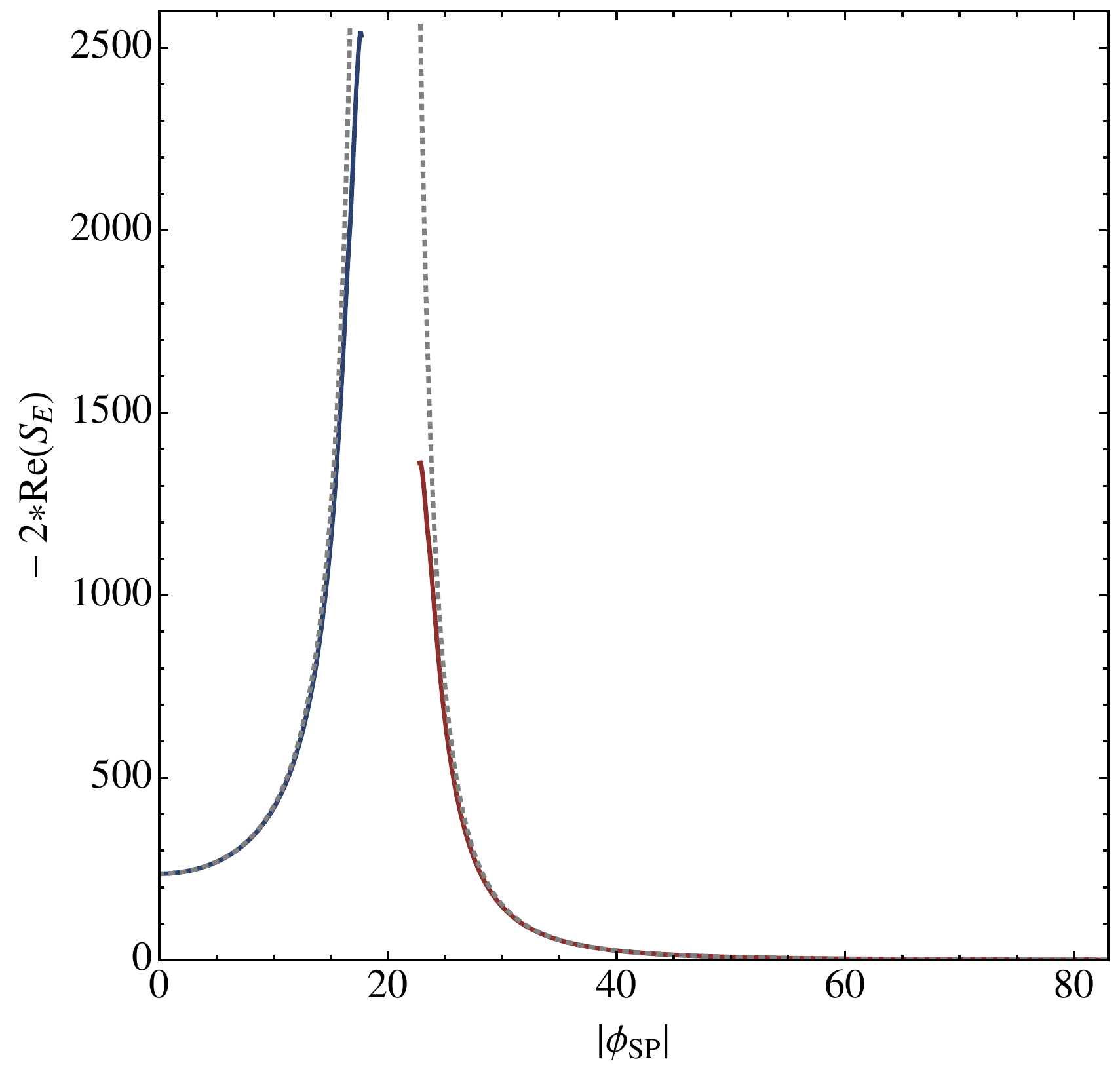}
\end{minipage}%
\caption{\label{fig:Higgs2} The logarithm of the relative probability for different values of $|\phi_{SP}|$. The dotted lines represent the approximation of the instanton by half of a 4-sphere. As is evident, low values of the potential (and thus small numbers of e-folds) are vastly preferred. Moreover, the probability for a history originating from the inflationary plateau is much higher than that of histories originating in the power-law region of the potential.}
\end{figure}

The relative probability for the various classical histories is given by $\Psi^\star \Psi \sim e^{-2Re(S_E)}.$ The logarithm of this (unnormalised, and thus relative) probability is plotted in Fig. \ref{fig:Higgs2}. As is evident from the figure, this quantity is well estimated by approximating the instanton with half of a Euclidean de Sitter 4-sphere, cf. Eq. \eqref{eq:deSitter}. The general trend is that smaller values of the potential are preferred. Thus, the plateau region (in which density perturbations of the observed amplitude can be generated at lower values of the potential) are preferred over the power-law region. Moreover, for the same reason a small number of e-folds is vastly preferred over a large number. One may argue against histories with too little inflation on anthropic grounds, however in that case the preference for a small number of e-folds would cause one to expect to see a significant contribution of spatial curvature in the current energy budget of the universe. Furthermore, one would not expect to detect any gravitational waves from the early universe, as their amplitude (which is determined by the height of the potential) would be far too small.

\begin{figure}[]
\centering
\begin{minipage}{\fullWidth}
\includegraphics[width=\fullWidth]{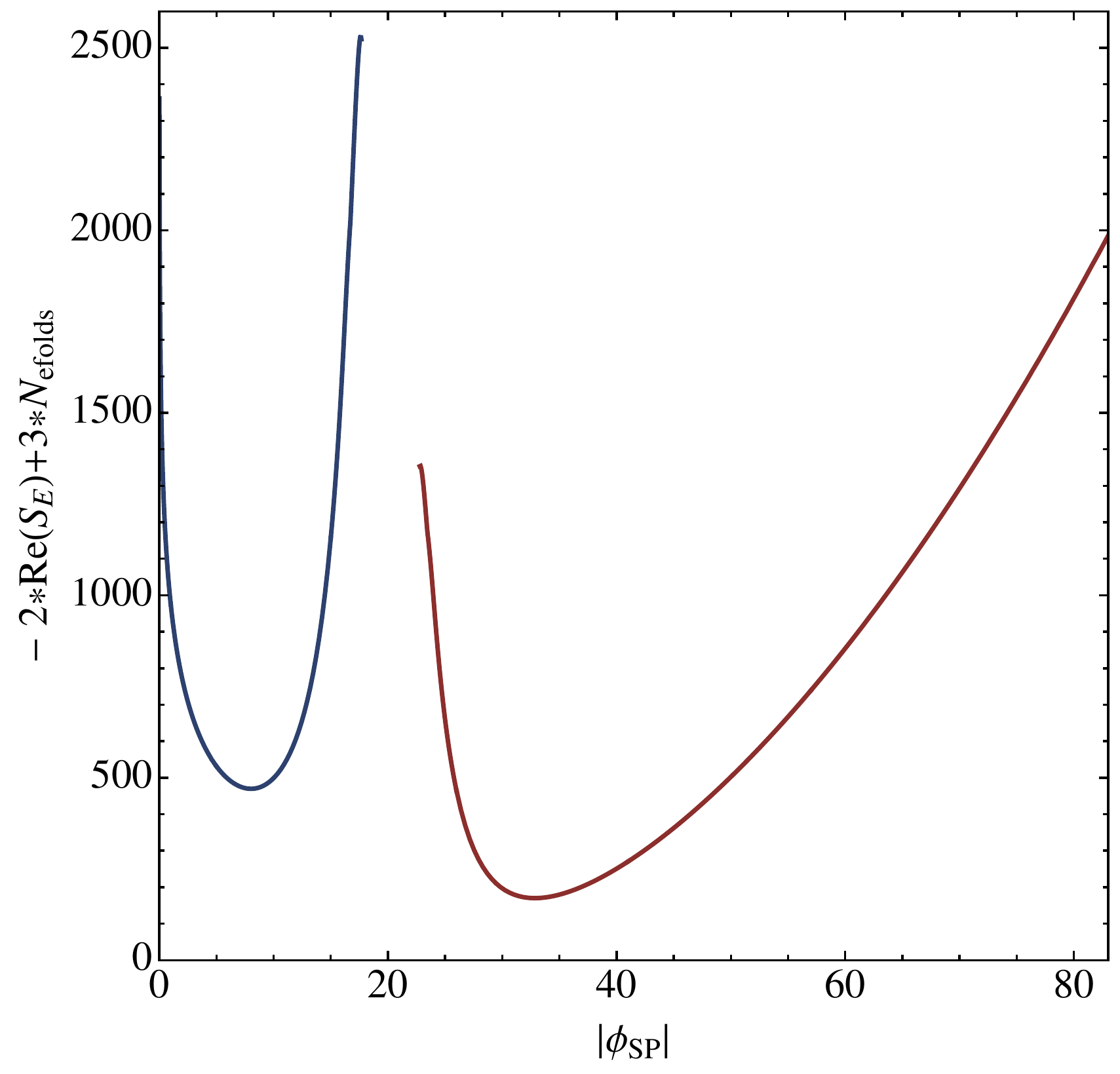}
\end{minipage}%
\caption{\label{fig:Higgs3} The logarithm of the volume-weighted probability distribution. The minimum probability occurs at the onset of the eternal inflationary regime on both sides of the potential.}
\end{figure}

In order to reverse the trend that small numbers of e-folds are preferred, it has been suggested that one should amend the formula for the relative probability by weighting it by the physical volume produced by the various histories \cite{Hartle:2007gi}. The idea is that if a universe is larger then there exist more regions where life can form and one should take this into account. The volume-weighted relative probability is then taken to be
\begin{equation}
P_{\textrm{volume weighted}} = e^{3 N_e} \Psi^\star \Psi = e^{-2Re(S_E) + 3 N_e} \;.
\end{equation}
For our example, the logarithm of the volume-weighted relative probability is plotted in Fig. \ref{fig:Higgs3}. The shape of the resulting probability curves can be understood by calculating the value of the field where the probability reaches its minimum: 
\begin{eqnarray}
\left( -2Re(S_E)+3N \right)_{,\phi} &&= \left( \frac{24\pi^2}{V} + 3 \int\frac{V}{V_{,\phi}} d\phi \right)_{,\phi} \nonumber \\ && = -\frac{24\pi^2 V_{,\phi}}{V^2} + 3 \frac{V}{V_{,\phi}} \nonumber \\ && = 0 \nonumber \\ && \rightarrow \quad\frac{V_{,\phi}^2}{V^3} = \frac{1}{8\pi^2}\;,
\end{eqnarray}
where we have assumed slow-roll. This result is very interesting: the relative probability reaches its {\it minimum} when the regime of slow-roll eternal inflation starts, i.e. when the amplitude of the generated density perturbations becomes equal to the background density. Only even further up the potential does the probability become larger again. Note that it is not clear that one should still trust the present treatment after the turn-around in probability: naively, the volume of space produced in the eternal inflation regime is infinite, and one may wonder whether one should now multiply the relative probability by infinity. However, one must bear in mind that this picture of eternal inflation relies on a number of extrapolations. In particular, it is assumed that near the Planck scale new quantum modes are continually ``created'' in the Bunch-Davies vacuum, {\it ad infinitum}. If only for this reason, we do not find the volume-weighted turn-around in probability convincing, and for the present paper we will stick to the bare probabilities provided by the wavefunction alone\footnote{It would also be interesting to study how false-vacuum eternal inflation can be treated within the present framework, and what the implications are.}.

Our conclusion is that in a pure inflationary context, the no-boundary wavefunction is only compatible with observations if one combines it with strong {\it ad hoc} (anthropic) arguments \cite{Hartle:2007gi,Hwang:2013nja}. As we will see below, in the ekpyrotic/cyclic context this situation is substantially improved.

\subsection{A Note on Numerical Precision}

The relative probabilities for various classical histories are determined by the real part of the Euclidean action. However, the latter is generally tricky to compute numerically. If one tries to determine the instanton corresponding to a classical history, then naively computes the Euclidean action along the $ \textrm{Re}(\tau) = X$ line, one sees that $ \textrm{Re}(S_E)$ generally starts increasing exponentially at late times, instead of remaining constant. There are two kinds of reasons for that:
\begin{itemize}
\item Finite numerical precision. Even if $ \theta$ is determined with infinite precision, the value $X$ corresponding to the classical history can only be determined with a finite precision $ \delta X$ and then:
\begin{equation} \label{eq:bigdeviation}
\textrm{Im}(a ^3) \sim \delta X\,e^{3Hy} \;.
\end{equation}
The quantity above enters in the computation of the real part of the Euclidean action. The finiteness in the precision on the optimal value of $ \theta$ also leads to a similar divergence.
\item Intrinsic deviation from classicality: this applies to those cases where classicality relies on the attractor nature of inflation. Given that the efficiency of the attractor mechanism is not infinite, there always remains a small imaginary part in the scalar field which, in turn, causes a backreaction on $a$ of order
\begin{equation}
\textrm{Im}(a ^3) \sim 1 \;.
\end{equation}
This error is therefore usually much smaller. However, it must be noted that the backreaction of this small complex part of the scalar field causes $a$ to deviate from its asymptotic behaviour $a \sim a_0 e^{Hy}$ by a term of order $1$, leading to an error in the estimation of $X$, hence to \eqref{eq:bigdeviation}.
\end{itemize}
For these reasons, if one optimises over $ \theta$ then computes $ \textrm{Re}(S_E)$ along the $ \textrm{Re}(\tau) = X$ line, then this quantity reaches a constant value for a while but eventually deviates from a constant again. Nevertheless, along this line the fields are very precisely real, and evolve as classical (real) solutions $( b_ \lambda, \chi_ \lambda)$. If one now computes the full no-boundary wavefunction, by optimising over $\phi_{SP}$ and $ \tau_f$, one finds that the real part of the Euclidean action is the same for all the values in the family $(b_ \lambda, \chi_ \lambda)$, confirming that these are classical histories. This is the way we compute the real part of $S_E$ presented in the plot.

\section{Ekpyrotic Instantons} \label{section:EkpyrInst} \label{section:ekpyrotic}

The fact that the no-boundary proposal has been studied only in the inflationary context until recently might have left the impression that inflationary instantons are the only possibility of instantons satisfying the no-boundary conditions. As demonstrated in our recent paper \cite{Battarra:2014xoa}, it is however possible to find a new type of {\it ekpyrotic instantons}, in theories with a steep and negative potential. Moreover, as we will discuss in detail below, the real part of the Euclidean action of these instantons can be very large and negative, rendering these instantons vastly more likely than all realistic inflationary instantons. This discovery thus leads us to revise the implications of the no-boundary state.

In the present paper we will restrict our attention to the simplest ekpyrotic potentials by taking
\begin{equation} \label{eq:ekpot}
V(\phi)=-V_0 e^{-c\kappa\phi} \;,
\end{equation}
with $V_0, c$ being positive constants \cite{Khoury:2001wf,Lehners:2008vx}. These potentials lead to an ekpyrotic phase of ultra-slow, high-pressure contraction. During this phase, as long as $c^2>6,$ the cosmological flatness puzzle is resolved as both homogeneous and anisotropic curvatures are suppressed. The requirement that $c^2 > 6$ corresponds to the condition that the pressure is larger than the energy density, and this condition will be seen to play an important role in what follows. Before describing the ekpyrotic instanton solutions, we should mention that simple extensions to two-field models allow for the generation of nearly scale-invariant curvature perturbations \cite{Lehners:2013cka,Li:2013hga,Qiu:2013eoa,Fertig:2013kwa,Ijjas:2014fja}, and thus an ekpyrotic phase provides an alternative model to inflationary models of the early universe. In order to obtain a complete cosmological history, one must also describe how the universe bounces from the contracting phase into the current expanding phase of the universe. This is a topic of much current research - some promising models are described in \cite{Turok:2004gb,Buchbinder:2007ad,Creminelli:2007aq,Easson:2011zy,Koehn:2013upa,Battarra:2014tga,Bars:2011aa,Bars:2013qna}, while a recent review is \cite{Battefeld:2014uga}. In the present paper, we will not include the bounce phase - we leave this important topic for future work.

Apart from an overall re-scaling of the potential, the exponential potentials that we study allow for an additional shift symmetry which considerably simplifies our search for ekpyrotic instantons. The action we consider is given by
\begin{equation}
S_E = - \int d ^4 x  \sqrt{g} \left( \frac{R}{2 \kappa^2} - \frac{1}{2} g ^{\mu \nu} \partial _{\mu} \phi\, \partial _{\nu} \phi + V_0 e^{-c\kappa\phi} \right) \;.
\end{equation}
Then one can perform the combined field shifts/re-scalings 
\begin{equation}
\phi  \equiv  \kappa ^{-1} \bar{ \phi} + \Delta \phi \;, \quad
g_{\mu\nu}  \equiv  \frac{e^{c \kappa \Delta \phi}}{\kappa^{2} V_0} \bar{ g}_{\mu\nu} \;,\label{eq:metricscaling}
\end{equation}
such that the action becomes
\begin{equation} \label{eq:actionrescaled}
S_E = -\frac{e^{c\kappa\Delta \phi}}{ \kappa^4 V_0} \int d ^4x \sqrt{ \bar{ g}} \left( \frac{ \bar{R}}{2} - \frac{1}{2} \bar{g} ^{\mu \nu} \partial _{\mu} \bar{ \phi} \partial _{\nu} \bar{ \phi} + e^{-c\bar\phi}\right) \;.
\end{equation}
Thus, if we have an instanton solution for which the scalar field at the South Pole takes the value $\bar\phi_{SP}$, we can find an entire family of instantons with South Pole values $\phi_{SP} =  \bar\phi_{SP} + \Delta \phi$ using Eqs. \eqref{eq:metricscaling}. Hence these instanton families depend only on $c,$ the steepness of the ekpyrotic potential. From now on we will drop the overbars and work with the re-scaled theory. We should point out that such a shift symmetry will of course not be available for more general steep and negative potentials. However, we expect that instanton solutions similar to the ones we will describe here will also exist in such theories. Finding all instantons will however be harder, as one will have to scan over all possible values of $\phi_{SP}$ in order to find all instantons. Incidentally, we will see an example where this shift symmetry is broken by the presence of a cosmological constant in section \ref{section:cyclic}, where we will consider potentials pertinent to the cyclic universe.

\subsection{An Example}

\begin{figure}[]%
\begin{minipage}{\smallWidthLeft} \flushleft
\includegraphics[width=\smallWidthRight]{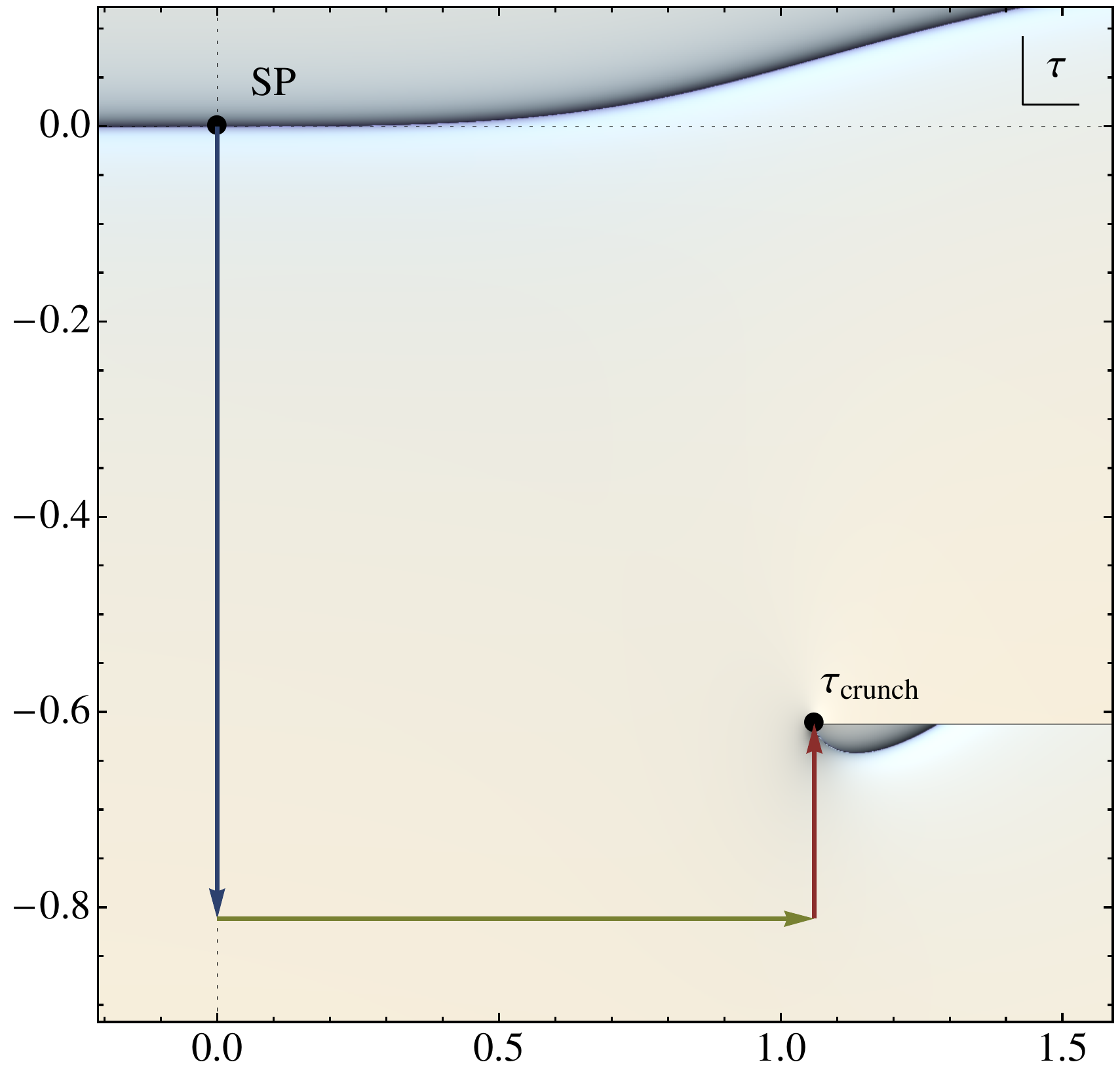}
\end{minipage}%
\begin{minipage}{\smallWidthRight} \flushleft
\includegraphics[width=\smallWidthRight]{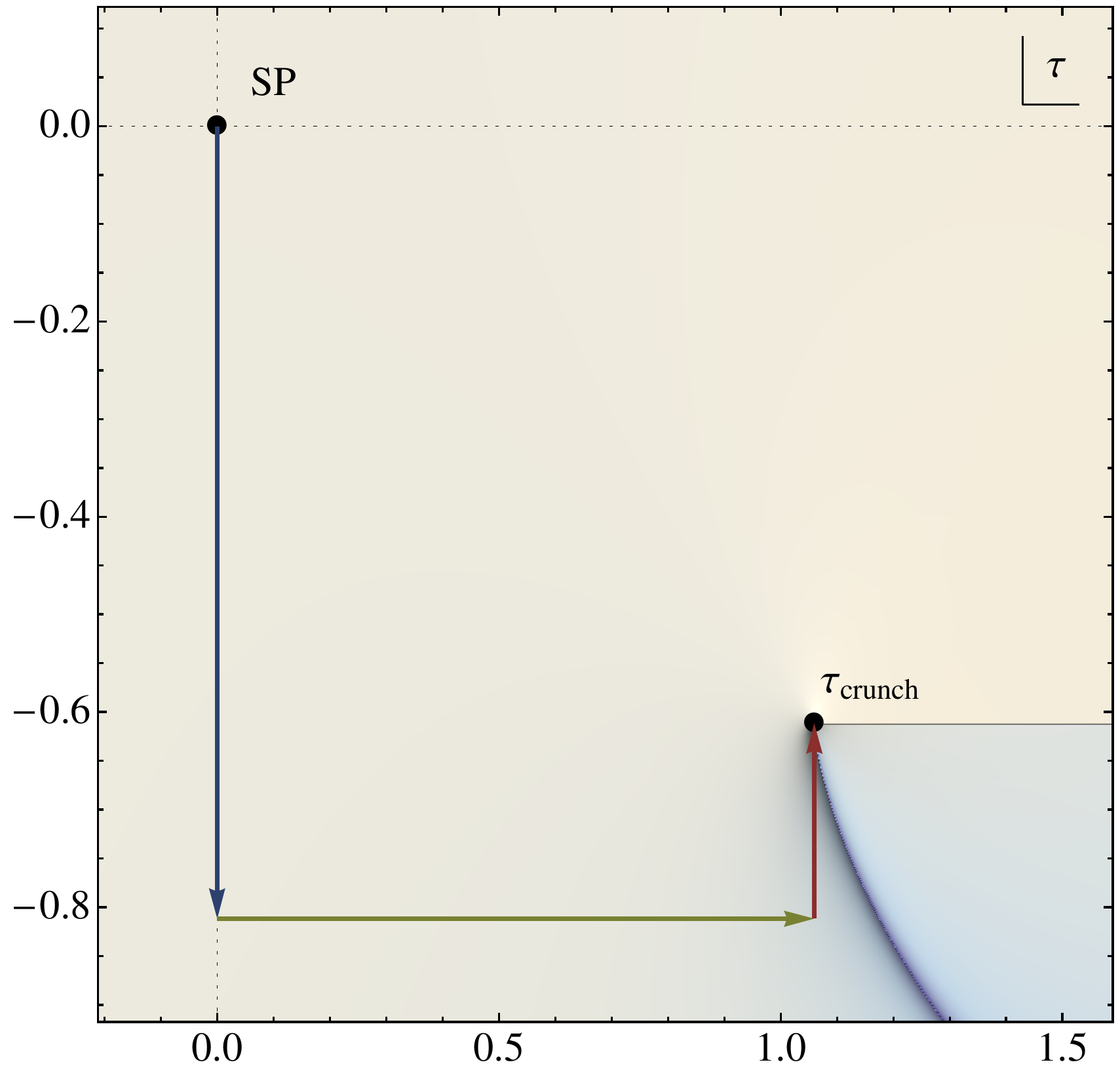}
\end{minipage}%
\caption{\label{Fig1} The dark lines indicate the locus where the scale factor $a$ (left panel) and scalar field $\phi$ (right panel) are real. To obtain the figure, we integrate the field equations along contours starting from the origin (South Pole) and running up/down in the vertical direction first, then along horizontal lines to the left and right. For this example, we have used the values $ \epsilon \equiv c^2/2 = 4$, $\phi_{SP}^R=0.$ The imaginary part $\phi_{SP}^I = -1.481$ has been tuned such that the scale factor is real in the approach to the crunch, i.e. such that $a_0$ in Eq. \eqref{eq:ekpyroticattractor1} is real. The ekpyrotic attractor then ensures that the lines of real scale factor and real scalar field both become vertical in the approach to the crunch. Note that there is a branch cut to the right of $\tau_{crunch},$ which is due to the fractional behaviour in the scale factor, see Eq. \eqref{eq:ekpyroticattractor1}. The coloured arrows indicate the integration contour that we have chosen to illustrate the shape of ekpyrotic instantons in Fig. \ref{Fig3}.}
\end{figure}

\begin{figure}[]%
\begin{minipage}{\smallWidthLeft} \flushleft
\includegraphics[width=\smallWidthRight]{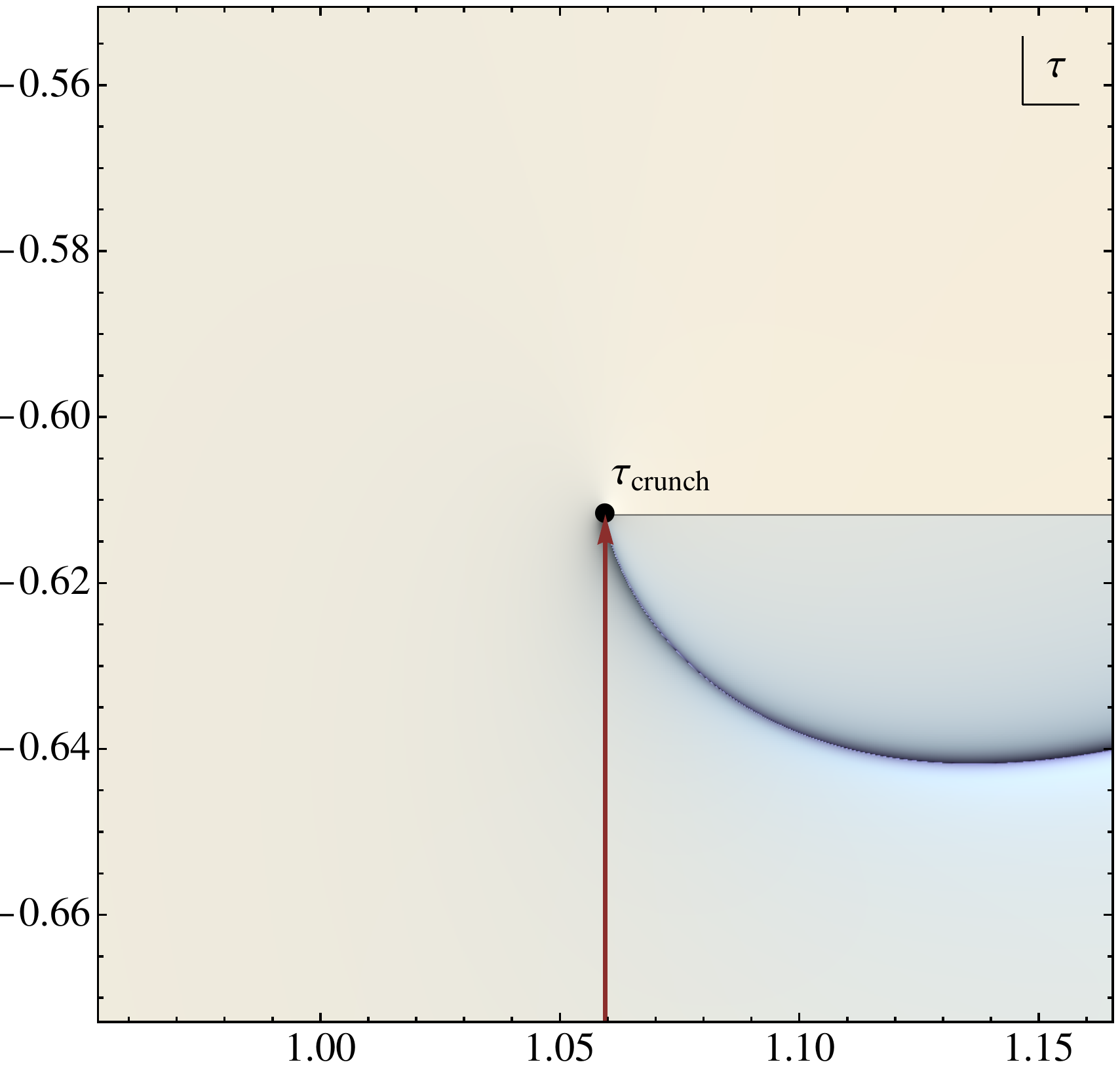}
\end{minipage}%
\begin{minipage}{\smallWidthRight} \flushleft
\includegraphics[width=\smallWidthRight]{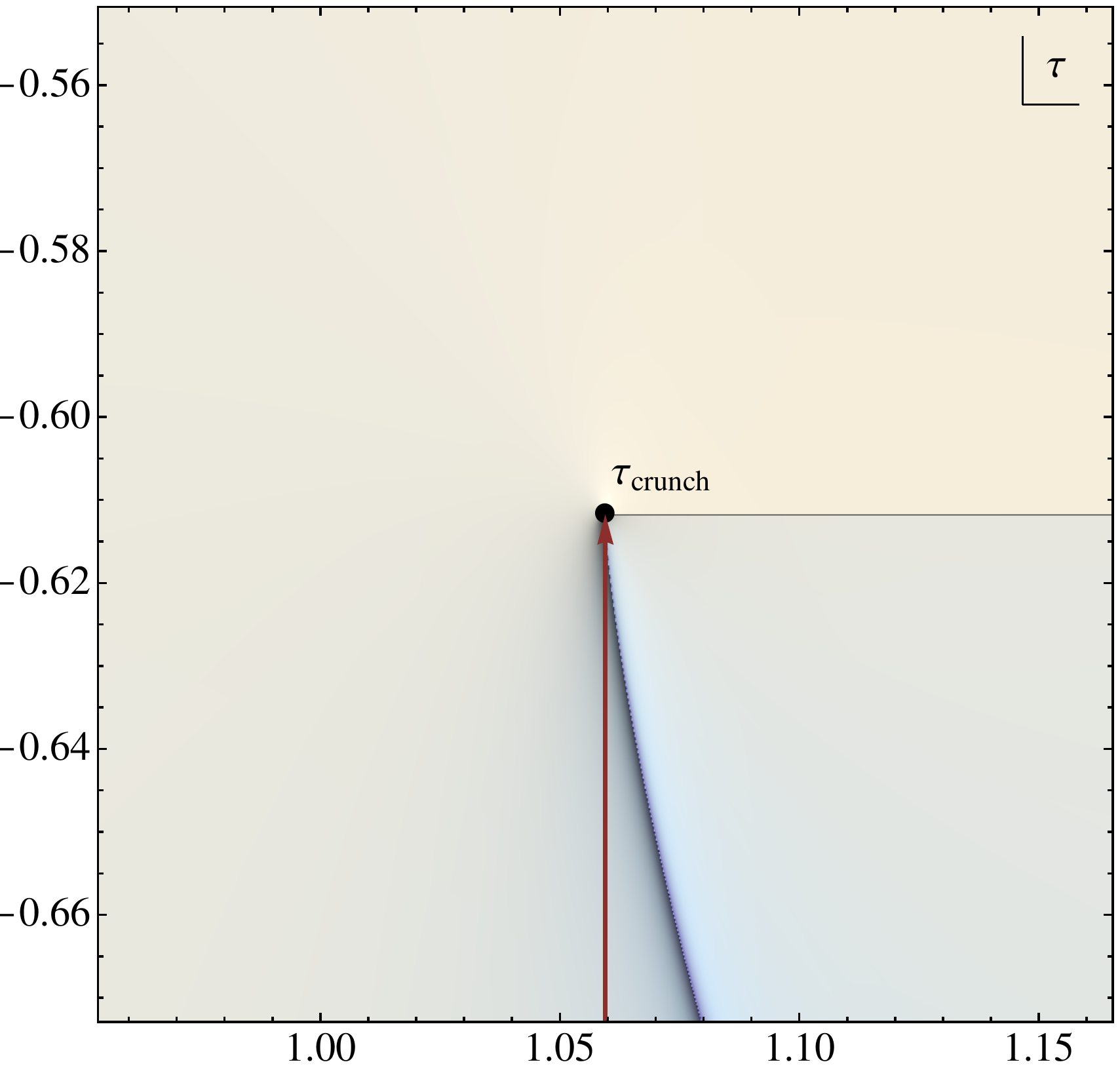}
\end{minipage}%
\caption{\label{Fig2} A zoom-in of Fig. \ref{Fig1} around the location of the crunch.}
\end{figure}

Given that the no-boundary conditions generally force the instantons to be complex-valued, obtaining a classical history turns out to be far from automatic. In the inflationary context, we saw that the phase $\theta$ of the scalar field at the South Pole must in fact be precisely tuned. Even then, as we discussed, one needs in addition an attractor mechanism that allows for a classical Lorentzian history to emerge. In the ekpyrotic context, the situation is similar. Here, we will label the instantons by the real and imaginary parts of $\phi_{SP},$
\begin{equation}
\phi_{SP} = \phi_{SP}^R + i \phi_{SP}^I.
\end{equation}
The shift symmetry \eqref{eq:metricscaling} can be used to fix $\phi_{SP}^R=0$ for now. Then, for general values of $\phi_{SP}^I,$ there will be a curved line in the complex $\tau$ plane along which the scalar field is real, such as in the right panel of Fig. \ref{Fig1}. Following this line, one will eventually reach a crunch singularity, where the scale factor has reached zero value. (Since we have not added any dynamics that could lead to a bounce, a crunch is inevitable.) However, generically the crunch will be complex, in the sense that the scale factor takes complex values all the way to the crunch. However, one can find a tuned value of $\phi_{SP}^I$ such that the scale factor becomes increasingly real (by which we mean $Im(\phi)/Re(\phi) \rightarrow 0$) in the approach to the crunch. Moreover, due to the ekpyrotic attractor the lines along which both the scale factor and the scalar field are real become progressively vertical (and coincident) in the approach to the crunch. Thus, for such a tuned value of $\phi_{SP}^I,$ a real, Lorentzian history is reached. This is illustrated in Figs. \ref{Fig1} and \ref{Fig2}, which show that in the no-boundary context a new type of {\it ekpyrotic instantons} (first presented in \cite{Battarra:2014xoa}) exists and is contained in the semi-classical approximation of the no-boundary state. 

\begin{figure}[]%
\begin{minipage}{\smallWidthLeft} \flushleft
\includegraphics[width=\smallWidthRight]{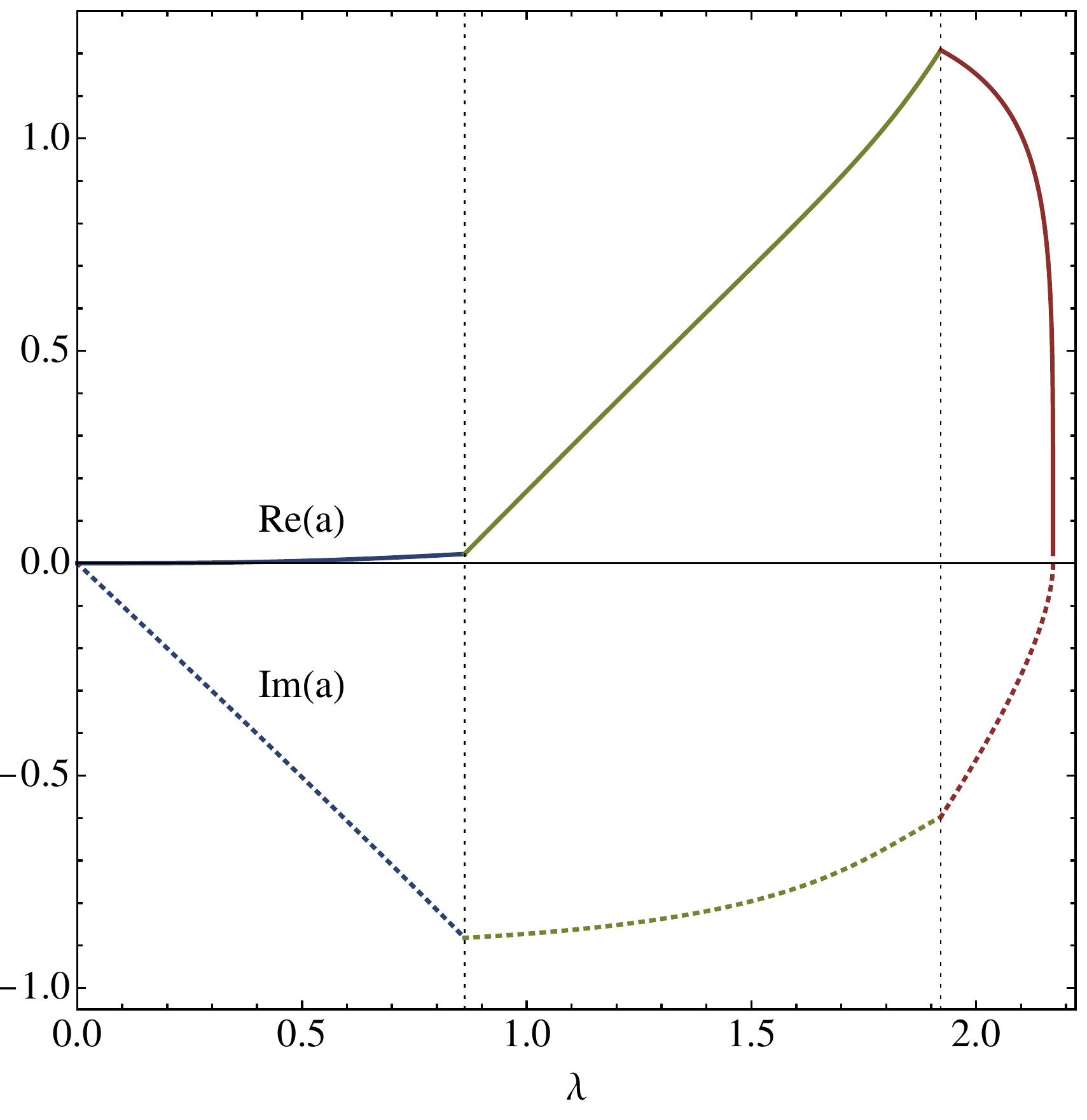}
\end{minipage}%
\begin{minipage}{\smallWidthRight} \flushleft
\includegraphics[width=\smallWidthRight]{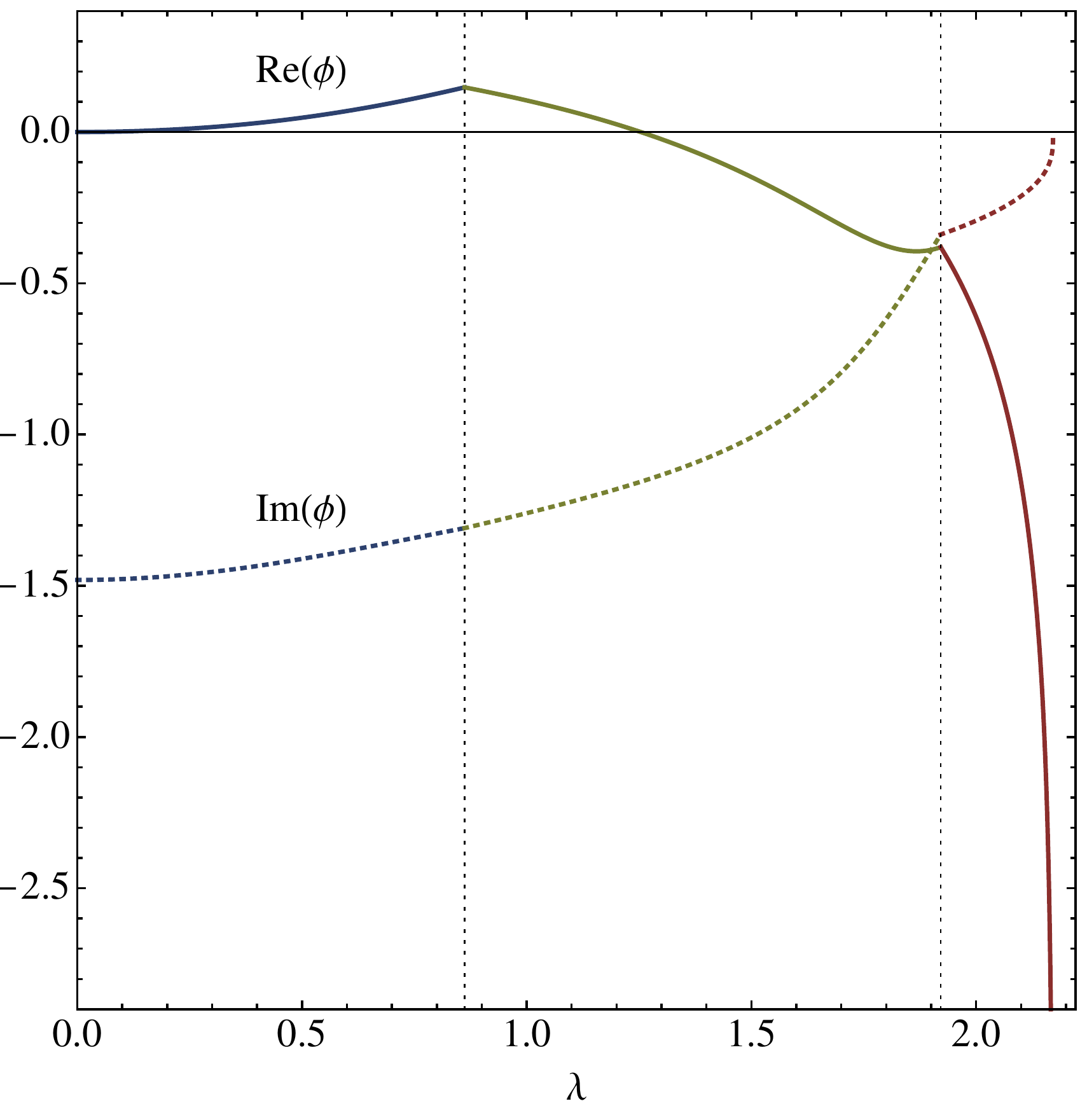}
\end{minipage}%
\caption{\label{Fig3} An example of an ekpyrotic instanton. To represent the instanton, we have used the contour indicated by the three arrows in Fig. \ref{Fig1}. The left panel shows the real and imaginary parts of the scale factor, while the right panel shows the real and imaginary parts of the scalar field. The dotted vertical lines indicate direction changes in the integration contour. Note that along the third segment, both $a$ and $\phi$ become real as the scalar field rolls down the ekpyrotic potential.}
\end{figure}

Below, we will discuss the approach to classicality of these new instantons in detail. For now, we would like to present the general shape of ekpyrotic instantons. For this purpose, we must choose an explicit integration contour in the $\tau$ plane. Our choice is represented by the three coloured arrows in Fig. \ref{Fig1}, and the resulting values for the real/imaginary parts of the scale factor and scalar field are shown in Fig. \ref{Fig3}. Along the first segment (vertically down from the origin, $d\tau = - id\lambda$), the scale factor is almost pure imaginary, with $a \propto i \lambda.$ This translates into a portion of (opposite-signature) Euclidean flat space,
\begin{equation}
ds^2 \approx - d\lambda^2 - \lambda^2 d\Omega_3^2 \;,
\end{equation}
and thus the bottom of the instanton is very flat. Along the second, horizontal segment ($d\tau = d\lambda$), both the scale factor and the scalar field are fully complex. The third segment is the most interesting one: as the contour runs back up vertically ($d\tau = i d\lambda$) towards the crunch at $\tau=\tau_{crunch},$ both the scale factor and scalar field become increasingly real, while the scalar field rolls down its steep and negative potential. Thus, in contrast with inflationary instantons, the scalar field is highly dynamical as the universe becomes classical. For illustrative purposes, we have also included a cartoon version of ekpyrotic instantons -- see Fig. \ref{fig:carafe}.

Now that we have found one explicit instanton, we can use the shift symmetry \eqref{eq:metricscaling} to transform it into other solutions with different values of $\phi_{SP}= \bar\phi_{SP} + \Delta \phi.$ Note that we have to restrict $\Delta \phi \in \mathbb{R}$ such that the new instanton also reaches a classical history with (asymptotically) $a,\phi \in \mathbb{R}.$ This implies that all members of the corresponding family of instantons will have the same value of the imaginary part $\phi_{SP}^I.$ The shift transformation has the consequence of re-scaling the action by a factor $e^{c\Delta \phi},$ cf. Eq. \eqref{eq:actionrescaled}. Thus the real part of the Euclidean action, whose asymptotic value determines the relative probability of the corresponding classical history, scales as 
\begin{equation}
Re(S_E)\mid_{\textrm{fixed } c} \,\, \propto \frac{V(0)}{V(\phi_{SP}^R)} = \frac{1}{|V(\phi_{SP}^R)|}.
\end{equation} 
Given that our numerical analysis shows that $Re(S_E) < 0,$ instantons that start out at smaller absolute values of the potential are preferred. This is in fact entirely analogous to the inflationary case, where small magnitudes of the potential are also preferred. However, it should be noted that for inflation this scaling translates into a preference for a small number of e-folds, while in the ekpyrotic case the preference is for a long ekpyrotic phase with many e-folds of slow, high-pressure contraction. It is interesting that for these preferred instantons, spacetime curvatures are small (except of course near the crunch), and thus the semi-classical approximation that is used throughout ought to be particularly good.

Before discussing how the properties of ekpyrotic instantons change when one alters the steepness of the potential, we will discuss the approach to classicality in more detail, as this strikes us as one of the most interesting features of the no-boundary approach.

\subsection{The Ekpyrotic Attractor and Classicality}

\begin{figure}[]
\centering
\begin{minipage}{\smallWidthRight}
\includegraphics[width=\smallWidthRight]{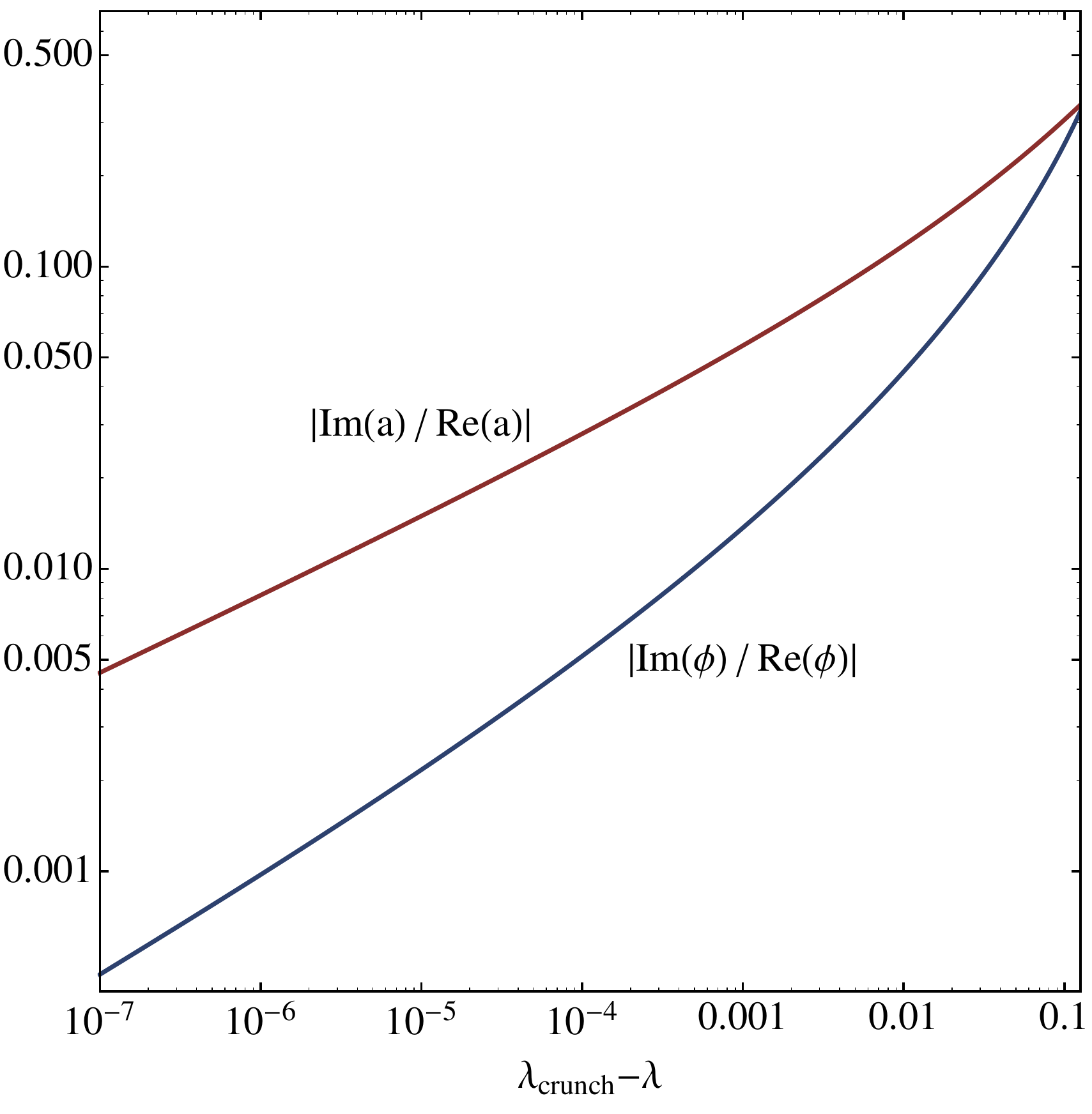}
\end{minipage}%
\caption{\label{Fig5} The scale factor and scalar field both become increasingly real as they decrease during the ekpyrotic phase. In this way a real, Lorentzian history of the universe is reached.}
\end{figure}

In the present section, we will study the manner in which the no-boundary wavefunction leads to classical, Lorentzian (and contracting) universes. We will do this by studying the properties of the geometry of the instantons that we described previously, as well as by studying the properties of the implied semi-classical wavefunction $\Psi(b,\chi)$ along a family of instantons. In the mini-superspace approximation, a clear-cut notion of classicality is provided by the WKB conditions, i.e. by the requirement that the amplitude of the wavefunction should vary slowly compared to its phase. Note that in quantum physics, one is often interested in further notions of classicality, in particular in the process of decoherence. But for decoherence to occur, one needs to include a description of observers who can interact with the system one is interested in. In cosmology, the role of such ``observers'' can sometimes be fulfilled by additional fields - see e.g. \cite{Battarra:2013cha}. Thus, if one were interested in decoherence, one would have to extend the present framework accordingly. For now, we will restrict our analysis to the mini-superspace model we have been considering thus far. In a sense, this is then similar to studying the issue of the squeezing of quantum fluctuations in cosmological spacetimes, only that here we are interested in the classicality of the background spacetime itself.

In Fig. \ref{Fig5} we can see the first hint of how classicality is reached in the approach of the crunch, as $Re(a)$ and $Re(\phi)$ are decreasing. In fact one can understand the approach to classicality in more detail by studying the asymptotic behaviour of the instanton. Denoting $ \lambda \equiv i (\tau_c - \tau)$, so that the crunch occurs as $\lambda \rightarrow 0^-,$ we find that in the approach to the crunch the instanton asymptotes to a scaling solution plus correction terms,
\begin{eqnarray} \label{eq:ekpyroticattractor1}
a( \tau) & = & a_0 (-\lambda) ^{1/ \epsilon} \left( 1 + \alpha_{1, a}\, (-\lambda) ^{1- 3/ \epsilon} + \alpha_{2, a}\, (-\lambda) ^{2 - 6/ \epsilon} + \ldots + \beta_{1, a} \,(-\lambda) ^{2-2/ \epsilon} + \ldots \right) \;, \\ \label{eq:ekpyroticattractor2}
\phi( \lambda) & = & \sqrt{ \frac{2}{ \kappa ^2 \epsilon}} \ln \left( - \sqrt{ \frac{ \epsilon^2 \kappa ^2 V_0}{ \epsilon- 3 }} \lambda \right) \nonumber \\ && + \alpha_{1, \phi}\, (- \lambda) ^{1 - 3/ \epsilon} + \alpha_{2,\phi}\, (-\lambda) ^{2- 6/ \epsilon} + \ldots + \beta_{1,\phi} \, (-\lambda) ^{2 - 2/ \epsilon} + \ldots \;,
\end{eqnarray}
where we have defined the fast-roll parameter
\begin{equation}
\epsilon \equiv \frac{c ^2}{2} \;.
\end{equation}
The field equations relate some of the (generally complex) $\alpha, \beta$ parameters to each other,
\begin{eqnarray}  \label{eq:relationPert1}
\alpha_{2, \phi} & = & \frac{45 - 33 \epsilon + 4 \epsilon ^2}{18(1 - 2 / \epsilon)} \frac{ \kappa}{ \sqrt{ 2 \epsilon}} \alpha_{1, \phi} ^2 \;, \\ \label{eq:relationPert2}
\alpha_{1, a} & = & \frac{ \sqrt{2 \epsilon}}{3} \kappa\,  \alpha_{1, \phi} \;,\\  \label{eq:relationPert3}
\alpha_{2, a} & = & \frac{1 - 4\epsilon + 9 / \epsilon}{36(1 - 2/ \epsilon)} \kappa ^2 \alpha_{1, \phi} ^2 \;,\\  \label{eq:relationPert4}
\beta_{1,a} & = & - \frac{1- 3 \epsilon}{3(1- \epsilon)} \frac{ \kappa}{ \sqrt{2 \epsilon}} \beta_{1, \phi} \;,
\end{eqnarray}
while the Friedmann equation fixes $ \beta_{1, \phi},$ 
\begin{equation}
\beta_{1, \phi} = - \frac{K}{a_0 ^2} \frac{ 3 \epsilon ^{5/2}}{ \sqrt{2} \kappa (2- \epsilon - 3 \epsilon ^2)} \;,
\end{equation}
with $K=1$ for our closed spatial slices. Thus we see that the correction due to spatial curvature is subdominant, even when compared to the second order correction proportional to $ \alpha_{1, \phi} ^2$. The important point is that all correction terms die off as long as the fast-roll parameter $\epsilon$ is large enough - more precisely, we must require 
\begin{equation}
\epsilon > 3 \quad \leftrightarrow \quad c^2 > 6 \;.
\end{equation} 
This condition is precisely the definition of an ekpyrotic phase, and it corresponds to the requirement that the pressure be larger than the energy density. The fact that the instanton reaches a classical history can then be viewed as a consequence of the ekpyrotic attractor mechanism \cite{Creminelli:2004jg}.

Now we can also look more closely at the on--shell Euclidean action in the approach to the crunch. Along the final segment of the integration contour, it is given by
\begin{equation}
S_E = \tilde{S}_{E} + \frac{4 \pi ^2 i}{ \kappa ^2} \int_\lambda^0 d\lambda^\prime\, ( -3 a + \kappa ^2 a ^3 V) \;,
\end{equation}
where $\tilde{S}_{E}$ denotes the action integral along the earlier parts of the contour. The first term in the integrand clearly poses no problem to the convergence of the integral for $ \lambda \rightarrow 0^-$ as it gives finite real and imaginary contributions, both scaling as $ (-\lambda) ^{1+1/ \epsilon}$. The asymptotic behaviour of $S_E$ can then be obtained by using the asymptotic series for $a$ and $ \phi$:
\begin{eqnarray}\nonumber
S_E & = & S_{E,crunch} - i\, \frac{4 \pi ^2 a_0 ^3}{ \kappa ^2 \epsilon} \left\{( - \lambda) ^{-1+3/ \epsilon} + \frac{3 (3- \epsilon)}{ \epsilon} \left( \alpha_{1, a} - \frac{ \sqrt{2 \epsilon}}{3} \kappa\, \alpha_{1, \phi} \right) \ln{(- \lambda)} \right. \\
&& \left. - \left(3 \alpha_{1, a} ^2 + 3 \alpha_{2, a} - 3 \sqrt{ 2 \epsilon}\, \kappa\, \alpha _{1, a} \alpha_{1, \phi} + \epsilon\, \kappa ^2 \alpha_{1, \phi} ^2 - \sqrt{2 \epsilon}\, \kappa\, \alpha_{2, \phi} \right) ( - \lambda) ^{1 - 3 / \epsilon}\right\} \;,
\end{eqnarray}
where $S_{E,crunch}$ is a constant. The term proportional to $ \ln{(- \lambda)}$ vanishes as a consequence of \eqref{eq:relationPert2}. Moreover, assuming the crunch is real, $ \textrm{Im}(a_0) = 0,$ and plugging in the relations (\ref{eq:relationPert1}, \ref{eq:relationPert3}, \ref{eq:relationPert4}) one gets to
\begin{equation}
\textrm{Re}(S_E) = S_{E, crunch} ^{R} +  \frac{(27 - 63 \epsilon + 42 \epsilon ^2 - 8 \epsilon ^3)\pi ^2 a_0 ^3  }{9 \epsilon\, (2- \epsilon)} \textrm{Im}( \alpha_{1, \phi} ^2) \, (- \lambda) ^{1 - 3/ \epsilon} \;.
\end{equation}
Hence the real part of $S_E$ approaches a finite constant as long as $\epsilon > 3$. By contrast, the imaginary part, which is proportional to $( - \lambda)$ to a negative power, is clearly divergent as $\lambda \rightarrow 0^-$. 

\begin{figure}[]%
\begin{minipage}{\smallWidthLeft} \flushleft
\includegraphics[width=\smallWidthRight]{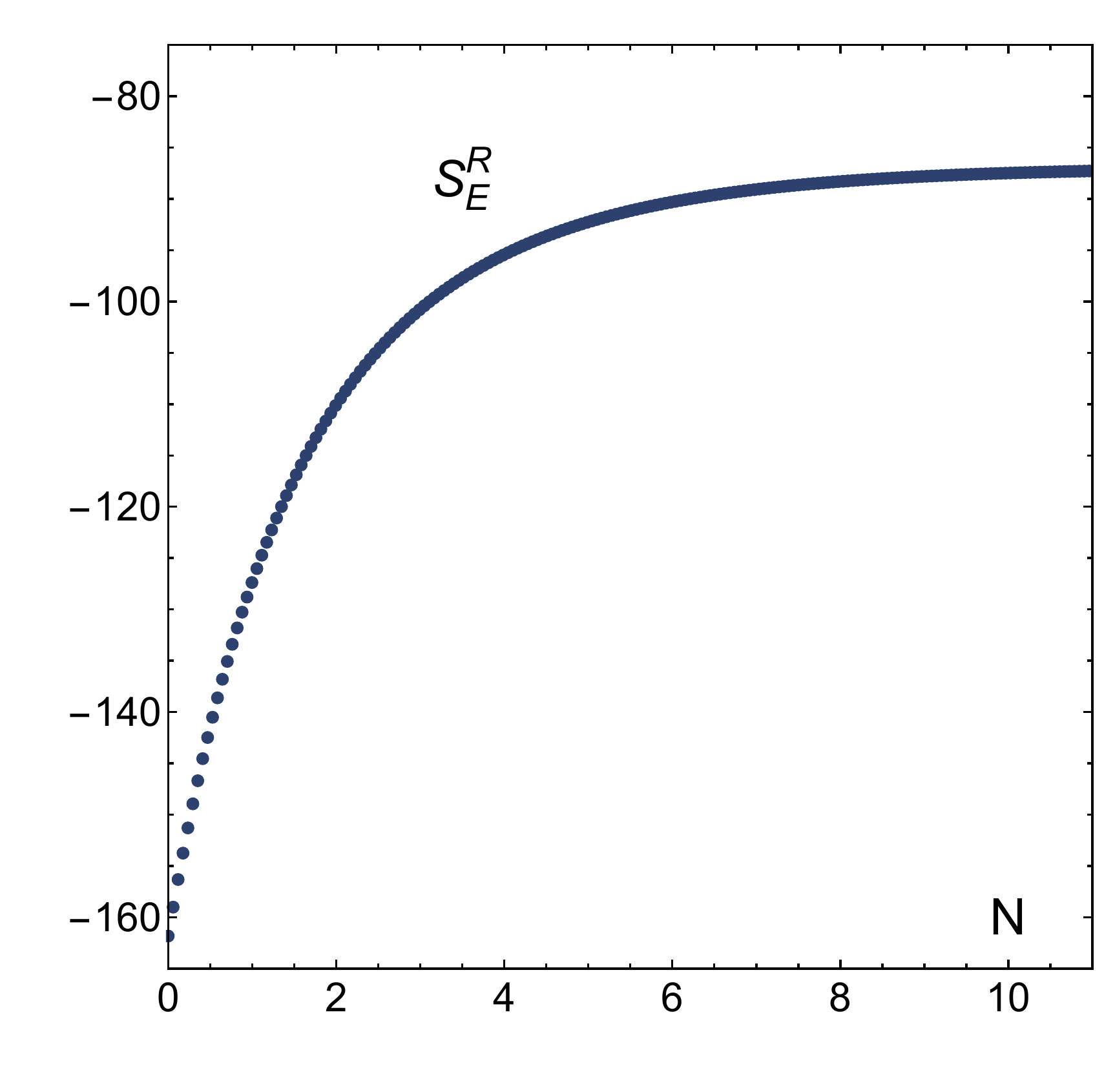}
\end{minipage}%
\begin{minipage}{\smallWidthRight} \flushleft
\includegraphics[width=\smallWidthRight]{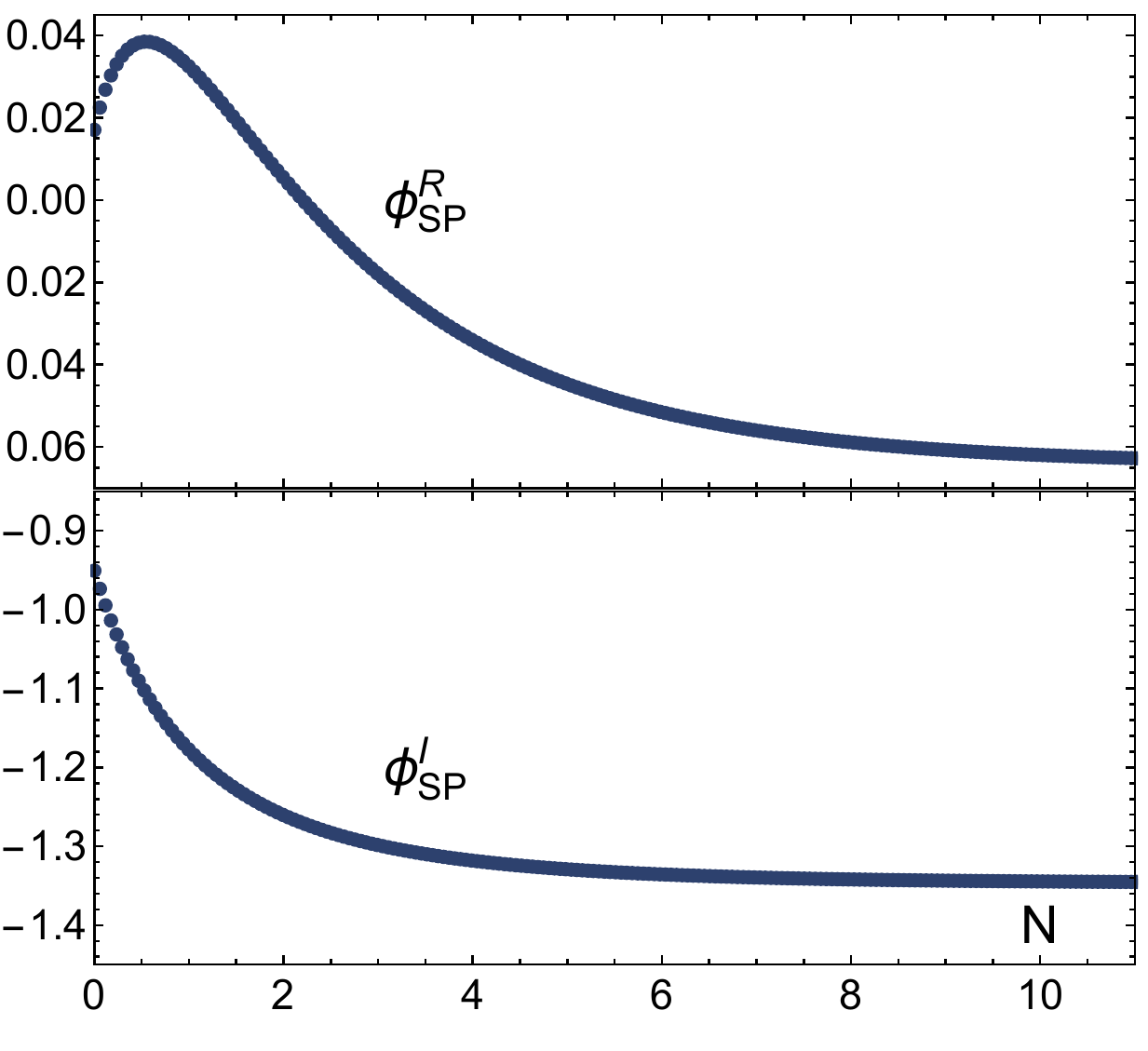}
\end{minipage}%
\caption{\label{FigReS} In order to assess the approach to classicality, we are interested in the no-boundary wavefunction $\Psi(b,\chi)$ along a series of $b,\chi$ values corresponding to a classical solution of the equations of motion. For this purpose, we have calculated $200$ instantons at equal $\chi$ intervals between the values $(b=3/2,\chi=0)$ and $(b=3/2 \times e^{-10/\sqrt{2\epsilon}},\chi=-10),$ with $\epsilon = 9/2.$ The left panel shows the corresponding real part of the action, as a function of the number of e-folds of ekpyrosis that have elapsed from our starting point. The right panel shows the corresponding values (real and imaginary parts) of the scalar field at the South Poles of the instantons. Both panels show indicate that a classical history is reached: the real part of the action asymptotes to a constant, while the value of $\phi_{SP}$ also stabilises as the ekpyrotic phase progresses.}
\end{figure}

So far we have shown in detail how the real part of the action approaches a constant for one particular instanton solution. In the no-boundary context, we are also interested in the classicality (in the WKB sense) of a family of instantons, with final values $(b_\lambda, \chi_\lambda)$ representing a real Lorentzian history of the universe, i.e. final values that correspond to a solution of the classical field equations,
\begin{equation}
a=a_0 (-\lambda)^{1/\epsilon}, \qquad \phi = \sqrt{\frac{2}{\kappa^2 \epsilon}} \ln \left( - \sqrt{\frac{\epsilon^2 \kappa^2 V_0}{\epsilon - 3}} \lambda\right)\,.
\end{equation}
We have calculated the no-boundary wavefunction along such a series of values, for $\epsilon = 9/2,$ from the initial wavefunction arguments $(b=3/2, \chi = 0)$ up to $(b=3/2 \times e^{-10/\sqrt{2\epsilon}}, \chi = -10)$\footnote{These values were chosen purely for numerical convenience - see also the comment at the end of the present subsection.}. Fig. \ref{FigReS} shows the relevant results, which we have plotted as a function of the number of e-folds $N$ of ekpyrosis,
\begin{equation}
dN \equiv d \ln |aH| \quad \rightarrow N = - (\epsilon - 1) \ln b + N_0,
\end{equation}
where we have defined $N=0$ to correspond to the start of our series, i.e. to $(b=3/2, \chi=0).$ The left panel in the figure shows that along the series, the real part of the action reaches a constant value. The corresponding values of the scalar field at the South Pole of the respective instantons are plotted in the right panel, where it is also evident that an asymptotic value is being neared. These are strong indications that we may in fact have approached a classical history. 

\begin{figure}[]%
\begin{minipage}{\smallWidthLeft} \flushleft
\includegraphics[width=\smallWidthRight]{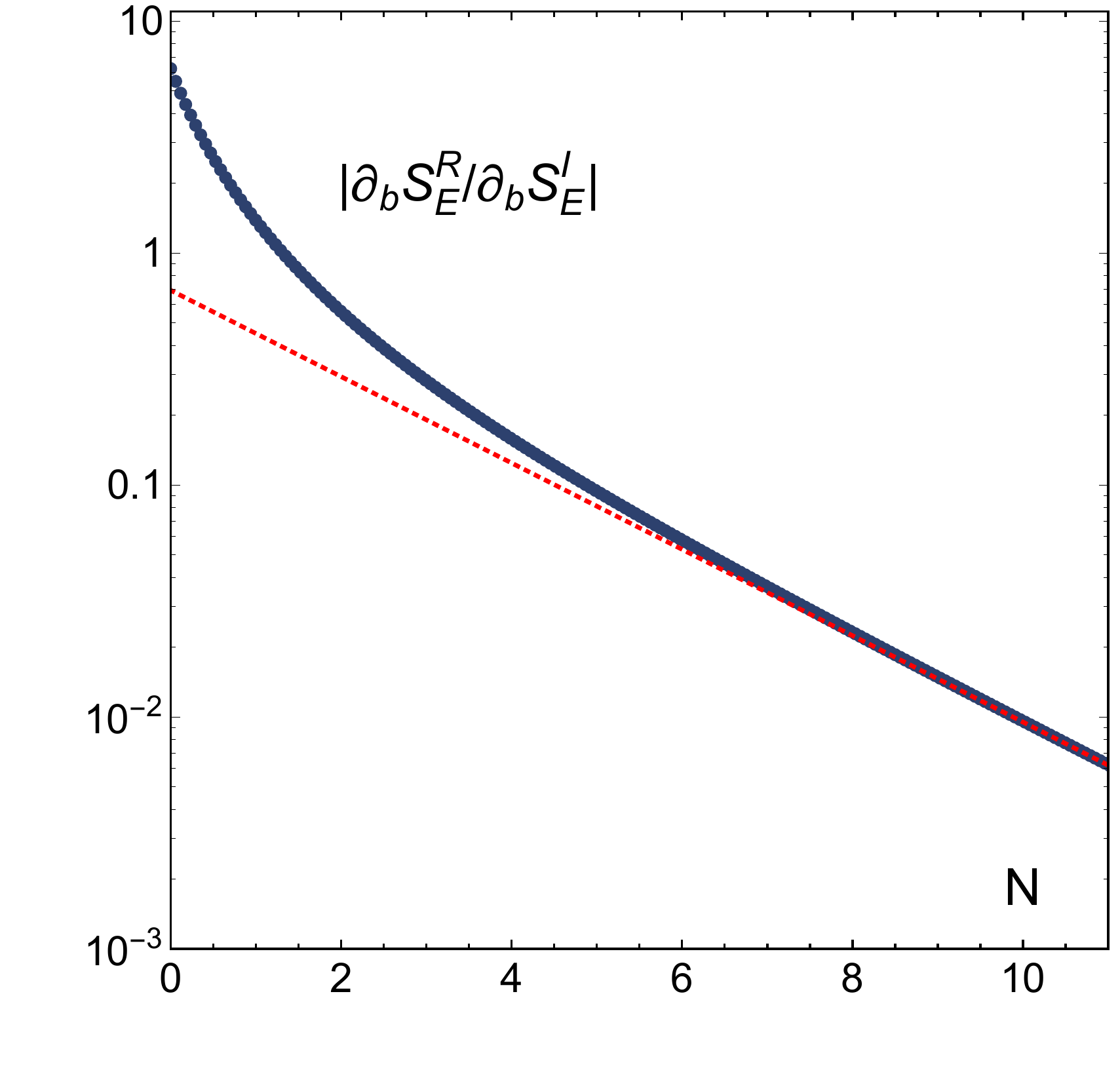}
\end{minipage}%
\begin{minipage}{\smallWidthRight} \flushleft
\includegraphics[width=\smallWidthRight]{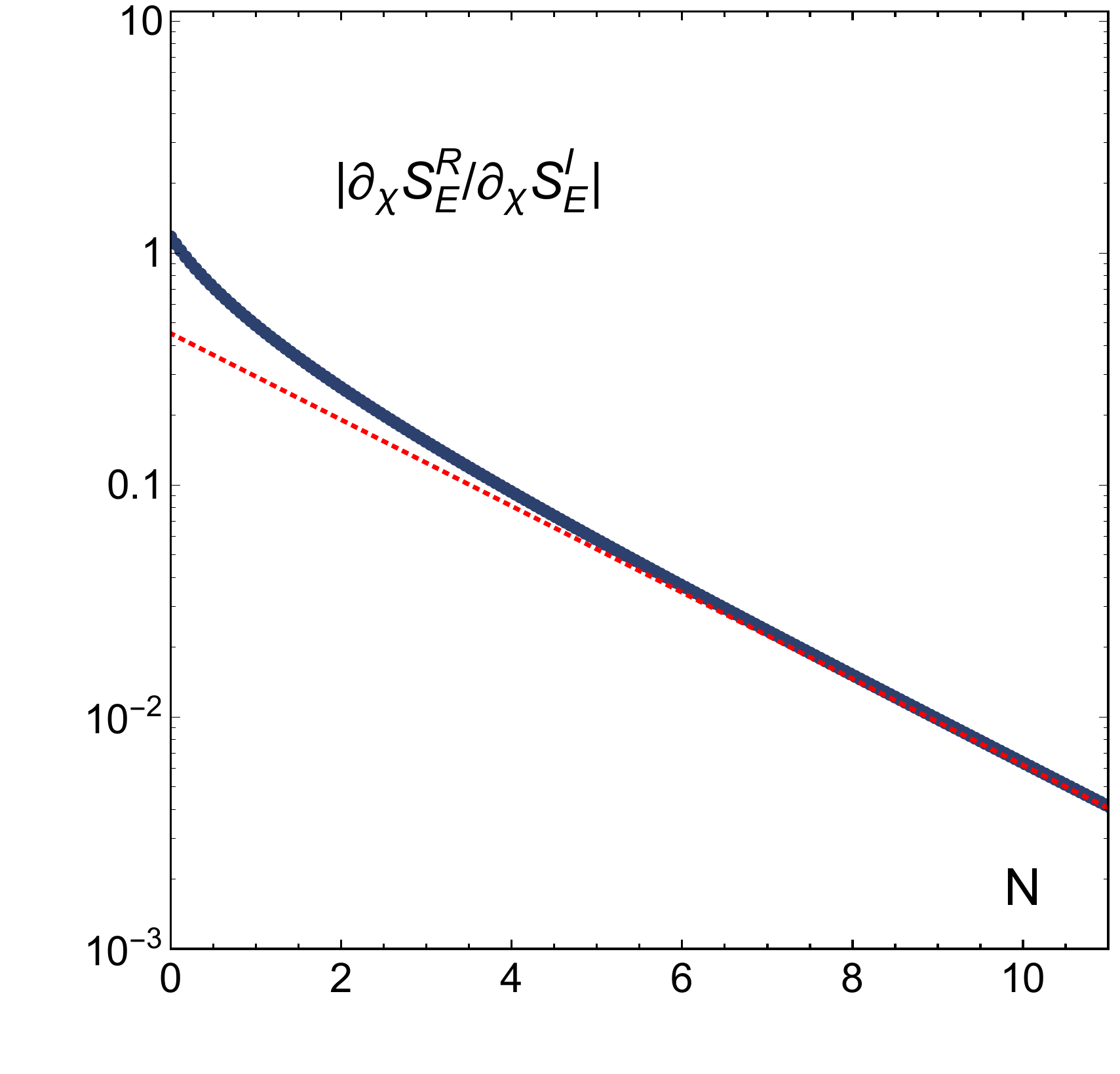}
\end{minipage}%
\caption{\label{FigWKB} The most precise criterion for classicality are the WKB conditions (\ref{eq:firstCondition}), (\ref{eq:secondCondition}) on the wavefunction. To verify these, we have evaluated the $b$ and $\chi$ derivatives of the Euclidean action along the same series of instantons as those shown in Fig. \ref{FigReS} by computing series of instantons with slightly shifted values of $b$ and $\chi$ (we have considered fractional shifts in $b$ and shifts in $\chi$ of sizes $10^{-6}, 10^{-5}$ and verified that finite difference estimates of the derivatives have converged to sufficiently high  precision) at each of the $200$ evaluated points shown in Fig. \ref{FigReS}. The results of these numerical computations show that both the $b$ and $\chi$ derivatives of the real part of the action become small compared to the respective derivatives of the imaginary parts. Moreover, the dotted red lines indicate a scaling $e^{-(\epsilon - 3)N/(\epsilon -1)},$ which can be seen to be approached by both WKB conditions as the ekpyrotic phase proceeds. In the main part of the text, we provide an analytic derivation of these scalings.}
\end{figure}

However, the most precise criterion for classicality (in the present, restricted context) stems from an analysis of WKB classicality. WKB classicality requires the modulus of the wavefunction to vary slowly compared to the phase of the wavefunction, as we evolve along this family of instantons. In other words, WKB classicality is synonymous with the following conditions being satisfied: 
\begin{eqnarray}
| \partial _{b} S_E ^{R}| & \ll & | \partial _{b} S_E ^{I}| \;,\label{eq:firstCondition} \\
| \partial _{ \chi} S_E ^{R}| & \ll & | \partial _{ \chi} S_E ^{I}| \;. \label{eq:secondCondition}
\end{eqnarray}
We have evaluated these derivatives numerically -- see Fig. \ref{FigWKB}. As the figure demonstrates, both WKB conditions are satisfied with increasing precision as the ekpyrotic phase proceeds. Moreover, the approach to classicality follows a precise asymptotic scaling, as both WKB conditions improve as $e^{-(\epsilon - 3)N/(\epsilon -1)}.$ As seen in the figure, after about 10 e-folds the derivatives of the real part of the action have reached a level of $10^{-2}$ of the derivatives of the imaginary part. This can be extrapolated to a ratio of $10^{-12}$ after $60$ e-folds, and to a value of about $10^{-23}$ after $120$ e-folds (as would be the relevant number in a cyclic model). This result clearly demonstrates that the no-boundary wavefunction predicts classical ekpyrotic histories \footnote{As indicated at the beginning of this section, these statements should be interpreted in the mini-superspace context considered here. It would be interesting to study more refined notions of the quantum-to-classical transition, such as decoherence, but such a study will require a more extended framework.}.

The scaling of the WKB conditions with the number of e-folds can in fact be understood analytically, as we will now show. For this purpose, we make use of the fact that the background reaches the ekpyrotic scaling solution asymptotically. First, we must find the dependence of the action on $b$ and $\chi.$ The shift/re-scalings in Eq. \eqref{eq:scaling} imply that (keeping $\kappa, V_0$ fixed here so as to remain in the same theory) the field equations are invariant under the transformations
\begin{eqnarray}
\bar{a} ( \bar{ \lambda}) & = & e^{c \Delta \phi / 2} \, a \left( e^{- c\, \Delta \phi /2} \bar{ \lambda} \right) \;, \label{Rescaling1}\\
\bar{ \phi}( \bar{ \lambda}) & = & \phi\left( e^{- c\, \Delta \phi /2} \bar{ \lambda} \right) + \Delta \phi \;. \label{Rescaling2}
\end{eqnarray}
These transformations then imply that, if
\begin{equation}
a = a_0\, (-\lambda) ^{1/ \epsilon} \;,\qquad
V( \phi) =  - \frac{ \epsilon - 3}{ \epsilon}  \frac{1}{ \lambda ^2}\;,
\end{equation}
then
\begin{equation} \label{eq:rescalinga0}
\bar{ a} = \bar{a}_0 \, (-\bar{ \lambda}) ^{1/ \epsilon} \;, \qquad \bar{a}_0 = \textrm{exp} \left( \frac{ \epsilon - 1}{ \epsilon}\frac{ c \, \Delta \phi}{2}  \right) \, a_0 \;,\qquad
V( \bar{ \phi}) =  - \frac{ \epsilon - 3}{ \epsilon} \frac{1}{ \bar{ \lambda} ^2} \;.
\end{equation}
Hence $a_0$ labels a continuous family of (asymptotic) ekpyrotic solutions. In fact, this label represents a particular constant of motion along each trajectory:
\begin{equation} \label{eq:labela0}
a_0 = a\,\left( -\frac{\epsilon}{\epsilon - 3}V \right)^{1/2 \epsilon} \;.
\end{equation}
The imaginary part of the Euclidean action along a classical trajectory scales as
\begin{equation}
S_E^I \sim i \, \int d \lambda\, a ^3\, V \sim - i \,a_0 ^3\, (-\lambda) ^{- 1 + 3/ \epsilon} \sim -i \, a_0 ^3\, |V| ^{ \frac{1}{2} - \frac{3}{2 \epsilon}} \;.
\end{equation}
Re--expressing the label $a_0$ as in \eqref{eq:labela0} then leads to
\begin{equation}
S_E^I \sim - i \, b ^3\, |V( \chi)| ^{1/2} \;.
\end{equation}
The scaling of the real part of the Euclidean action implies that 
\begin{equation}
\bar{S}_E ^{R} = e^{ c\, \Delta \phi} S_R = \left( \frac{ \bar{a}_0}{a_0} \right) ^{2 \epsilon/( \epsilon-1)} S_E ^{R} \;,
\end{equation}
and  hence
\begin{equation}
S_E ^{R} \sim a_0 ^{ \frac{2 \epsilon}{\epsilon - 1}} \sim b ^{\frac{2 \epsilon}{\epsilon - 1}} |V( \chi)| ^{1/ ( \epsilon-1)} \;.
\end{equation}
The sign is unspecified by these arguments, but turns out to be negative as the numerical analysis shows. We are now in a position to understand the asymptotic scaling of the WKB classicality conditions \eqref{eq:firstCondition} and \eqref{eq:secondCondition} in the approach to the crunch, along curves of constant $a_0,$ i.e. along wavefunction arguments corresponding to solutions of the classical equations of motion. The second condition \eqref{eq:secondCondition} is clearly satisfied, since the derivatives w.r.t. $\chi$ only add a constant prefactor and the ratio $| S_E ^{R} / S_E ^{I}|$ decreases along each classical history as
\begin{equation}
\frac{S_E^R}{S_E^I} \sim \frac{b ^{\frac{2 \epsilon}{\epsilon - 1}} |V( \chi)| ^{1/ ( \epsilon-1)}}{b ^3\, |V( \chi)| ^{1/2}} \sim b^{\epsilon -3} \sim e^{-(\epsilon - 3)N/(\epsilon - 1)},
\end{equation}
where in the second to last step we have used Eq. (\ref{eq:labela0}) once more. As for the first condition, we can evaluate
\begin{eqnarray}
\partial _{b} S_E ^{I} &\sim& b ^2 |V( \chi)| ^{1/2} \sim (-\lambda) ^{- \frac{ \epsilon - 2}{ \epsilon}} \;,\\
\partial _{b} S_E ^{R} &\sim& b ^{ \frac{ \epsilon + 1}{ \epsilon-1}} |V( \chi)| ^{1/( \epsilon-1)} \sim (-\lambda) ^{- 1/ \epsilon} \;,
\end{eqnarray}
so that 
\begin{equation}
\left| \frac{ \partial _{b} S_E ^{R}}{ \partial _{b} S_E ^{I} }\right| \sim (-\lambda) ^{\frac{ \epsilon - 3}{ \epsilon}} \sim b^{\epsilon - 3} \sim e^{-(\epsilon - 3)N/(\epsilon - 1)} \;.
\end{equation}
Thus we can see that both WKB conditions are satisfied exponentially fast with the number of e-folds $N.$ As discussed above in relation to our numerical results in Fig. \ref{FigWKB}, this asymptotic scaling with $N$ is reached very precisely after a few e-folds already. Note also the $\epsilon$ dependence of the WKB conditions: they will be satisfied as long as $\epsilon > 3.$ Thus we see once more that the ekpyrotic attractor, which only exists when $\epsilon > 3,$ allows the universe to become classical.

At the risk of repeating ourselves, we would like to emphasise one further point: the numerical calculations have all been performed in the re-scaled theory \eqref{eq:actionrescaled}. Using the re-scalings in Eq. \eqref{eq:scaling} one can relate these to any desired (physically meaningful) values of the scale factor $b$ and scalar field $\chi$ via the relations 
\begin{equation}
\chi  =  \kappa ^{-1} \bar{ \chi} + \Delta \chi \;, \quad
b  =  \frac{e^{c \kappa \Delta \chi/2}}{\kappa V_0^{1/2}} \bar{ b} \;,
\end{equation}
where the corresponding instantons then have the South Pole value $\kappa^{-1} \bar\phi_{SP} + \Delta \chi$ and the crunch occurs at $\frac{e^{c \kappa \Delta \chi/2}}{\kappa V_0^{1/2}} \bar\tau_{crunch}.$ The values of the scale factor and scalar field used in generating Figs. \ref{FigReS} and \ref{FigWKB} were chosen for numerical convenience, and the small numerical values of $b$ do not indicate that we are in a sub-Planckian regime. Interestingly, the scaling symmetry implies that classicality will be reached in an equal manner for all histories related by the transformations above.

\subsection{Families of Ekpyrotic Instantons}

\begin{figure}[]
\begin{minipage}{\smallWidthLeft} \flushleft
\includegraphics[width=\smallWidthRight]{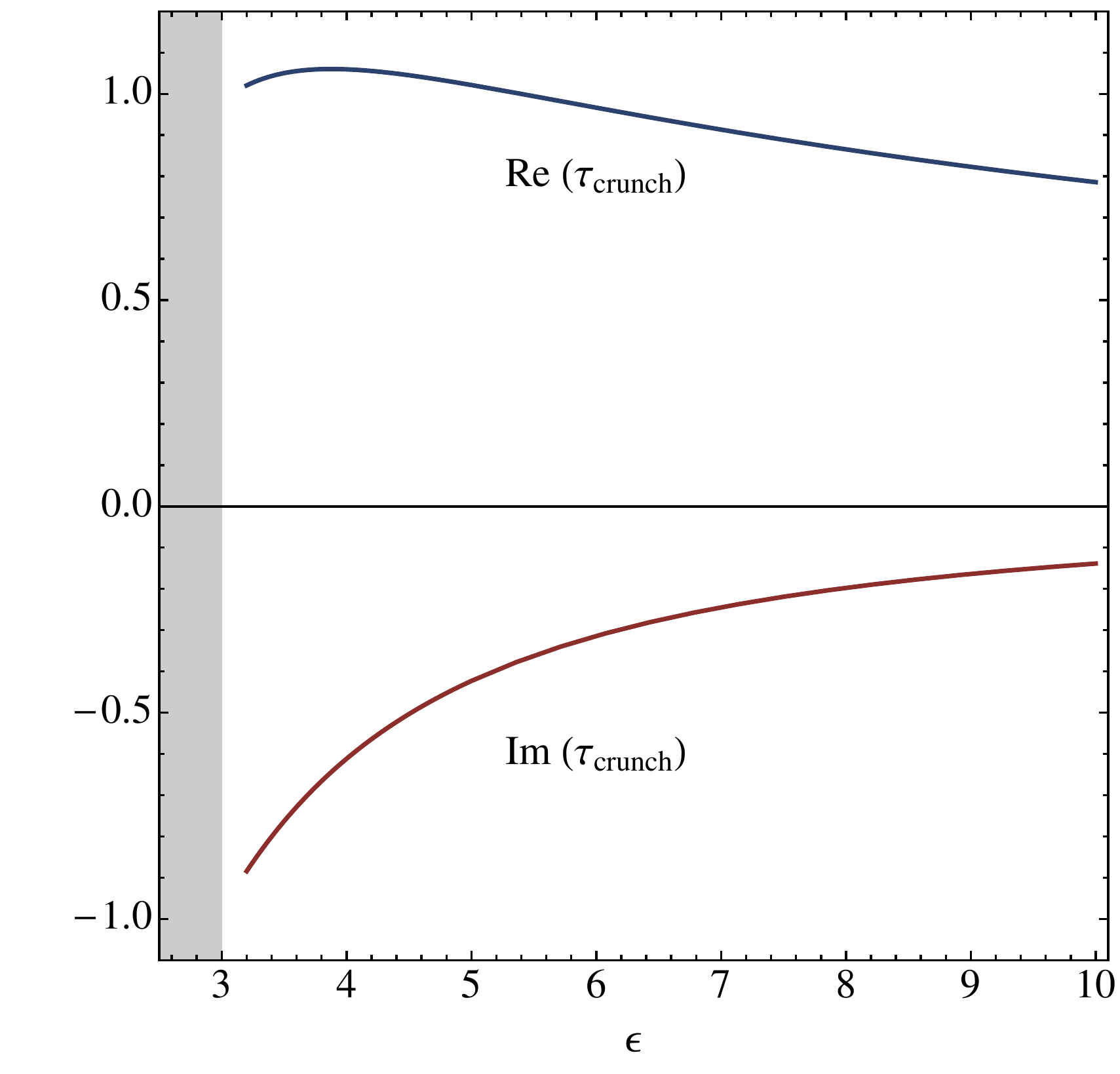}
\end{minipage}%
\begin{minipage}{\smallWidthRight} \flushleft
\includegraphics[width=\smallWidthRight]{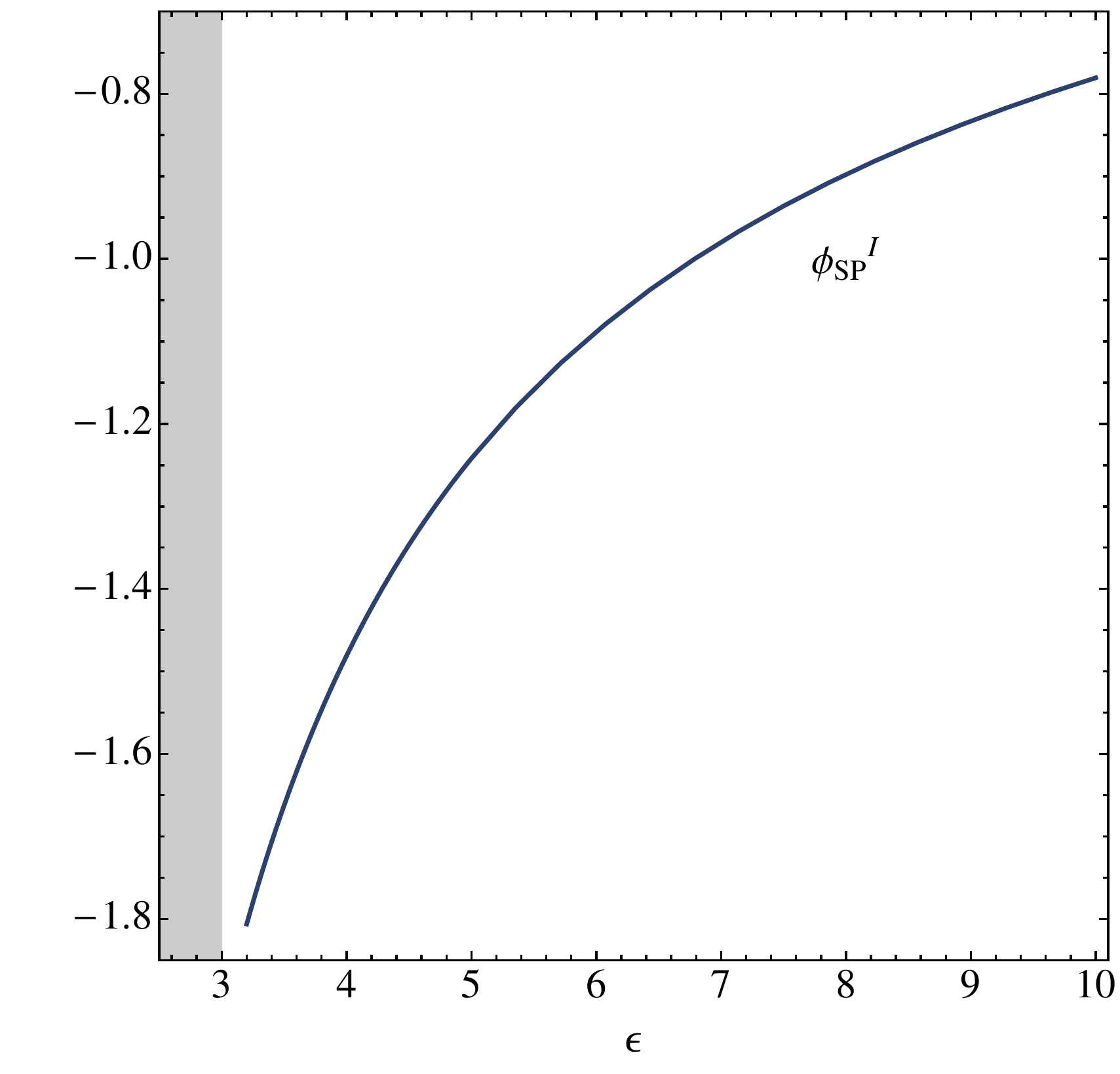}
\end{minipage}%
\caption{\label{Fig6} Left panel: The location of the crunch in the complex $\tau$ plane as a function of the fast-roll parameter $\epsilon.$ Right panel: The imaginary part of the scalar field at the South Pole of the instanton, as a function of the fast-roll parameter $\epsilon.$ The value of $\phi_{SP}^I$ is tuned such that classical histories arise.}
\end{figure}

\begin{figure}[]
\centering
\begin{minipage}{\smallWidthRight}
\includegraphics[width=\smallWidthRight]{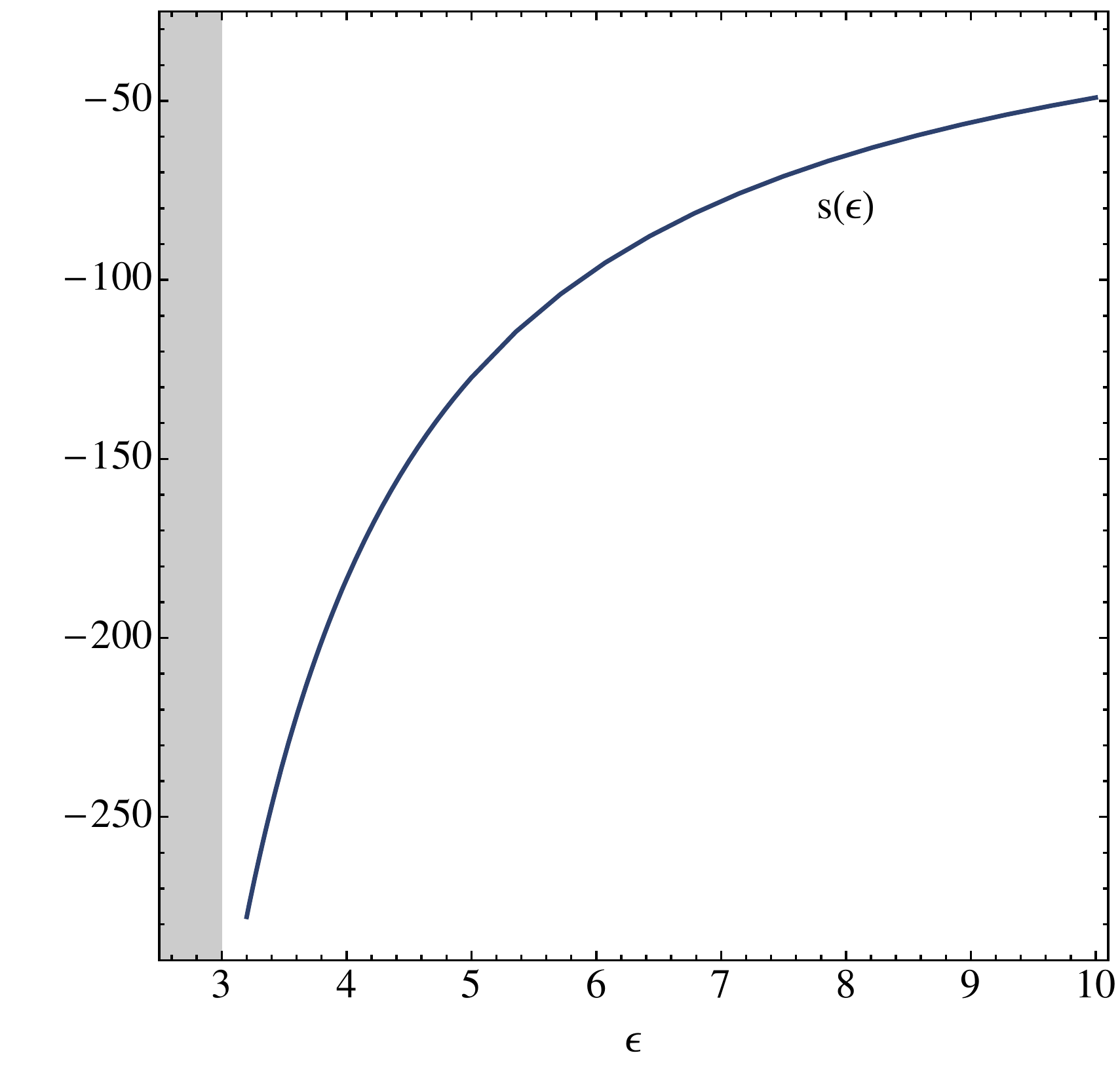}
\end{minipage}%
\caption{\label{Fig8} The real part of the Euclidean action as a function of the fast-roll parameter $\epsilon,$ once a classical history has been reached. Here we have taken $\phi_{SP}^R=0,$ and thus $|V(\phi_{SP}^R)|=1,$ so that this graph also represents $s(\epsilon),$ cf. Eq. \ref{eq:ekprob}. All else being equal, smaller values of $\epsilon$ are seen to be preferred.}
\end{figure}

As a consequence of the shift symmetry \eqref{eq:metricscaling}, which relates instantons with different $\phi_{SP}^R$ but for the same potential \eqref{eq:ekpot}, ekpyrotic instantons come in families that can be characterised solely by the steepness $c$ of the potential. Thus we may revert back to setting $\phi_{SP}^R=0$ and study the effect of changing the steepness $c,$ or equivalently, the fast-roll parameter $\epsilon=c^2/2.$ The discussion of the approach to classicality suggests that it is impossible to find ekpyrotic instantons with a late-time classical behaviour when $\epsilon < 3.$ Our numerical studies confirm this expectation. Fig. \ref{Fig6} shows how the location of the crunch in the $\tau$ plane, as well as the imaginary part $\phi_{SP}^I$ at the origin, change when $\epsilon$ is varied. In fact nothing spectacular happens, in the sense that all these instantons (and the solutions related by a shift $\phi_{SP}^R + \Delta \phi$ ) are qualitatively similar: in particular, all of them have a significant imaginary component $\phi_{SP}^I,$ and the crunch always occurs in the bottom right quadrant, $\textrm{Re}(\tau_{crunch})>0, \,\textrm{Im}(\tau_{crunch})<0$. As Fig. \ref{Fig8} shows, all of these instantons settle to a negative value for the real part of the Euclidean action, once a classical history is reached. If we denote $Re(S_E(\phi_{SP}^R=0)) \equiv s(\epsilon)$ and then re-insert the dependence of the action on $\phi_{SP}^R,$ we obtain the following formula for the relative probability of the various implied ekpyrotic histories:
\begin{equation} \label{eq:ekprob}
\Psi^\star\Psi \sim \mathrm{Exp}\left[-2Re(S_E)\right] = \mathrm{Exp}\left[\frac{-2s(\epsilon)}{|V(\phi_{SP}^R)|}\right].
\end{equation}
Thus, small values of $\epsilon$ come out as preferred, alongside small magnitudes for the potential $|V(\phi_{SP}^R)|.$ Hence, in a potential energy landscape allowing different ekpyrotic regions, the preference will be for the shallowest ekpyrotic potential (though of course we always must have $\epsilon > 3$), with a starting value as far up the potential as possible. 

It is of interest to compare this formula for the relative probabilities with that obtained in inflation. As we discussed earlier, the relative probability of inflationary instantons can be estimated reliably by approximating the instanton with half of a Euclidean 4-sphere. In that case, the above formula Eq. \eqref{eq:ekprob} would also hold, but with $s(\epsilon) = -12 \pi^2$ being a constant. Note that $s(\epsilon)$ is of the same order of magnitude in the ekpyrotic case, and hence the dominant factor in estimating the relative probability of the various possible instantons comes from the dependence on the starting value of the potential, $|V(\phi_{SP}^R)|.$ The overall preference for a small magnitude of the potential implies that, if a potential contains both inflationary and ekpyrotic regions, ekpyrotic instantons will be vastly preferred. Thus, the implications of the no-boundary proposal have to be updated, as it is now clear that the most likely origin of the universe is in an ekpyrotic contracting phase.

\section{The No-Boundary Wavefunction for a Cyclic Universe} \label{section:cyclic}

In the cyclic universe, a small constant is added to the potential, such that a dark energy plateau is obtained at large positive values of $\phi,$
\begin{equation}
V(\phi) = \frac{3 H^2}{\kappa^2} \left( 1- e^{-c\kappa \phi}\right),
\end{equation}
where $H$ is a positive constant (corresponding to the asymptotic value of the Hubble rate of dark energy expansion at large $\phi$). The cyclic scenario proposes that we currently find ourselves at the onset of this dark energy phase, and that at some time in the far future, the potential will diminish and become negative. This will cause the universe to revert from expansion to contraction. During the ensuing phase of ekpyrotic contraction, anisotropies are suppressed and quantum fluctuations can be amplified into classical density fluctuations. The ekpyrotic phase is then followed by a bounce into a radiation dominated phase of the universe (where reheating occurs at the bounce), and the standard phases of radiation and matter dominated expansion follow. Eventually, the universe becomes dominated by dark energy again, and the entire cycle starts afresh. For a detailed review of the cyclic picture, see \cite{Lehners:2008vx}. Let us just highlight one important feature, which is that from cycle to cycle the universe grows by a large amount. In other words, although the evolution of the Hubble rate is strictly cyclic, that of the scale factor is not: each new cycle is vastly bigger than the previous one. Thus, going back in time, any finite region was sub-Planckian in size a finite number of cycles ago. In this sense, the cyclic universe still requires initial conditions. In the present section, our aim is to explore the consequences of applying the no-boundary proposal, seen as a theory of initial conditions, in the cyclic context.

\begin{figure}[]
\centering
\begin{minipage}{\fullWidth}
\includegraphics[width=\fullWidth]{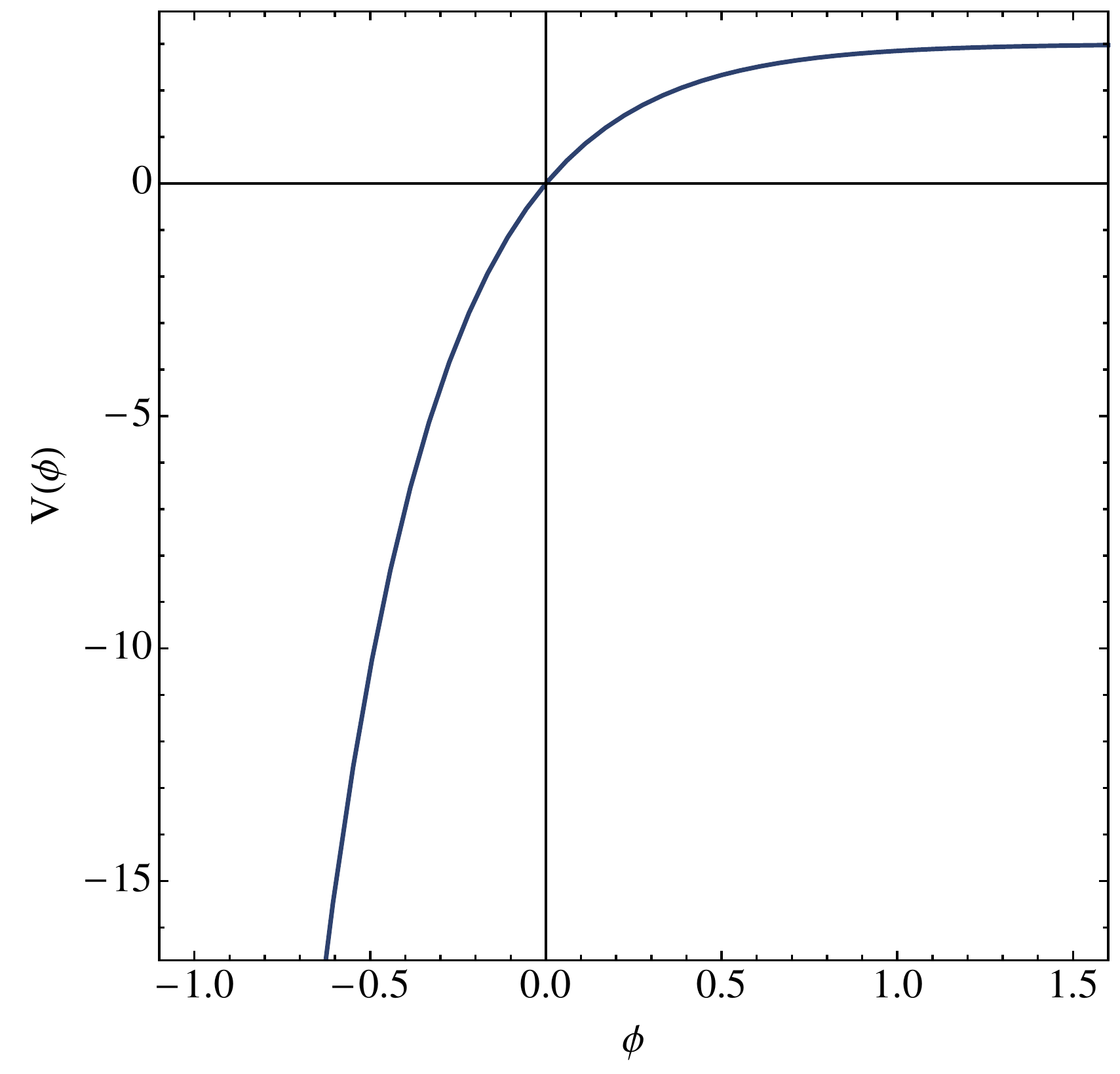}
\end{minipage}%
\caption{\label{Fig:CycPot} The cyclic universe potential combines an ekpyrotic phase (at negative $\phi$) with a dark energy plateau (at positive $\phi$). We find that on the plateau both ekpyrotic- and inflationary-type instantons co-exist.}
\end{figure}

An interesting aspect can immediately be deduced from looking at the shape of the potential in Fig. \ref{Fig:CycPot}: there is a region of steep ekpyrotic fall-off at negative $\phi,$ while the potential flattens out at positive $\phi.$ Thus, we may expect ekpyrotic instantons to exist on the left, and inflationary instantons to exist on the dark energy plateau. Our numerical analysis confirms both expectations, but in addition we find that ekpyrotic instantons continue to exist with starting values on the dark energy plateau. Thus, at positive $\phi$ we have the interesting situation that both types of instantons co-exist (with the same values of $\phi_{SP}^R$ but different imaginary parts $\phi_{SP}^I$). We will now present examples of these solutions.

\begin{figure}[t]%
\begin{minipage}{\smallWidthLeft} \flushleft
\includegraphics[width=\smallWidthRight]{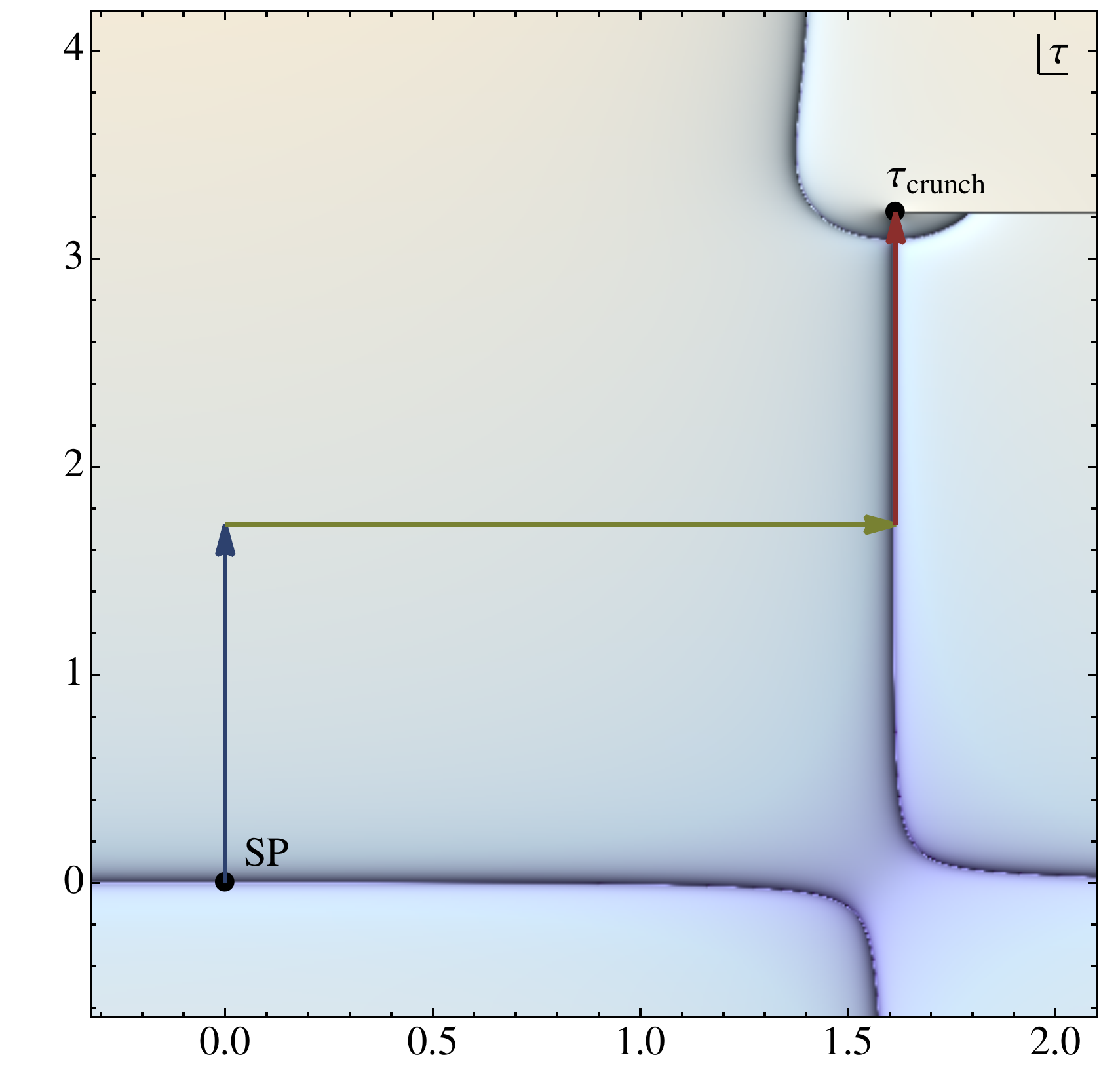}
\end{minipage}%
\begin{minipage}{\smallWidthRight} \flushleft
\includegraphics[width=\smallWidthRight]{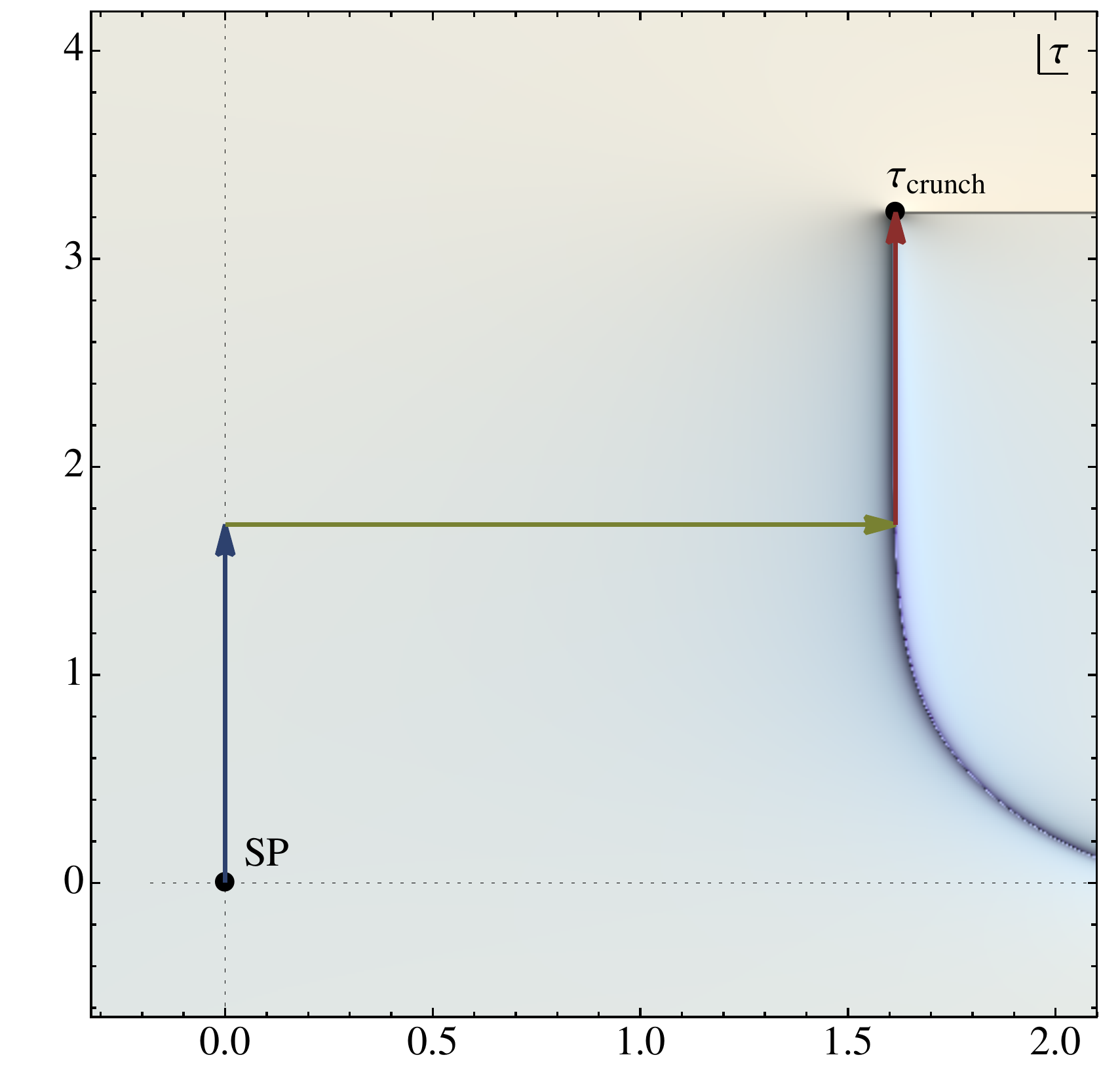}
\end{minipage}%
\caption{\label{Fig:CyclicdS1} A de Sitter-like instanton on the dark energy plateau, with $H=1$, $c=3$, $\phi_{SP}^R=1$, $\phi_{SP}^I\approx - 0.2276$ and $\tau_{crunch}\approx 1.615 + 3.222 i.$ This figure is obtained in an analogous manner to Fig. \ref{Fig1}.}
\end{figure}

\begin{figure}[]%
\begin{minipage}{\smallWidthLeft} \flushleft
\includegraphics[width=\smallWidthRight]{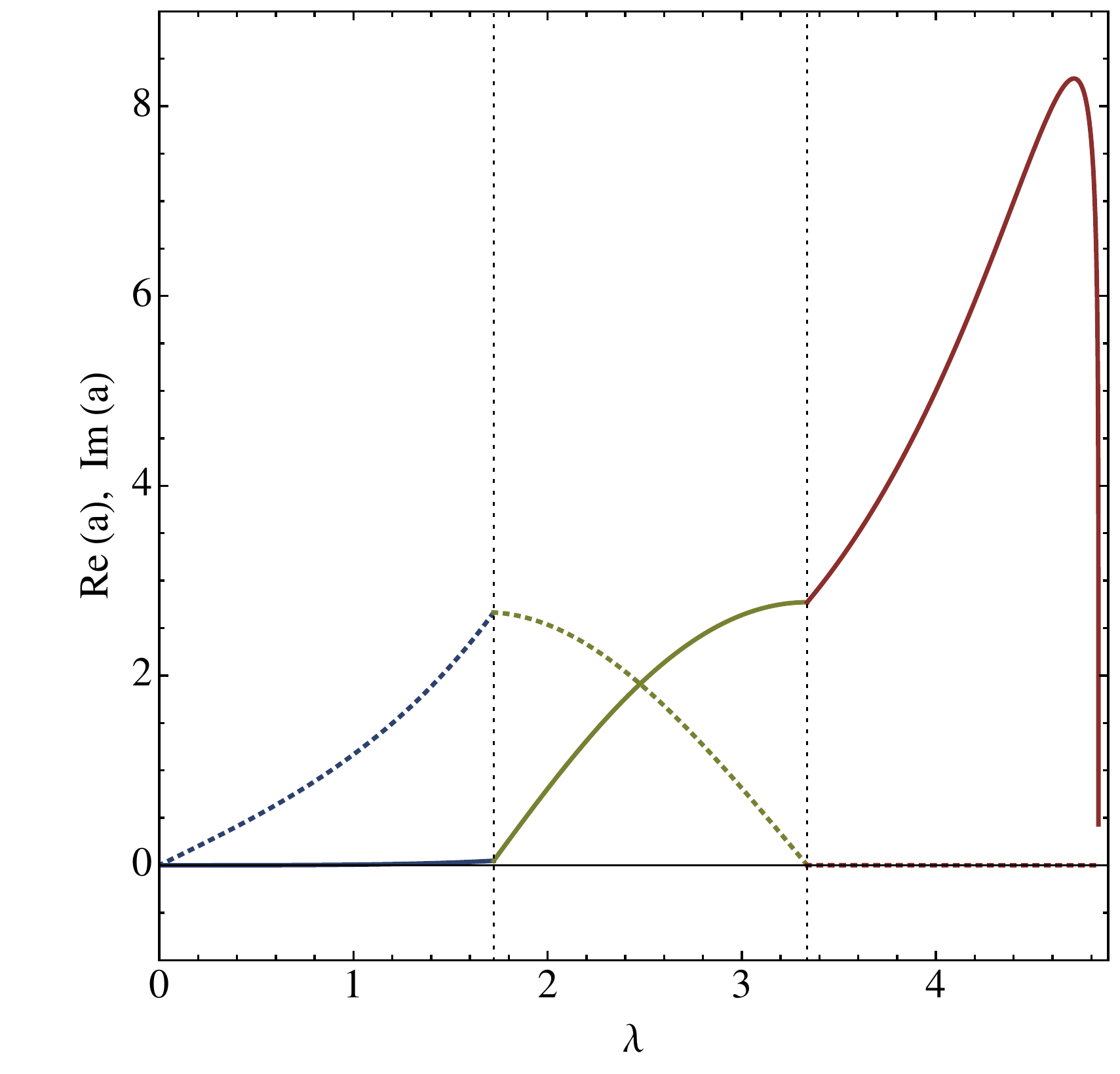}
\end{minipage}%
\begin{minipage}{\smallWidthRight} \flushleft
\includegraphics[width=\smallWidthRight]{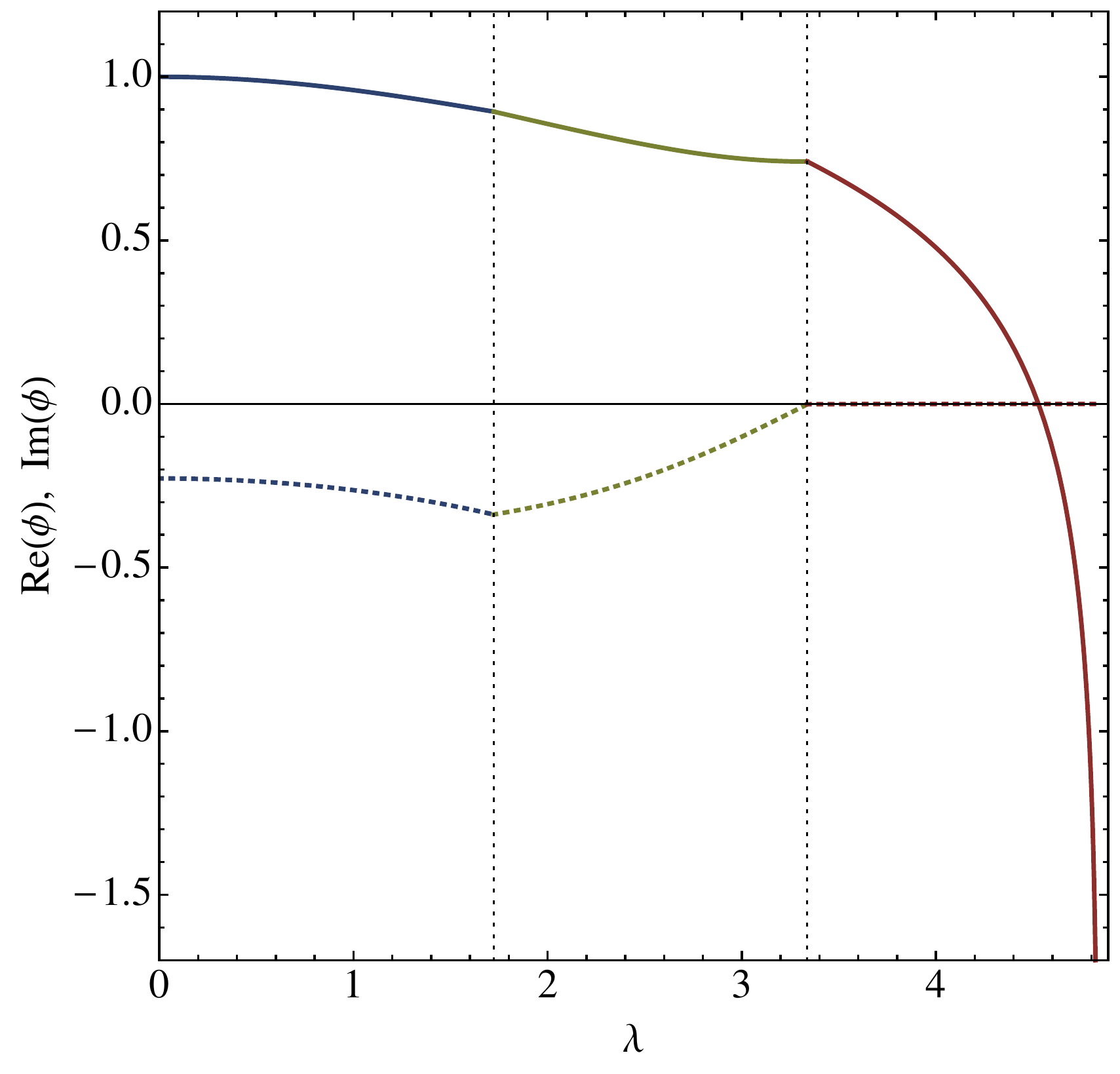}
\end{minipage}%
\caption{\label{Fig:CyclicdS2} The real (solid) and imaginary (dotted) parts of the scale factor (left panel) and scalar field (right panel) for the instanton depicted in Fig. \ref{Fig:CyclicdS1}, with an integration contour indicated by the arrows in that figure.}
\end{figure}

First, we will discuss the inflationary/de Sitter type instantons at positive $\phi$. Intuitively, one may expect instantons to exist which reach a classical history with the scalar field slowly rolling down the dark energy plateau, and then, once the universe is already classical, to evolve into the ekpyrotic contraction phase as the potential becomes negative. Fig. \ref{Fig:CyclicdS1} shows an example of such an instanton for $\phi_{SP}^R=1$, where the value of the imaginary part $\phi_{SP}^I$ has been tuned such that a classical history is achieved (meaning that the vertical lines of real scale factor and scalar field coincide). Again, since we have not added any dynamics that could lead to a bounce, all such histories end up in a crunch singularity at $\tau_{crunch}.$ By choosing an explicit integration contour, we can plot a particular representation of the instanton. Our choice of contour is indicated by the three coloured arrows in Fig. \ref{Fig:CyclicdS1}, and the real and imaginary parts of the scale factor and scalar field along this contour are shown in Fig. \ref{Fig:CyclicdS2}. Along the first segment, running vertically up from the origin, the spacetime is given approximately by a region of (opposite-signature) Euclidean Anti-de Sitter space (as discussed in more detail in \cite{Hertog:2011ky}). The second, horizontal segment is a fully complex geometry interpolating between the Euclidean Anti-de Sitter space and a Lorentzian de Sitter space corresponding to the phase of dark energy expansion. This second segment reaches the vertical line emanating vertically downwards from the location of the eventual crunch, and at the junction with the third integration segment the scale factor and scalar field are already real to very high precision. This is because of the inflationary-type attractor that evidently also applies to the dark energy phase. As one can see from Fig. \ref{Fig:CyclicdS2}, along the third segment the scale factor $a$ then expands at an accelerated rate until the scalar field $\phi$ has rolled to negative values of the potential, at which point the scale factor halts its expansion and starts shrinking. The ensuing ekpyrotic phase ends in a crunch after a finite time interval.

\begin{figure}[t]%
\begin{minipage}{\smallWidthLeft} \flushleft
\includegraphics[width=\smallWidthRight]{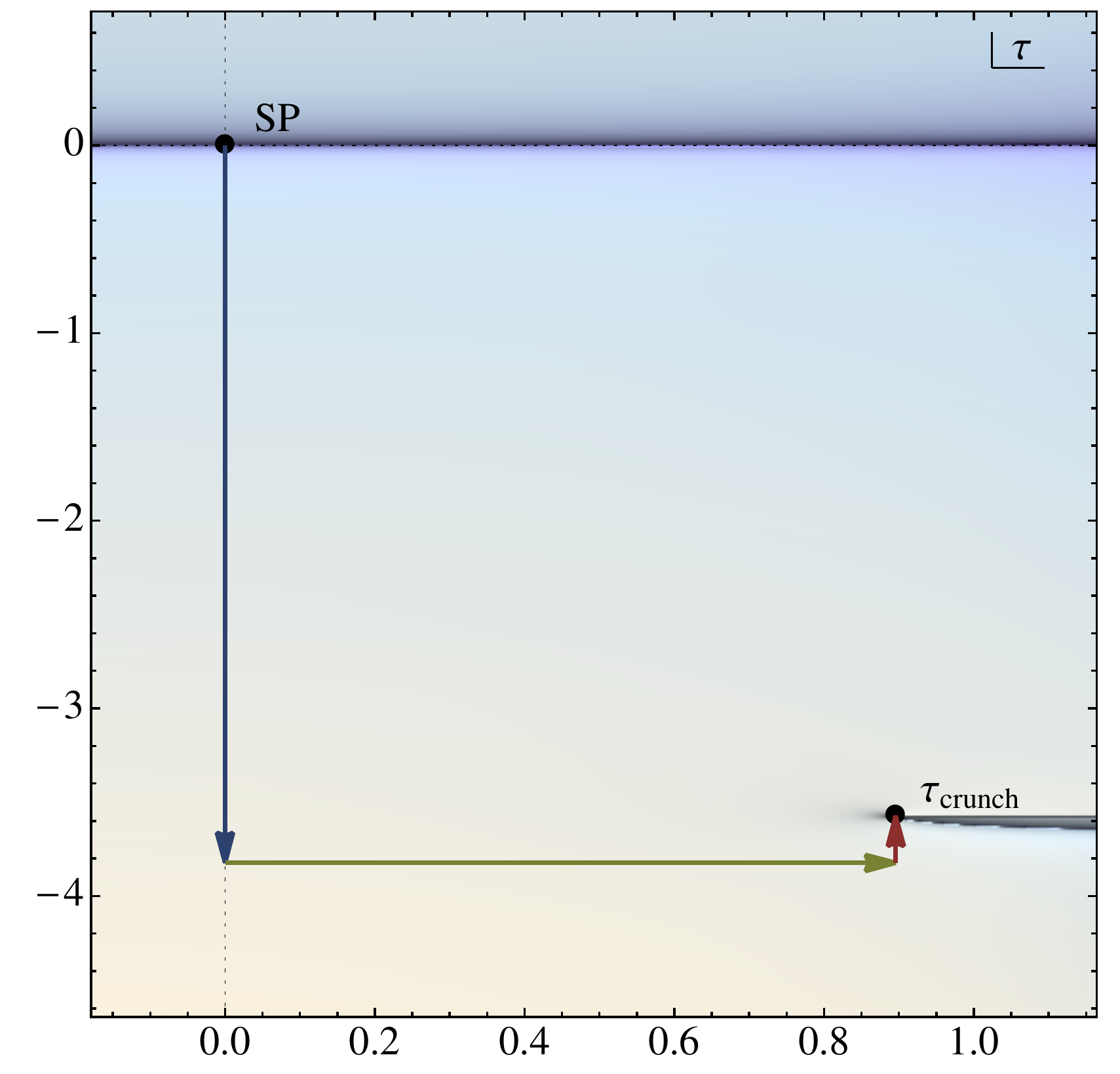}
\end{minipage}%
\begin{minipage}{\smallWidthRight} \flushleft
\includegraphics[width=\smallWidthRight]{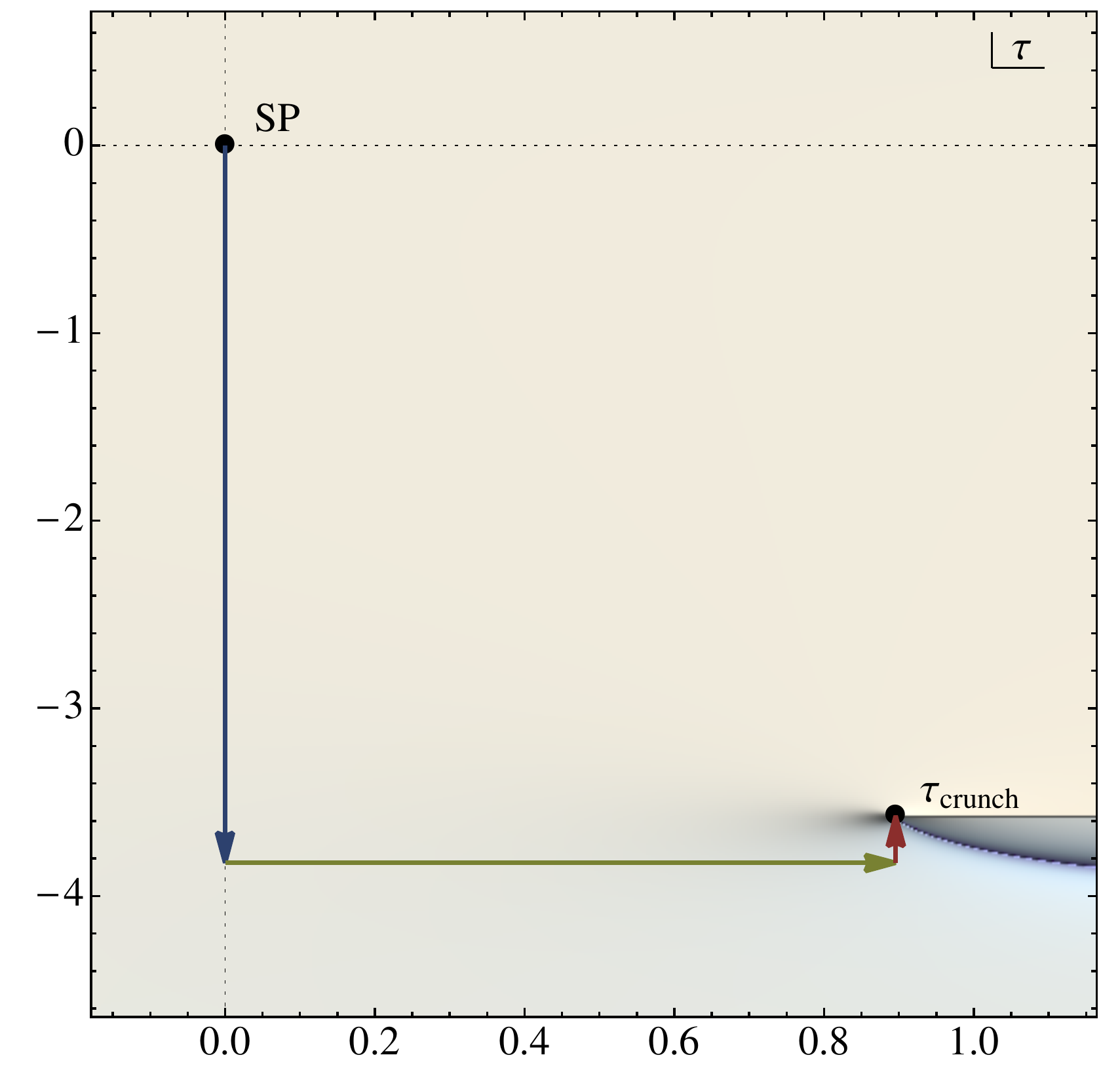}
\end{minipage}%
\caption{\label{Fig:CyclicEkp1} An ekpyrotic instanton on the dark energy plateau, with $H=1$, $c=3$, $\phi_{SP}^R=1$, $\phi_{SP}^I \approx -1.974$ and $\tau_{crunch} \approx 0.8956 - 3.573 i.$}
\end{figure}

\begin{figure}[h]%
\begin{minipage}{\smallWidthLeft} \flushleft
\includegraphics[width=\smallWidthRight]{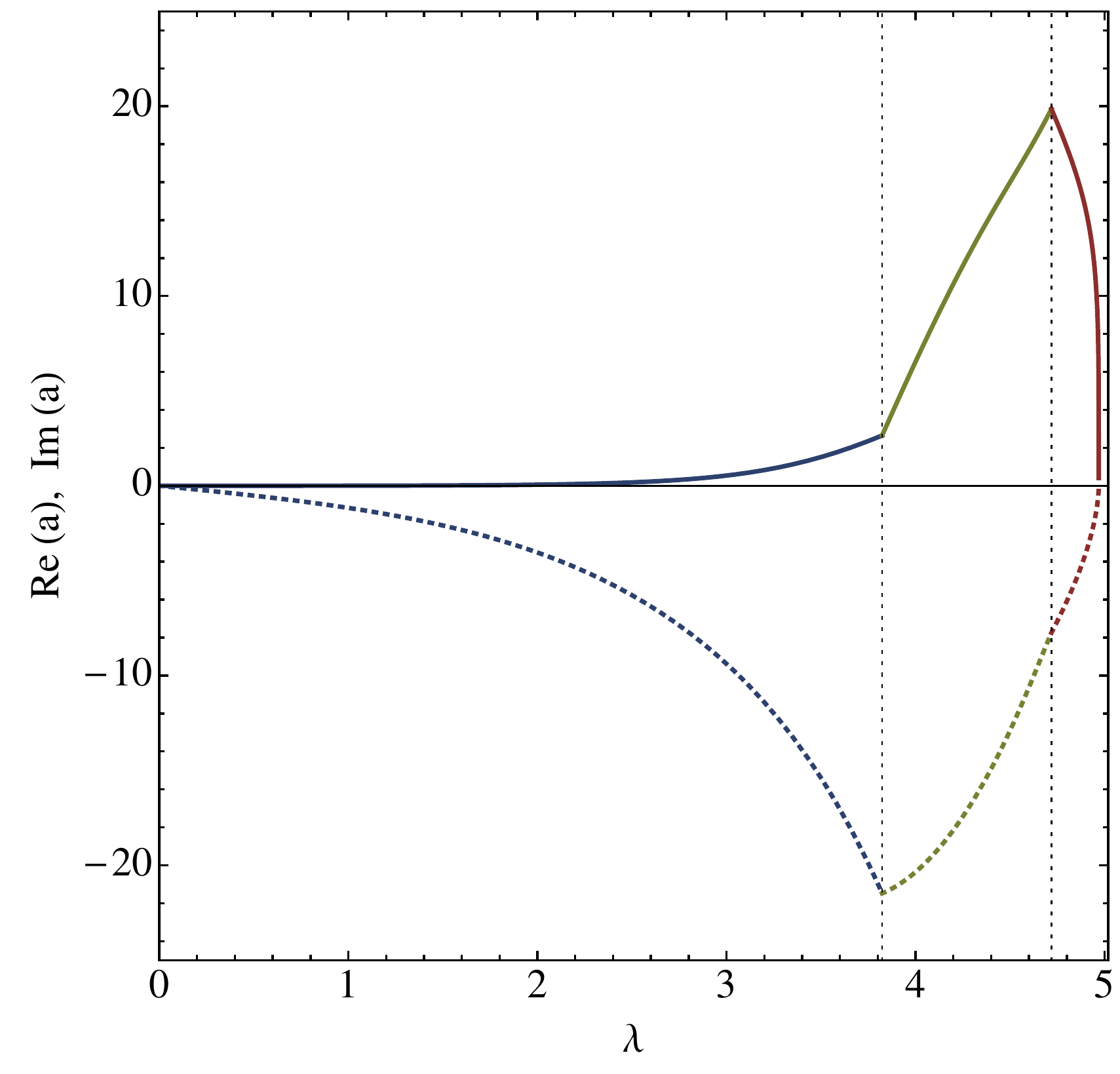}
\end{minipage}%
\begin{minipage}{\smallWidthRight} \flushleft
\includegraphics[width=\smallWidthRight]{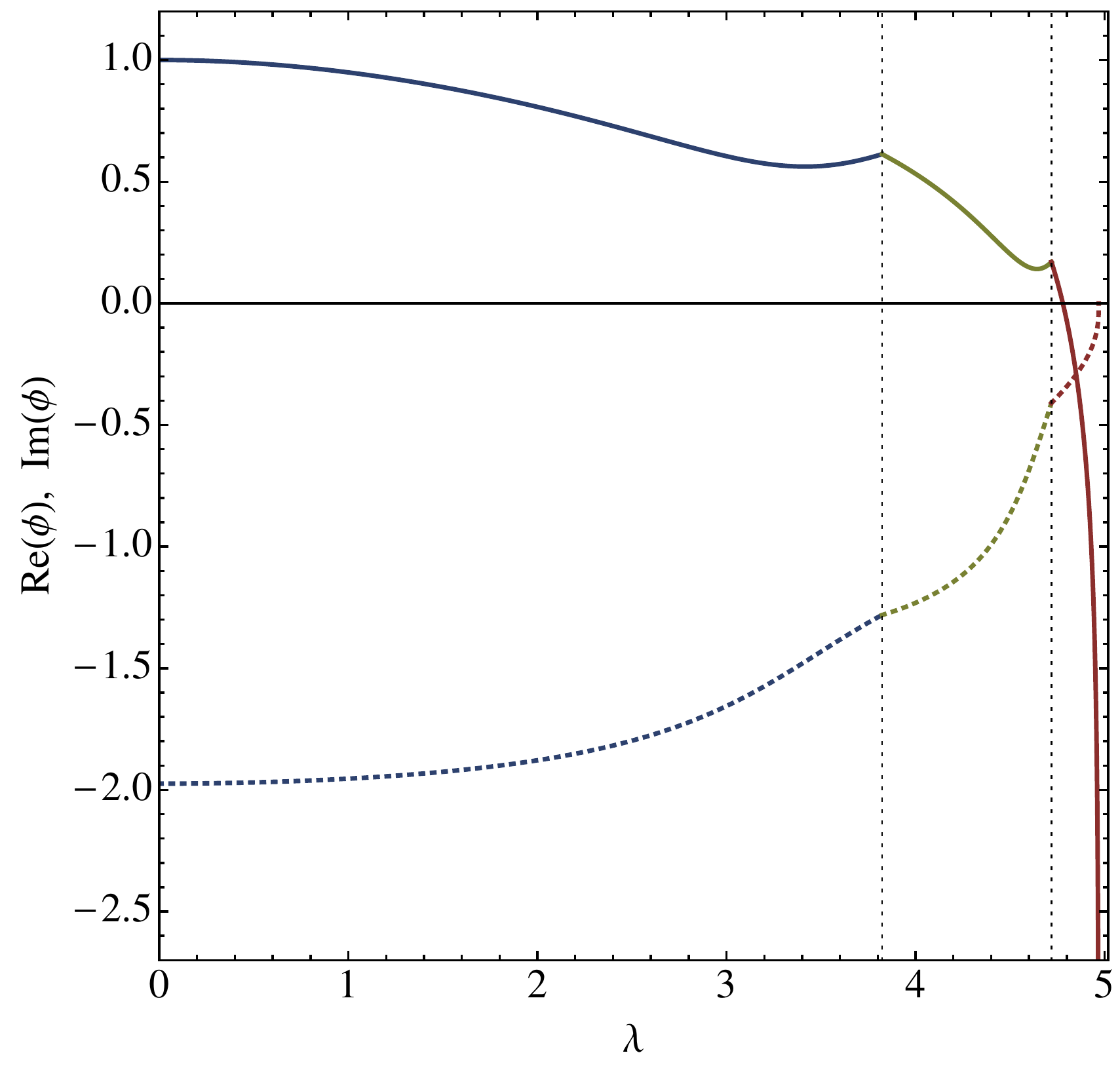}
\end{minipage}%
\caption{\label{Fig:CyclicEkp2} The real (solid) and imaginary (dotted) parts of the scale factor (left panel) and scalar field (right panel) for the instanton depicted in Fig. \ref{Fig:CyclicEkp1}, with an integration contour indicated by the arrows in that figure.}
\end{figure}

So much for the inflationary-type instantons, whose existence is unsurprising. Given the results of the previous section, it is now equally unsurprising that ekpyrotic instantons exist at negative $\phi,$ where the addition of a small constant to the potential is essentially irrelevant. The ekpyrotic instantons in this region of the potential are virtually identical to those discussed in section \ref{section:ekpyrotic}, and thus we need not discuss these further here. However, what is more surprising is that ekpyrotic-type instantons continue to exist in the region where the potential becomes positive. Fig. \ref{Fig:CyclicEkp1} shows graphs of such an instanton for $\phi_{SP}^R=1,$ i.e. for the same value of $\phi_{SP}^R$ used above to illustrate the inflationary-type instantons. Of course, the value of the imaginary part $\phi_{SP}^I$ is different in the ekpyrotic case. As Fig. \ref{Fig:CyclicEkp1} shows, the lines of real scale factor and real scalar field only start overlapping shortly before the crunch. In other words, the universe becomes real in the approach to the crunch by virtue of the ekpyrotic attractor, even though this instanton also traverses the dark energy plateau. By specifying an integration contour, we can again represent the evolution of the real and imaginary parts of $a$ and $\phi$ -- this is done in Fig. \ref{Fig:CyclicEkp2}, along the contour indicated by the coloured arrows in Fig. \ref{Fig:CyclicEkp1}. Along the first segment, running vertically down from the origin, the spacetime is again a portion of Euclidean Anti-de Sitter space, in analogy with the co-existing inflationary-type instanton. The second (horizontal) segment is once more fully complex, and it is only along the third segment that the universe approaches a real classical history. This happens in exactly the same way as it did in a pure exponential potential in section \ref{section:ekpyrotic}: the instanton approaches a real scaling solution up to complex-valued correction terms which die off as long as $\epsilon>3,$ cf. Eqs. \eqref{eq:ekpyroticattractor1} - \eqref{eq:ekpyroticattractor2}.

\begin{figure}[t]
\begin{minipage}{\thirdWidthLeft}
\includegraphics[width=\thirdWidthRight]{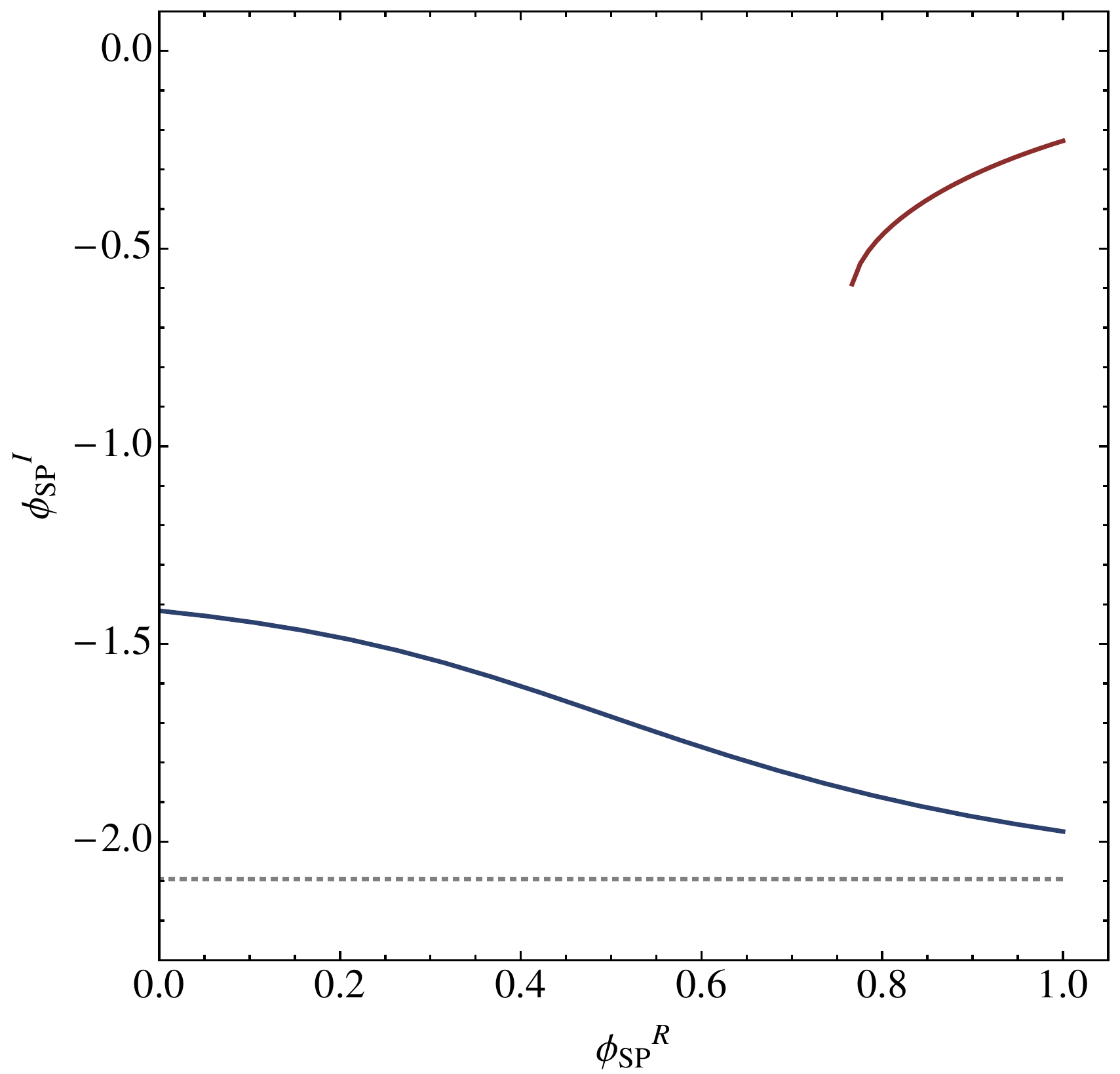}
\end{minipage}%
\begin{minipage}{\thirdWidthLeft}
\includegraphics[width=\thirdWidthRight]{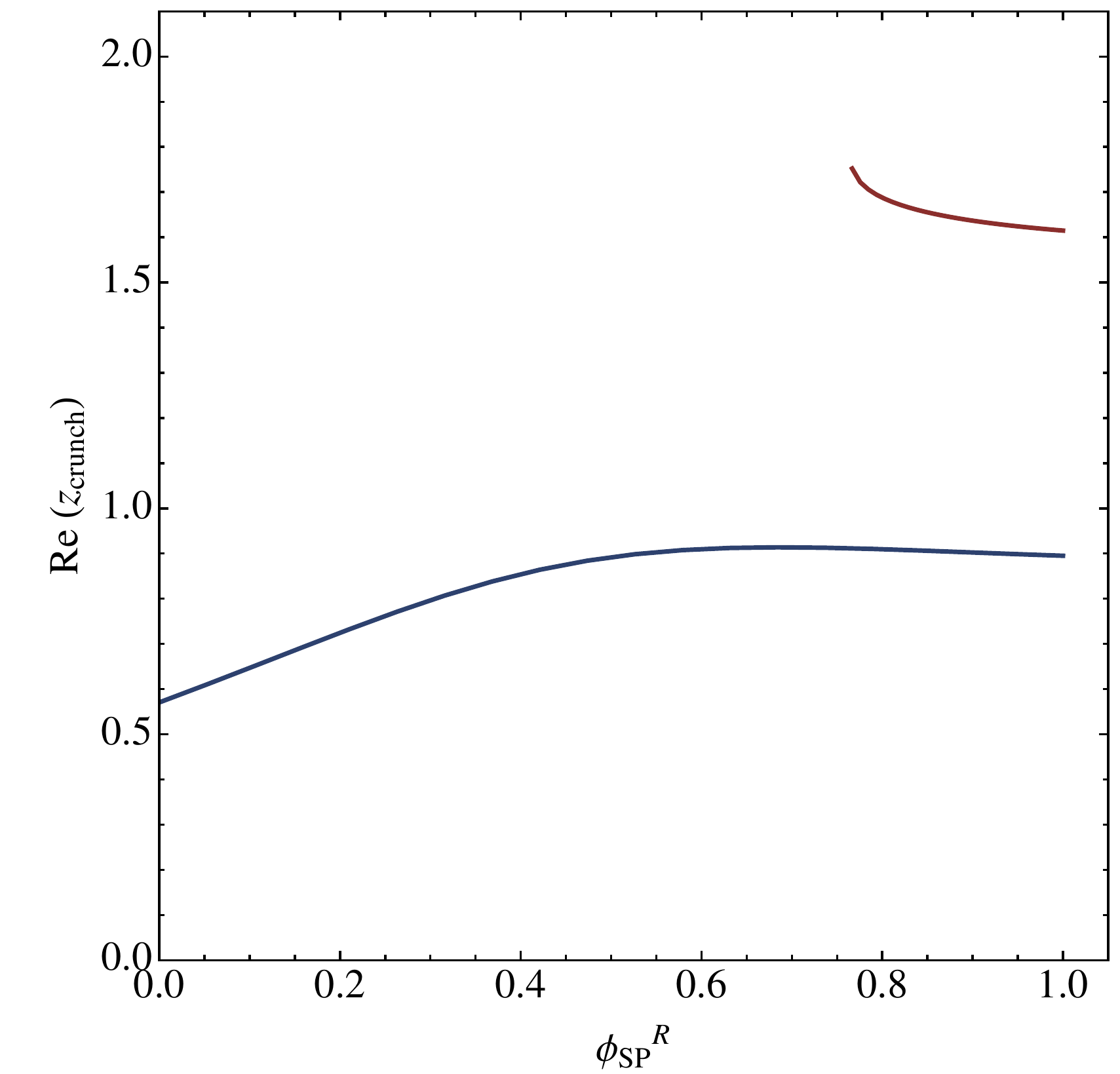}
\end{minipage}%
\begin{minipage}{\thirdWidthRight}
\includegraphics[width=\thirdWidthRight]{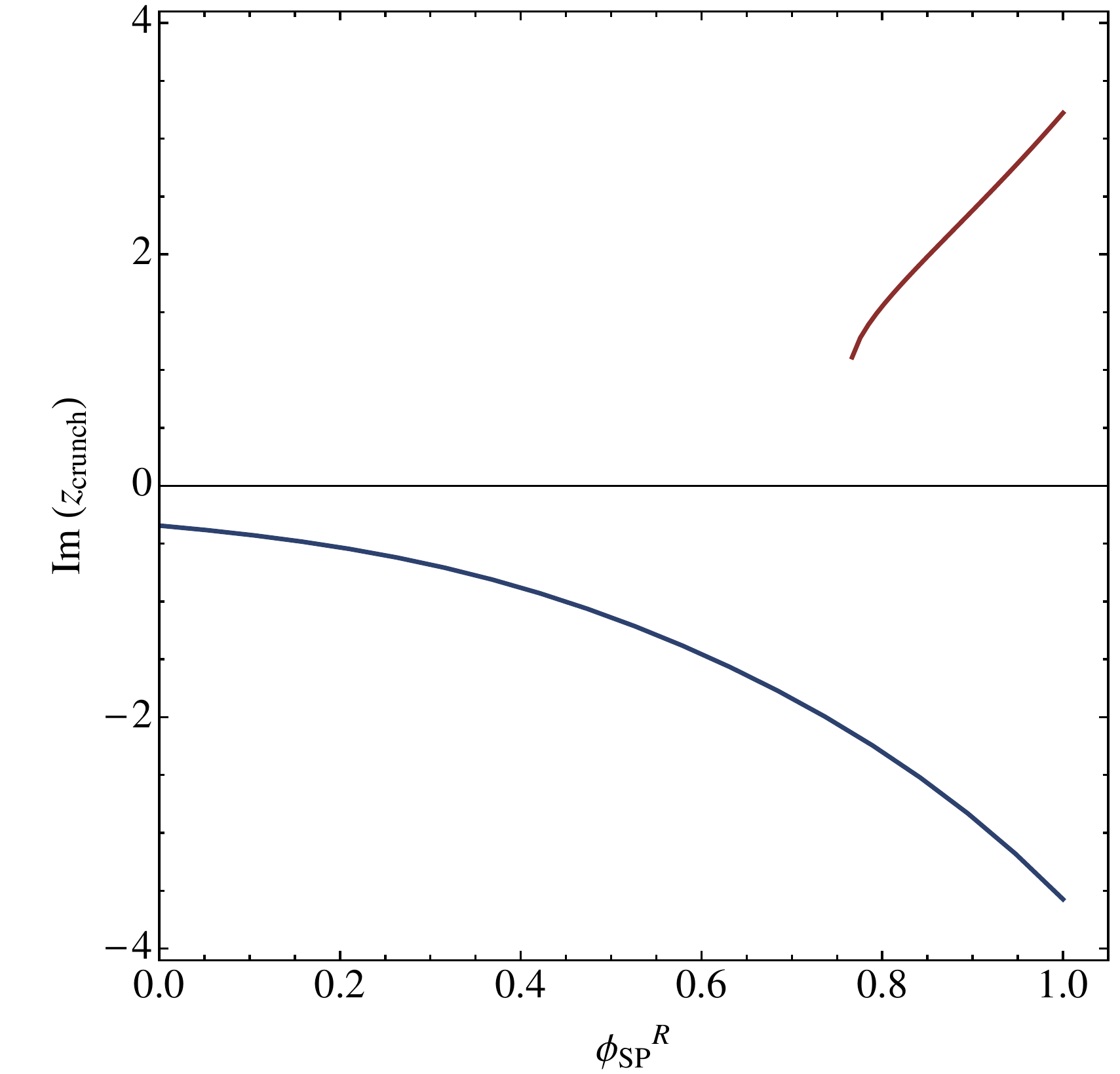}
\end{minipage}%
\caption{ \label{Fig:Cyclic3} The imaginary part $\phi_{SP}^I$ and the location of the crunch for ekpyrotic (blue) and inflationary (red) instantons in a cyclic potential, all as functions of $\phi_{SP}^R$. As the potential grows, $\phi_{SP}^I$ appears to approach the new asymptotic value of $-2\pi/c$ (dotted grey line).}
\end{figure}

The two instantons that we have just presented are not isolated examples. In fact, in our numerical analysis we have not found any obstruction at large $\phi_{SP}^R$ that prevents ekpyrotic instantons from occurring -- however, at large $\phi$ finding ekpyrotic-type instantons numerically becomes increasingly delicate. Fig. \ref{Fig:Cyclic3} shows the tuned values for $\phi_{SP}^I$ as well as the crunch locations in the range $0 \leq \phi_{SP}^R \leq 1,$ both for inflationary (red curves) and ekpyrotic-type (blue curves) instantons. Note that the inflationary-type instantons cease to exist for $\phi_{SP}^R \lesssim 0.77,$ which is due to the fact that below that value the phase of accelerated expansion becomes too short for the universe to reach classicality. There is an interesting feature to the graph of the imaginary part $\phi_{SP}^I,$ namely that as the potential becomes positive, on the ekpyrotic branch $\phi_{SP}^I$ is no longer constant (as it was for the families of pure ekpyrotic instantons in section \ref{section:ekpyrotic}). Rather, $\phi_{SP}^I$ appears to approach a new asymptotic value at large $\phi.$ In fact, the values of $\phi_{SP}^I$ seem to approach $-2\pi/c = -2\pi/3,$ which implies that for these solutions the potential is very nearly real at the South Pole of the instanton. This large $\phi$ regime clearly deserves further study. 

\begin{figure}[]
\centering
\begin{minipage}{\fullWidth}
\includegraphics[width=\fullWidth]{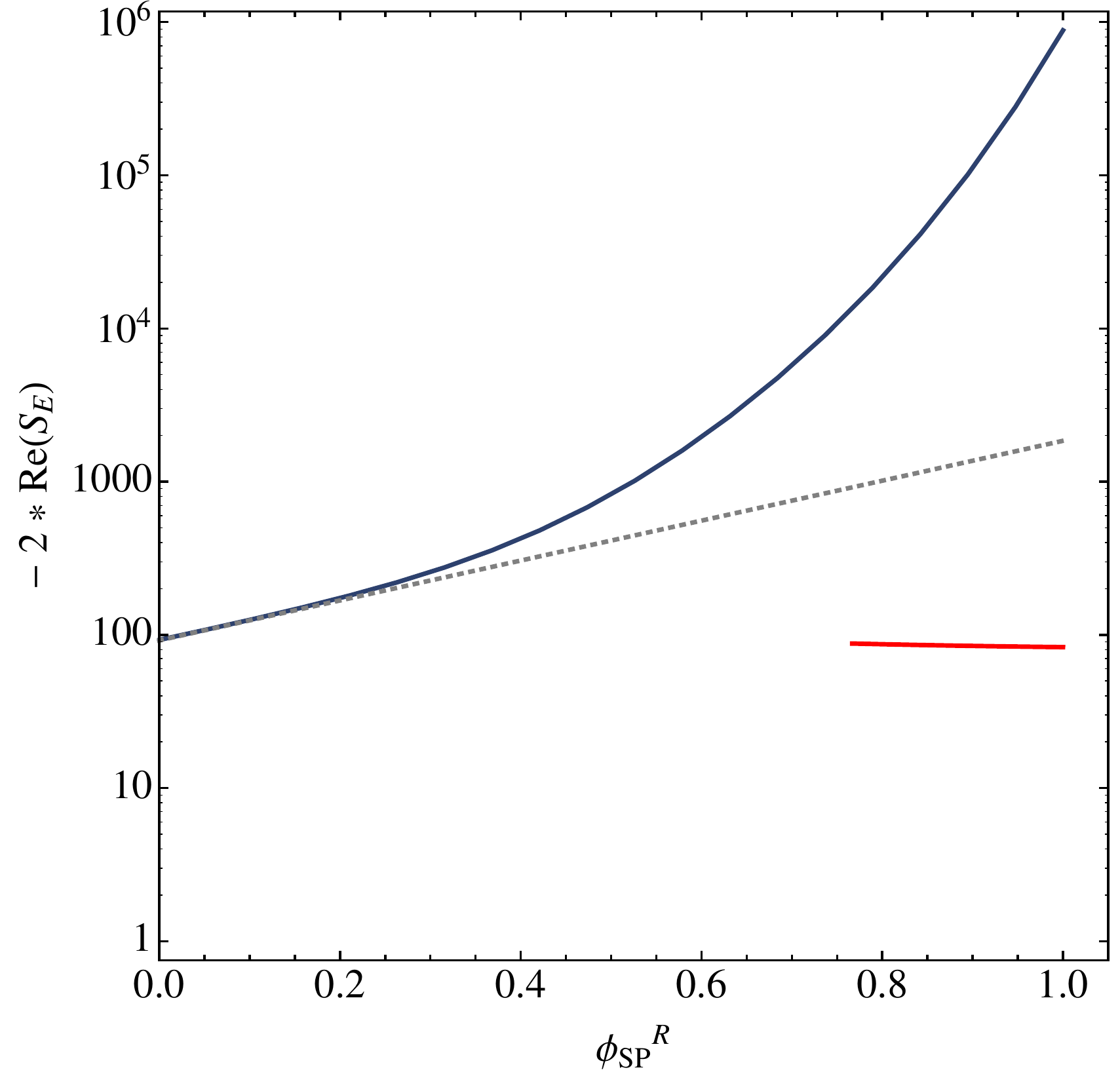}
\end{minipage}%
\caption{\label{Fig:Cyclic4} The logarithm of the relative probability for the two families of instantons, as a function of $\phi_{SP}^R$: ekpyrotic-type instantons are depicted in blue while inflationary ones are in red. The probability for ekpyrotic instantons can be seen to be vastly higher. Moreover, with increasing $\phi_{SP}^R$ their probability grows even faster than the naive $e^{c\Delta \phi}$ scaling that would apply in the pure ekpyrotic case (and that we have indicated by the dotted grey line).}
\end{figure}

In Fig. \ref{Fig:Cyclic4} we plot the logarithm of the relative probabilities $-2 Re(S_E)$ for these two families of instantons. From section \ref{section:inflation} we know that in the inflationary case the real part of the action is very well approximated by the action of half a Euclidean 4-sphere, $-2 Re(S_E) \approx 24 \pi^2/V(\phi_{SP}^R)$, and we have used this approximation to obtain the red curve. The blue curve applies to the ekpyrotic-type instantons, and it can be seen that in the region of overlap, these instantons yield a much higher probability than the inflationary ones. In fact, the probability grows even faster with increasing $\phi$ than the naive $e^{c\phi}$ scaling suggested by the shift symmetry in the pure ekpyrotic case (and indicated by the grey dotted line). This may be due to the fact that the dark energy plateau causes the instantons to grow even bigger than in the pure ekpyrotic case, in turn rendering the action larger in magnitude. However, an analytic understanding of the large $\phi$ behaviour is currently lacking, and represents an interesting question for the future.

\section{Discussion} \label{section:Discussion}

Two of the central goals of cosmology are to understand the initial conditions and the classicality of our universe. The no-boundary proposal provides an attractive framework for discussing both of these issues. In our earlier paper \cite{Battarra:2014xoa}, we identified a new class of instantons, called ekpyrotic instantons, that arise in theories with a steep and negative potential. These instantons can explain how the universe became classical by virtue of the ekpyrotic attractor mechanism, and they describe the emergence of a universe in the ekpyrotic contracting phase. In the present paper, we have discussed their properties in detail, and shown that in the cyclic universe potential, ekpyrotic instantons also arise, even on the dark energy plateau. In fact, on the dark energy plateau ekpyrotic instantons co-exist with inflationary-type instantons (where the universe becomes classical due to the dark energy/quintessence attractor), with the ekpyrotic-type instantons having a much higher probability of occurring.

Our results imply that in a potential energy landscape containing a variety of inflationary, ekpyrotic and cyclic regions (cf. also \cite{Lehners:2012wz}), the most likely initial conditions are represented by ekpyrotic instantons, describing the emergence of a classical universe in an ekpyrotic contracting phase  - whether a pure ekpyrotic or a cyclic region comes out as preferred depends on the details of the potential, in particular on the possible field ranges for the ekpyrotic and dark energy regions. We should point out that there is no contradiction in having a universe arise ``out of nothing'' in a contracting phase, as the ekpyrotic contracting phase is preceded by a complex/Euclidean geometry where space and time are ``created'' in the first place, cf. Fig. \ref{fig:carafe} for a heuristic picture.

There are many possible extensions of the present work: the most obvious one is that of adding a bounce, so that the ekpyrotic phase goes over into an expanding, radiation dominated phase of the universe. In our current work, we did not consider any dynamics that could have led to such a bounce. We see no obstacle in principle for doing so, although the numerical analysis may be complicated. Moreover, as the results of section \ref{section:cyclic} suggest, there may be a new asymptotic regime for ekpyrotic instantons originating on the dark energy plateau of a cyclic universe potential. It would certainly be of interest to obtain an analytic understanding of this regime. Finally, we have not included any fluctuations in the present work. Given their importance in testing cosmological theories, the inclusion of fluctuations represents another worthwhile goal for future studies. Our work has highlighted the importance of having an attractor mechanism in order to obtain a classical background spacetime. In light of this, it will be desirable to focus on the recently described stable mechanisms for producing density perturbations \cite{Li:2013hga,Qiu:2013eoa,Fertig:2013kwa,Ijjas:2014fja}, rather than on the older unstable entropic scenario \cite{Finelli:2002we,Notari:2002yc,Lehners:2007ac,Koyama:2007mg,Koyama:2007ag,Lehners:2007wc,Lehners:2009ja} that led to the so-called phoenix universe \cite{Lehners:2008qe,Lehners:2009eg,Lehners:2011ig}.


\acknowledgements

We thank Michael Koehn for his careful reading of our manuscript. We gratefully acknowledge the support of the European Research Council via the Starting Grant Nr. 256994 ``StringCosmOS''.

\bibliographystyle{utphys}
\bibliography{NBWFBib}

\end{document}